\documentclass[prb,twocolumn,showpacs,preprintnumbers]{revtex4}
\usepackage{graphicx}
\usepackage{dcolumn}

\newcommand{\beq}{\begin{equation}}
\newcommand{\eeq}{\end{equation}}
\newcommand{\beqa}{\begin{eqnarray}}
\newcommand{\eeqa}{\end{eqnarray}}

\begin{document}

\title{The doping-driven evolution of the superconducting state
of a doped Mott insulator:\\
a key for the high temperature superconductivity.}
\author{M. Civelli$^1$}
\affiliation{$^1$ Theory Group, Institute Laue Langevin, Grenoble,
France}

\begin{abstract}
High-temperature superconductors at zero doping can be considered
strongly correlated two-dimensional Mott insulators. The
understanding of the connection between the superconductor and the
Mott insulator hits at the heart of the high-temperature
superconducting mechanism. In this paper we investigate the
zero-temperature doping-driven evolution of a superconductor
towards the Mott insulator in a two dimensional electron model,
relevant for high temperature superconductivity. To this purpose
we use a cluster extension of dynamical mean field theory. Our
results show that a standard (BCS) d-wave superconductor,
realized at high doping, is driven into the Mott insulator via an
intermediate state displaying non-standard physical properties.
By restoring the translational invariance of the lattice, we give
an interpretation of these findings in momentum space. In
particular, we show that at a finite doping a strong
momentum-space differentiation takes place: non-Fermi liquid and
insulating-like (pseudogap) character rises in some regions
(anti-nodes), while Fermi liquid quasiparticles survive in other
regions (nodes) of momentum space. We describe the consequence of
these happenings on the spectral properties, stressing in
particular the behavior of the superconducting gap, which reveals
two distinct nodal and antinodal energy scales as a function of
doping. We propose a description of the evolution of the
electronic structure while approaching the Mott transition and
compare our results with tunneling experiments, photoemission and
magnetotransport on cuprate materials.
\end{abstract}

\date{\today}
\maketitle
\section{Introduction}
\label{sec1}

Since the discovery in 1986 of the high temperature (H-TC)
superconductivity in copper oxide based
materials\cite{bednorz86}, much effort has been devoted by the
scientific community to understand  the physics of this
phenomenon, but its key-ingredients remain still unknown. On the
experimental side, the complexity of these materials, which
present a rich phase diagram with many competing instabilities,
has made impossible to perform resolutive experiments. On the
theoretical side, many theories have been proposed, but a general
consensus has not been achieved yet because of the lack of tools
capable of perform reliable calculations. The strongly-interacting
many-body nature of these systems makes in fact standard
techniques hardly applicable. The recent discovery of
high-temperature superconductivity in a new family of
materials\cite{kamihara08} (for a perspective see e.g.
ref.\cite{day08}), with a general composition
LaFeAsO$_{1-x}$F$_{x}$, different from the typical Cu-O plane
structure of all known H-TC compounds, has revitalized the
attention on the origin of the H-TC mechanism.

In early times, P.W. Anderson\cite{anderson} suggested that H-TC
superconductivity is the result of doping a Mott insulator with a
small number of carriers. The understanding of the way a
superconductor can transform into a Mott insulator has been a
longstanding open problem and its connection with a
strongly-interacting superconductivity has been long sought. The
aim of this article is to investigate the doping-driven evolution
of the superconducting state of a doped Mott insulator in a two
dimensional lattice model of strongly correlated electrons. To
this purpose we employ a cluster extension of Dynamical Mean
Field Theory (DMFT) (for a review see ref.\cite{revmodmft}).

\subsubsection{Dynamical Mean Field Theory and its Cluster
extensions to study H-TC superconductivity}

DMFT has in recent years proved very successful in revealing the
physics of the Mott transition in three dimensional compounds,
like e.g. Vanadium Oxide. DMFT is a self-consistent mean field
method, which maps a lattice problem onto a single-impurity
embedded in a bath of free electrons. It is exact in the limit of
infinite dimension, where the physics is purely local and DMFT
virtually provides the complete solution. Its strength stands in
the ability of treating on the same footing high and low energy
physics, capturing in one framework both the Mott insulating
state as well as the metallic state, and hence allowing a
non-perturbative investigations. A good summary of the physics
captured by DMFT is described in the DMFT phase diagram. For
example, single-site DMFT in conjunction with electronic
structure methods has provided results in excellent agreement
with experiments on numerous three dimensional compounds (with
valence electrons in the  sp shell\cite{zein05}, 3d shell, 4d
shell\cite{lich02}, 4f shell\cite{held01,haule05} and 5f
shell\cite{nature01}). For some recent reviews of this field see
e.g. ref.\cite{rmp06}.

In spite of these successes in genuine strongly correlated
materials, DMFT has also well understood limitations, mainly due
to its purely local character. It does not capture for example
the feedback of collective modes, such as magnetic fluctuations
on the single particle quantities. One consequence of this
shortcoming is that the effective mass diverges as the Mott
transition is approached, while in finite dimensions exchange
effects should remove this enhancement in the region where the
renormalized kinetic energy is of the order of the
super-exchange-interaction or the temperature. These effects are
negligible whenever strong frustration, finite temperature or
orbital degeneracy help justifying a local approximation. The DMFT
approach, however, breaks down in the cuprate-based
superconductors, whose physics is two-dimensional and which have
a large super-exchange. As the Mott transition is approached in
fact, experimental evidence shows that the effective mass remains
finite and that physical properties are strongly momentum
dependent (see for example Angle Resolved Photoemission
Spectroscopy ARPES results\cite{damascelli,campuzano} ). Being a
local theory, DMFT is not able by construction to describe a
strong momentum dependence, hence it is not suitable to study real
finite-dimensional H-TC superconductors.

Recently, extensions of DMFT capable to go beyond the local
approximation have been developed. For a review of different
cluster extensions of DMFT see e.g.
ref.\cite{maier00,lich00,jarrell-rmp04,tremblay06,biroli04}. The
Cellular Dynamical Mean Field Theory (CDMFT\cite{cdmft}) is a
cluster method which retains the real space interpretation of
single-site DMFT but incorporates non-perturbative
momentum-dependence. It replaces the single site of DMFT by a
cluster of impurities, offering the possibility of well
describing short-ranged spatial correlation and providing a first
correction to momentum-dependent properties. It also allows to
naturally describe phases characterized by a spatially dependent
order parameter, like for example antiferromagnetism (AF) or
d-wave superconductivity (dSC). The first cluster-DMFT studies of
the Hubbard model in two dimensions were carried out on a 2
$\times$ 2 plaquette in ref.\cite{lich00,maier00}. These groups
have shown the existence of antiferromagnetism, pseudogap and
superconductivity, focusing on relatively small values of the
on-site interaction (mostly U =4t), for which the Fye-Hirsch
Quantum Monte Carlo (QMC) \cite{fye-hirsch86,fye-hirsch89} method
is applicable in implementing the cluster DMFT. One school of
thought, for example in
ref.\cite{jarrell05,jarrell-scalapino06,senechal04}, advocates
the study of large clusters, at present possible for relatively
small interactions. In this latter case, however, it is not yet
known what aspects of the Mott physics are captured. Cluster DMFT
implemented on the 2$\times$2 plaquette for stronger values of U
(the Mott regime) were the subject of several
publications\cite{bpk,marce05,bumsoo06,venky05} on the Hubbard
Model.

\subsubsection{A 2$\times$2 plaquette Cellular DMFT study}
Following the cluster DMFT studies mentioned above, we focus in
this paper on a two-dimensional 2$\times$2 cluster-plaquette.
According to our viewpoint, elucidating the physical content of
the mean field theory on a plaquette is a very important step to
accomplish before proceeding to realistic studies of the actual
instabilities that govern the phase diagram of the model. The
work in
references\cite{senechal05,massimo06,aichhorn06,aichhorn06b} on
small clusters have in fact shown that there are several competing
phases and possible phase-separation, which could lead to
complicate patterns in real space depending on the boundary
conditions or on various terms not explicitly included in the
Hamiltonian. A proper investigation of those states in the
framework of DMFT-based methods, requires therefore an
understanding of the pure phases of the simple cluster-plaquette.

Implementation of CDMFT on a 2$\times$2 plaquette for large values
of the interaction has already revealed several interesting
effects not present in single-site DMFT, indicative of a very rich
physics. For example, publications\cite{bpk,marce05} reveal that
the approach to the Mott transition as a function of doping occur
in a very anisotropic way in momentum space. In the Hubbard model
(relevant for the description of hole-doped cuprates), the
spectral weight disappears much more rapidly in the antinodal
than in the nodal region of momentum space, where quasiparticles
survive close to the Mott transition at a fixed temperature.
Evidence for the formation of a pseudogap in the one electron
spectra can be seen in other cluster-DMFT
studies\cite{huscroft-jarrell01,stanescu03,senechal05}. CDMFT
allows a natural interpolation of the nodal and antinodal spectral
function, and it opens the way to study the $k$-dependence of the
one-electron spectra in the Mott insultor\cite{bumsoo06}. The
origin of the pseudogap can be traced to the growth of the  self
energy in certain parts of $k$-space, where eventually at zero
temperature lines of poles of the self-energy (i.e. zeroes of the
one-particle Green's function) appear\cite{tsvelik,tudor,berthod}.
In this latter case, the Fermi arcs observed in ARPES are
interpreted as the result of a binding of segments of Fermi
surface and of a lines of zeros of the Green's function. Results
along those lines beyond mean field theory can be found in
ref.\cite{stanescu-2007-75}. Within CDMFT not only the
under-doped regime is anomalous. At optimal doping, where the
maximum of the critical temperature occurs, a maximum in the one
electron scattering rate and the presence of power laws in the
optical conductivity are found\cite{haule06}.

In this paper we study the superconducting state which arises upon
doping a Mott insulator by implementing the 2$\times$2 plaquette
CDMFT with exact diagonalization\cite{krauth-caffarel} (ED-CDMFT)
at zero temperature (Lanczos method). Our work is complementary
to the 2$\times$2 plaquette CDMFT study of
ref.\cite{haule-ctqmc}, where the continuous time quantum Monte
Carlo (CTQMC) at finite temperature was used as impurity solver.
As compared with QMC methods, ED allows to extract in an unbiased
way real frequency quantities, which can be more easily
physically interpreted. This will turns out to be fundamental in
interpreting our CDMFT results and making important connection
with experiments (as we will widely present throughout the paper).
The Lanczos method, on the other hand, is limited by the finite
size of the system used to describe an effective impurity model,
associated with the original lattice model (see the following
discussion on the ED-CDMFT procedure). If rightly implemented,
however, it is able to well capture the difficult physics of the
finite dimensional Mott transition (see for example
references\cite{marce,rmp06,civelli-2007}). We will in fact show
that the physics described by our results well compares (at least
at the qualitative level) with the CTQMC-CDMFT results of
reference\cite{haule-ctqmc}, where finite size limitations are
absent.

Earlier studies\cite{venky05} with ED-CDMFT have compared the
phase diagram of the two dimensional Hubbard Model with the one of
cuprate-based materials, studying the interplay between the
superconducting and antiferromagnetic instabilities, either in
the hole and electron doped sides. Here we complete the work
presented in a previous short publication\cite{marce08}, and we
focus on doping-driven evolution of the superconductor towards the
Mott insulator. In particular we show that, in a small region
around optimal doping, novel properties appear in the
superconducting state which are not ascribable to the standard
Bardeen Cooper Schrieffer (BCS) theory of superconductivity. The
most striking phenomenon, as widely presented in
publication\cite{marce08}, is the rising of two distinct
doping-dependent energy scales, which do not fit neither into the
framework of BCS approaches nor within the most popular theories
of H-TC superconductivity, like for instance the resonating
valence bond theories (RVB\cite{anderson}, for a recent review see
e.g.\cite{patrickrmp}). We interpret our cluster results
extracting the momentum-dependence (i.e. restoring the
translational invariance of the lattice) of the electronic
structure as a function of doping. According to our physical
picture, the Fermi liquid description holds at low energies in
the nodal region of momentum space, where the d-wave
superconducting gap is zero. We show that this fact is best
described in momentum space introducing a periodizing scheme
based on the local (within cluster)
self-energy\cite{biroli02,bpk}. On the other hand, in the
antinodal region, where the d-wave gap is maximal, besides the
superconducting contribution, a further contribution to the
one-particle gap appears also in the normal component at the
finite critical doping. In this case we show that the
periodization of another local quantity, the irreducible
two-point cumulant, offers a more adequate
description\cite{tudor,tudor06}. We introduce therefore a
mixed-periodization scheme, which was foreshadowed in a
phenomenological approach to the transport properties of cuprate
materials in the normal state\cite{PSK} and in a previous CDMFT
publication\cite{marce08}, and confront the resulting picture of
the electronic evolution as a function doping with
photo-emission, scanning tunneling experiments and
magnetotransport on cuprate materials.

\subsubsection{Set-up of the paper}

The paper consists of two main parts. The first comprises sections
\ref{sec1}-\ref{sec4}. In section \ref{sec1}, after this
introduction, we present the model and briefly explain the
ED-CDMFT method. In section \ref{sec2} we present the raw cluster
DMFT results, mainly stressing their evolution with doping and
showing the appearance (at small doping) of two distinct energy
scales. It is not however straightforward to interpret these
cluster quantities in terms of physical observables. Therefore, in
section \ref{sec3}, we cope with the problem of extracting
lattice quantities (which can be more easily compared with
experimental results) from the corresponding cluster ones, i.e. we
present and justify different periodization-methods. We show that
nodal and antinodal regions of momentum space turn out having
different physical properties which require different
periodization schemes to be rightly described. In section
\ref{sec4} we characterize the different properties in the nodal
and antinodal regions, according to the periodization schemes
introduced, clearly linking them with spectroscopy experiments.

The second part starts from section \ref{sec5}, where, in order
to be able to compare our results with experiments resolved in
momentum space, we introduce a more general mixed-periodization
scheme. This latter, interlacing the nodal and antinodal
properties in all the momentum space, allows us to propose a
description of the evolution of quasiparticle spectra in
approaching the Mott transition. Spectra, which are directly
comparable with photo-emission on the cuprates, are presented in
detail in section \ref{sec6}. In section \ref{sec7} we present
peculiar characteristics of the spectra, the so called "kink"
feature, also comparing with photoemission results. Finally, in
section \ref{sec8}, by applying a phenomenological Boltzmann
approach on our CDMFT mixed-periodization result, we derive the
Hall resistivity, which is a direct probe of the charge carries
in the system. We compare its evolution as a function of doping
with magnetotransport experiments and draw conclusions on a
topological phase transition of the Fermi surface, which takes
place at low doping within the mixed-periodization scheme
introduced in section \ref{sec5}. We finally derive our
conclusions in section \ref{sec9}.

%------------------------------------------------------------
\subsection{The Model}

We consider the one-band two dimensional Hubbard Model on a
square lattice:
\begin{equation}
\mathcal{H} = -\sum_{\langle i,j\rangle,\sigma} t_{ij}\,
(c^{\dagger}_{i,\sigma} c_{j,\sigma} + h.c.) + U \sum_i
n_{i\uparrow}n_{i\downarrow} -\mu\sum_i n_i, \label{hamiltonian}
\end{equation}
which is universally considered a minimal description of
cuprate-based materials\cite{hubbard63}. Here $c_{i,\sigma}$
($c^{\dagger}_{i,\sigma}$) are destruction (creation) operators
for electrons of spin $\sigma$, $n_{i\sigma} =
c^{\dagger}_{i\sigma} c_{i\sigma}$ is the density of electrons,
$\mu$ is the chemical potential tuning doping and $t_{ij}$ are
the orbital hopping integrals. For convenience's sake we consider
only the nearest neighbor amplitude $t=1$, and a next nearest
neighbor hopping $t^{\prime}= -0.3 t$. We set the on-site
repulsion $U=12t$, larger than the band-width $8t$, to be in the
Mott regime.

%====================================================================
%====================================================================
\subsection{ED-CDMFT procedure} Similarly to single-site DMFT
\cite{revmodmft}, in CDMFT the original model (eq.
\ref{hamiltonian}) is described in terms of an effective action
containing a Weiss dynamical field $\hat{\cal G}_{0}(\tau)$
describing the degrees of freedom outside the cluster (in the
bath) as a time dependent hopping within the cluster
\begin{equation}
S_{\rm eff}= \int_{0}^{\beta }d\tau d\tau'\Psi_{\tau }^{\dagger }
\left[ \hat{\cal G}_{0 \tau-\tau'}^{-1} \right] \Psi_{\tau'}+ U
\sum_{\mu }\int_{0}^{\beta } n_{\mu \uparrow }n_{\mu \downarrow
}d\tau. \label{Seff}
\end{equation}
$\mu=1..N_c$ ($N_c=4$ in the 2$\times$2 plaquette) labels the
degrees of freedom inside the cluster. For the case of a
$2\times2$ plaquette considered in this paper, a convenient
Nambu-spinor notation has been introduced:
\begin{equation}
\Psi_{}^{\dagger } \equiv
(c_{1\uparrow}^{\dagger},\dots,c_{4\uparrow}^{\dagger}
,c_{1\downarrow},\dots,c_{4\downarrow}) \label{spinor}
\end{equation}
With this notation the Weiss field $\hat{\cal G}_{0}$ is a
$8\times 8$ matrix with both normal (particle-hole) and anomalous
(particle-particle) components \cite{venky05}. Physically, this
action describes a cluster embedded in a self-consist bath of
free electrons with dSC correlations. In the CDMFT procedure, a
starting guess of the Weiss field $\hat{\cal G}_{0}$ is given as
input. Then the cluster single-particle propagator $\hat G_c$ is
computed trough the effective action eq. (\ref{Seff}) and the
cluster self-energy is determined through the Dyson's equation
\begin{equation}
\hat \Sigma_c = \hat {\cal G}_0^{-1} - \hat G_c^{-1}
\label{selfconsistency}
\end{equation}
Here,
\begin{equation}
\hat G_c\left( \tau ,\tau ^{\prime }\right) =\left(
\begin{array}{cc}
\hat G_{\uparrow }\left( \tau ,\tau ^{\prime }\right) & \hat
F\left( \tau ,\tau
^{\prime }\right) \\
\hat F^{\dagger}(\tau ,\tau ^{\prime }) & -{\hat G}_{\downarrow
}\left( \tau ^{\prime },\tau \right)
\end{array}
\right) \label{nambugreen}
\end{equation}
is an 8 X 8 matrix, $G_{\mu\nu,\sigma}\equiv\langle -T c
_{\mu\sigma}(\tau) c^{\dagger}_{\nu\sigma}(0)\rangle$ and
$F_{\mu\nu}\equiv\langle -T
c_{\mu\downarrow}(\tau)c_{\nu\uparrow}(0)\rangle$,
($\mu,\nu=1...N_{c}$ label sites of the cluster) are the normal
and anomalous cluster-Green's functions respectively.  From the
cluster self-energy $\hat \Sigma_c$, we use the CDMFT
self-consistency condition to re-compute the local cluster
Green's function $\hat{G}_{loc}(i\omega_n)= \sum_{K} \,\hat
G(K,i\omega_n) $, where
\begin{equation}
\hat G(K,i\omega_n)=\left[i\omega_n + \mu - \hat t(K) - {\hat
\Sigma_c}(i\omega_n)\right]^{-1} \label{Gloc}
\end{equation}
In eq.(\ref{Gloc}) $\hat t(K)$ is the Fourier transform of the
%superlattice (i.e. the lattice formed by the clusters)
hopping matrix defined on the lattice formed by the clusters and
the sum over $K$ is therefore performed over the Brillouin zone
reduced by the partition in clusters of the lattice\cite{venky}.
We finally re-derive a new Weiss field
\begin{equation}
\hat \mathcal{G}^{new}_0(i\omega_n)^{-1}=
\hat{G}^{-1}_{loc}(i\omega_n) + \hat \Sigma_c(i\omega_n)
\label{CDMFT-selfconsistency}
\end{equation}
and iterate until convergence is reached.

In practice, as we mentioned in the introduction, in order to
solve the cluster impurity problem, in this work (as in
ref.\cite{venky05,marce08}) we use the Exact Digonalization method
\cite{krauth-caffarel}.  A parametrized Anderson-impurity
Hamiltonian describes the action eq.(\ref{Seff}) and couples the
cluster impurity with a discrete number $N_{b}$ of bath orbitals
(we have fix throughout this work $N_b=8$, which is the limit in
practice accessible with standard computational resources) :
\begin{eqnarray}
\mathcal{H}_{\rm imp}&=&\sum_{\mu \nu \sigma}E_{\mu \nu
\sigma}c_{\mu \sigma }^{\dagger }c_{\nu \sigma}+  U \sum_{\mu}
n_{\mu\uparrow}n_{\mu\downarrow} + \nonumber \\
&&+\sum_{k \sigma}\epsilon _{k \sigma} a_{k\sigma
}^{\dagger}a_{k\sigma }  +\sum_{k\mu \sigma} V_{k\mu
\sigma}a_{k\sigma }^{\dagger}
c_{\mu \sigma} + {\rm h.c.}+ \nonumber \\
&&+\sum_{k\mu \sigma} V^{\rm sup}_{k\mu \sigma}a_{k\sigma
}^{\dagger} c^{\dagger}_{\mu \bar{\sigma}} +
 \sum_{k\mu \sigma} V^{\dagger \rm sup}_{k\mu \sigma} c_{\mu \bar{\sigma}} a_{k\sigma }
\label{Ham-imp-full}
\end{eqnarray}
Here $E_{\mu \nu \sigma}= -\mu \delta_{\mu\nu}$,  $E_{\mu \nu
\sigma}= -t \delta_{\mu, \nu\pm 1}$. Under the self-consistency
constrain eq.(\ref{Gloc}) and (\ref{CDMFT-selfconsistency}), the
bath-parameters $\epsilon_{k\sigma}$, $V_{k\mu\sigma}$ and $V^{\rm
sup}_{k\mu\sigma}$ are determined at each CDMFT-iteration by
fitting the Anderson-impurity Weiss field (eq.
\ref{CDMFT-selfconsistency}) with a $N_{b}$-pole bath function
$\hat{\mathcal{G}}^{new}_{N_b}(i\omega_n)= \imath
\mathbf{1}\omega- \hat{E}- \hat{\Delta}$
\begin{eqnarray}
& & \hat{\Delta} = \sum_{k}^{N_b}\, \mathbf{V}_{k}^{\dagger} (
\imath
\mathbf{1}\omega- \hat{\varepsilon}_{k} )^{-1} \mathbf{V}_{k}  \\
& &\mathbf{V}_{k}(2 \times 8)= \left(
\begin{array}{cc}
V_{k\mu\uparrow}   &   V^{sup}_{k\mu\uparrow}              \\
-V^{sup}_{k\mu\downarrow}    & -V_{k\mu\downarrow}      \\
\end{array}
\right)_{\mu=1,N_c} \\
& & \mathbf{E}_{k}(2 \times 2) = \left(
\begin{array}{cc}
\varepsilon_{k\uparrow}   &      0          \\
0                          & -\varepsilon_{k\downarrow}      \\
\end{array}
\right)
\end{eqnarray}
The fitting is obtained via a conjugate gradient minimization
algorithm, which uses a distance function\cite{marce}
\begin{equation}
f= \sum_{\mu\nu} \, \left| \hat{\mathcal{G}}^{new}_0(i\omega_n)-
\hat{\mathcal{G}}^{new}_{N_b}(i\omega_n) \right|_{\mu\nu} /
\omega_n \label{fdist}
\end{equation}
that emphasizes the lowest frequencies and it is computed on the
imaginary frequency axis $\omega_n= (2n-1)\pi/\beta$. This
introduces an effective inverse temperature, which is a fitting
parameter (it is not the real temperature which is $T=0$ in our
study) and is set $\beta=300 t$ (much higher than the $\beta=50t$
used in publication of ref.\cite{venky05}) throughout the whole
paper. This parameter determines the energy resolution accessed
in this work (see also appendix \ref{apxA}). On a practical level,
to start the ED-CDMFT procedure it is most useful to introduce a
reduced parameterization of the bath, which enlightens the
symmetries of the input-guess $\hat \mathcal{G}_0(i\omega_n)$.
The constrain on the bath parameters can be then relaxed in a
second step. Further details are given in appendix \ref{apxB}.
%-------------------
%=======================================================================

\section{Cluster Results}\label{sec2}
We start this section by considering raw cluster quantities, which
directly output from the cluster-impurity solution.

\subsection{A d-wave Superconducting state}
%====================
\begin{figure}[!!tbh]
\begin{center}
\includegraphics[width=9cm,height=4.5cm,angle=-0]{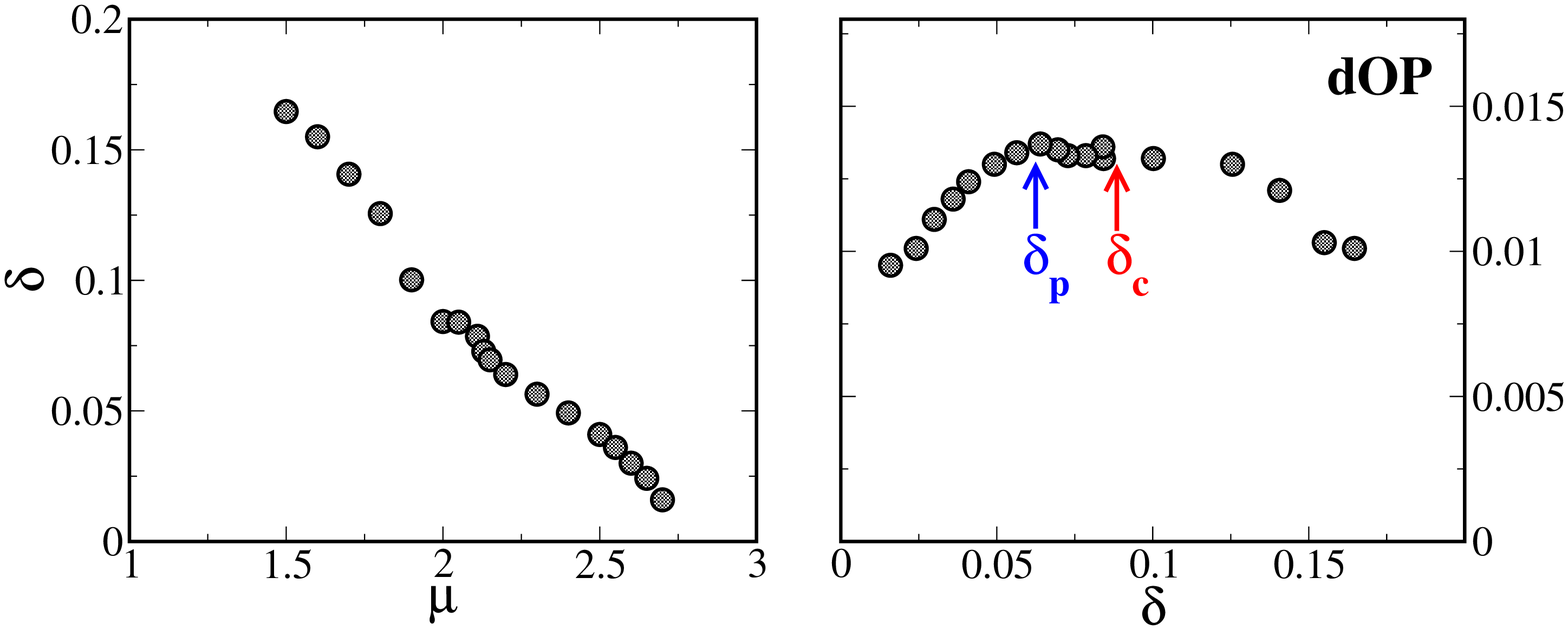}
\caption{(Color online). Left panel: doping $\delta=1- < n >$
versus chemical potential $\mu$. Right panel: cluster d-wave
order parameter dOP= $<<
c_{\mu\uparrow}c_{\mu+1\downarrow}>>_{(\tau=0)}$ as a function of
doping $\delta$. } \label{dOP}
\end{center}
\end{figure}
%==============================================================
As we explained above, within CDMFT a superconducting state can
be studied by allowing particle-particle pairing terms $V^{\rm
sup}_{k\mu \sigma} a_{k\uparrow}c_{q\downarrow}$ in the impurity
Hamiltonian eq.(\ref{Ham-imp-full}) (or equivalently upon a
unitary rotation, pairing bath terms
$a_{\mu\uparrow}a_{k\downarrow}$, see Appendix \ref{apxB}).
CDMFT-self-consistency condition may accept non-zero values of
the $V^{\rm sup}_{k\mu \sigma}$ terms, resulting in a non-zero
superconducting-pair Green's function $F_{\mu,\nu}(\tau)=
T_{\tau} < c_{\mu\uparrow}(\tau)c_{\nu\downarrow}>$. In drawing a
parallel with the classical mean field example of the Ising
Model\cite{revmodmft}, this is equivalent to assume a non-zero
on-site magnetization $m=<S^{z}_{i}>$ as starting hypothesis in
building an effective Hamiltonian, and to determine then $m$
self-consistently. As reported in previous work \cite{venky05},
CDMFT supports indeed a d-wave superconducting state in a region
of doping that precedes the Mott transition. This is shown in
Fig. \ref{dOP}, where we show the doping $\delta=1-< n_{i} >$
versus the chemical potential $\mu$ and the d-wave order
parameter (dOP), defined as $F_{\mu,\mu\pm1}(\tau=0)$, also as a
function of $\delta$. As expected, $\delta$ versus $\mu$ is
monotonically decreasing. The dOP has a dome-like shape and its
sign is alternating by exchange of the $x$-$y$ bonds on the
cluster plaquette. We can locate the maximum  only with some
degree of uncertainty around $\delta_p\sim 0.06 <\delta<
\delta_c\sim 0.08$. The two doping values $\delta_p\sim 0.06$ and
$\delta_c\sim 0.08$ present as two special points, which mark
changes in the physical properties of the system. We will come
back on these two points in throughout the paper. Here we stress
that at $\delta_c$ in particular, our dynamical mean field result
seems to branch two different lines of solution. This is evident
either in the $\delta-\mu$ and the dOP plots. In mean field
approaches this behavior may be the signature of a phase
transition. It is therefore intriguing that we find such a
behavior close to the maximum of the dOP dome.

Following the nomenclature typical for cuprate materials, we will
call hereafter the region $\delta< \delta_p$ under-doped,
$\delta> \delta_c$ over-doped and $\delta_p\sim
0.06<\delta<\delta_c\sim 0.08$ optimal doping region. It is clear
we do not intend to draw a quantitative parallel with
cuprate-based system, where the typical optimal doping
(unambiguously defined as the maximum of critical temperature
$T_C$) is around $\delta_c\sim 0.15$, but we rather follow the
qualitative aspects of the physics of these material, marking a
correspondences with our CDMFT results on the Hubbard Model in
two dimensions.
%-------------------
\subsection{Cluster self-energies}

\begin{figure}[!!th]
\begin{center}
\includegraphics[width=8cm,height=10.0cm,angle=-0]{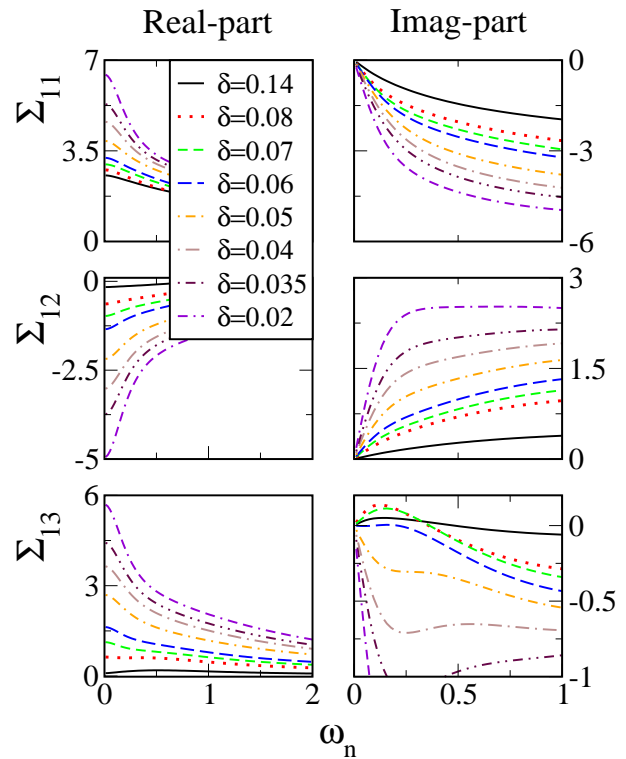}
\caption{(Color online). Normal components of the cluster
self-energy $\Sigma_{\mu\nu}$ vs the Matsubara frequency
$\omega_n$ as a function of doping $\delta$. In the left column
the real parts are displayed, in the right column the imaginary
parts. From the top row to the bottom we show the local
$\Sigma_{11}$, the next neighbor $\Sigma_{12}$, and nearest next
neighbor $\Sigma_{13}$ self-energies.} \label{Sig-nor-cluster}
\end{center}
\end{figure}

It is worth to investigate in detail the cluster-outputs. The
typical output of the CDMFT-scheme is a cluster-self-energy
$\Sigma_{\mu\nu}$ (eq. \ref{selfconsistency}), which can be
expressed as a 2$\times$ $N_c\times N_c$ ($N_c=4$ for the
2$\times$2 plaquette) matrix with normal and anomalous components
in the Nambu notation introduced in eq. (\ref{spinor}). We first
look at the normal components, which are shown on the
Matsubara-frequency axis in Fig. \ref{Sig-nor-cluster}.
%====================
%-------------------
\begin{figure}[!!th]
\begin{center}
\includegraphics[width=8cm,height=10.0cm,angle=-0]{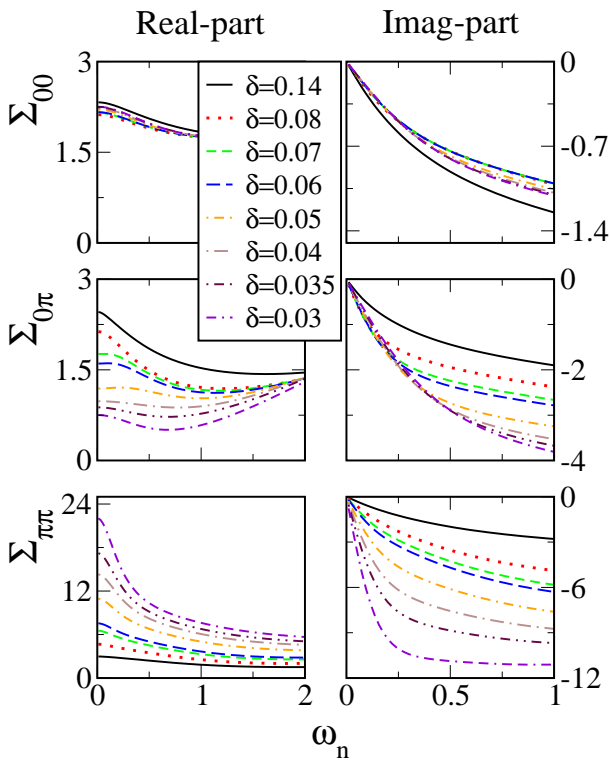}
\caption{(Color online). Eigenvalues of the normal component of
the cluster self-energy $\Sigma_{\mu\nu}$ vs the Matsubara
frequency $\omega_n$ as a function of doping $\delta$. In the
left column the real parts are displayed, in the right column the
imaginary parts. From the top row to the bottom we show the
self-energy corresponding to the points $k= (0,0)$, $(0,\pi)$ and
$(\pi,\pi)$ in the first quadrant of the Brillouin Zone (see
section \ref{sec3}).} \label{Eig-Sig-nor-cluster}
\end{center}
\end{figure}
%-------------------
%==============================================================
The real parts are displayed on the left and the imaginary parts
on the right column from high (in our solution) doping
$\delta=0.14$ (the over-doped side) until close to the Mott
transition for $\delta=0.02$ (in the under-doped side). From the
top row to the bottom we have the local self-energy
$\Sigma_{11}$, the next neighbor $\Sigma_{12}$ and the nearest
next neighbor $\Sigma_{13}$. At high doping, the local components
is dominant. We expect in this region that the single site DMFT
is already a good approximation. By decreasing $\delta$, however,
also the non-local components $\Sigma_{12}$ and $\Sigma_{13}$
grow considerably, becoming comparable with the local one. This
effect is totally missed by single site DMFT and it is only
captured by using a cluster extension. The growing of the non
local components of $\Sigma_{\mu\nu}$ determines the physical
properties of the system in approaching the Mott transition in a
fundamental way, as we explain in the following. The imaginary
parts (right column) display a low-energy Fermi-liquid-like
behaviour Im$\Sigma \sim \alpha \omega_n$. The slope $\alpha$ is
connected with the quasiparticle residuum (see the following).
Here we notice that, with respect to the other components, the
energy range of linear behavior in the nearest next neighbor
Im$\Sigma_{13}$ is narrower and that its slope $\alpha$ also
changes behavior in passing through the doping $\delta_c\sim0.08$
(it grows for $\delta<\delta_c$, decreases until becoming
negative for $\delta>\delta_c$), while it also changes sign at
$\delta_c\sim0.06$. This behavior is the only mark we find in the
cluster self-energy of the special nature of the points
$\delta_c$ and $\delta_p$ that we identified in the Fig.
\ref{dOP}. The local and next neighbor components show instead a
smooth continuous behavior as a function of doping $\delta$.
%==============================================================

As in ref.\cite{marce05,tudor06}, it is convenient to look at the
eigenvalues of the cluster-self-energy matrix, which can be
directly related to the corner points of the quadrant of the
Brillouin Zone (as we will explain in more detail in the following
section). A convenient way is to re-express the Nambu-spinor
notation eq.(\ref{spinor}) by grouping two by two the up and down
spin construction/destruction operators on each cluster-site:
\begin{equation}
\Psi_{}^{\dagger } \equiv \left[
(c_{1\uparrow}^{\dagger},c_{1\downarrow})\dots,(c_{4\uparrow}^{\dagger},c_{4\downarrow})
\right] \label{spinor-egn}
\end{equation}
The cluster self-energy matrix assumes the form:
\begin{equation}
\mathbf{\Sigma}_{c}=\, \left(
\begin{array}{cccc}
 \hat{\Sigma}_{0}  &  \hat{\Sigma}_{1x}   &  \hat{\Sigma}_{2}  &   \hat{\Sigma}_{1y}  \\
 \hat{\Sigma}_{1x}  &  \hat{\Sigma}_{0}   &  \hat{\Sigma}_{1y}  &   \hat{\Sigma}_{2}   \\
 \hat{\Sigma}_{2}  &  \hat{\Sigma}_{1y}   &  \hat{\Sigma}_{0}  &   \hat{\Sigma}_{1x}  \\
 \hat{\Sigma}_{1y}  &  \hat{\Sigma}_{2}   &  \hat{\Sigma}_{1x}  &   \hat{\Sigma}_{0}    \\
\end{array}
\right)  \\  \label{Sig-cluster}
\end{equation}
where the $2 \times 2$ matrices are function of
cluster elements $\Sigma_{\mu\nu,\sigma}=
\Sigma_{|\mu-\nu|,\sigma}$:
\begin{eqnarray}
%----------------
\hat{\Sigma}_{0}=\, \left(
\begin{array}{cc}
\Sigma_{0\uparrow}   &  0              \\
0            &  -\Sigma_{0\downarrow}     \\
\end{array}
\right) &
\hat{\Sigma}_{1x}=\, \left(
\begin{array}{cc}
\Sigma_{1\uparrow}   &  \Sigma_{ano}    \\
\Sigma_{ano}  &  -\Sigma_{1\downarrow}     \\
\end{array}
\right) \nonumber \\
%-----------------------
\hat{\Sigma}_{2}=\, \left(
\begin{array}{cc}
\Sigma_{2\uparrow}   &  0              \\
0            &  -\Sigma_{2\downarrow}     \\
\end{array}
\right) &
 \hat{\Sigma}_{1y}=\, \left(
\begin{array}{cc}
\Sigma_{1\uparrow}   &  -\Sigma_{ano}    \\
-\Sigma_{ano}  &  -\Sigma_{1\downarrow}     \\
\end{array}
\right)
%-------------------------------
\end{eqnarray}
We follow the procedure used to diagonalize the
cluster-self-energy matrix in ref.\cite{haule06b}:
\begin{equation}
\mathbf{\Sigma}_{c}=\, \left(
\begin{array}{cccc}
 \hat{\Sigma}_{00}  &  0           &  0         &   0  \\
  0         & \hat{\Sigma}_{\pi\pi}    &  0         &   0   \\
  0         & 0            & \hat{\Sigma}_{0\pi}  &   0  \\
  0         & 0            & 0          &  \hat{\Sigma}_{\pi 0}    \\
\end{array}
\right)  \label{Sig-diag}
\end{equation}
The diagonal elements are linear combination of the original
cluster self-energies matrices:
\begin{eqnarray}
\hat{\Sigma}_{00}&=&\, \hat{\Sigma}_{0}+ \tilde{\Sigma}_{1}+
\hat{\Sigma_{2}}  \nonumber \\
\hat{\Sigma}_{0\pi}&=&\, \hat{\Sigma}_{0}+ \tilde{\Sigma}_{a}-
\hat{\Sigma_{2}} \nonumber \\
\hat{\Sigma}_{\pi 0}&=&\, \hat{\Sigma}_{0}- \tilde{\Sigma}_{a}-
\hat{\Sigma_{2}} \nonumber \\
\hat{\Sigma}_{\pi\pi}&=&\, \hat{\Sigma}_{0}- \tilde{\Sigma}_{1}-
\hat{\Sigma_{2}} \label{Sig-rel}
\end{eqnarray}
For convenience's sake we have defined :
\begin{eqnarray}
\tilde{\Sigma}_{1}=\, 2 \left(
\begin{array}{cc}
  \Sigma_{1\uparrow}       &  0                     \\
  0                 & - \Sigma_{1\downarrow}
\end{array}
\right) &
%---------------------
\tilde{\Sigma}_{a}=\,2 \left(
\begin{array}{cc}
0                  &   \Sigma_{ano}                    \\
 \Sigma_{ano}      &   0
\end{array}
\right)
\end{eqnarray}
Notice that from eq. (\ref{Sig-rel}) the anomalous self-energy
appears only in the $(0,\pi)$ and $(\pi,0)$ components, where the
dSC gap is expected to open. The $2 \times 2$ eigenvalue-matrices
of the cluster-self-energy are interpreted as describing the four
momentum-space points $(0,0)$, $(0,\pi)$, $(\pi,0)$ and
$(\pi,\pi)$ (hence the choice of the labels), as we will explain
in detail in the following section \ref{sec3}.

%-------------------
\begin{figure}[!!t]
\begin{center}
\includegraphics[width=8cm,height=6.0cm,angle=-0]{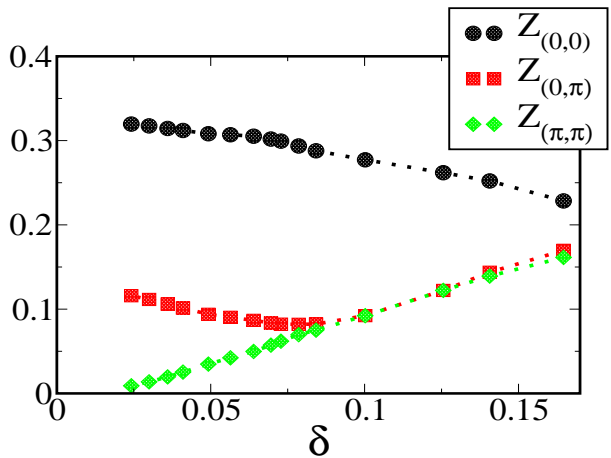}
\caption{(Color online). Cluster quasiparticle residua, associated
with the eigenvalues of the cluster self-energy matrix, as a
function of doping $\delta$.} \label{Zcluster}
\end{center}
\end{figure}
%------------------
The normal components of the eigenvalues of $\hat{\Sigma}_c$ are
shown in Fig. \ref{Eig-Sig-nor-cluster}, in a fashion similar to
Fig. \ref{Sig-nor-cluster}. The first striking difference with
respect to the Fig. \ref{Sig-nor-cluster} is that the eigenvalue
$\Sigma_{00}$ changes little as a function of doping (either the
real and the imaginary parts). $\Sigma_{0\pi}$ and
$\Sigma_{\pi\pi}$ both grow in reducing doping $\delta$, but the
growth in $\Sigma_{\pi\pi}$ is on order of magnitude bigger. This
eigenvalue is indeed the one that drives the system into the Mott
insulating state as $\delta\to0$, as evident from the big values
of both the real part and of the slope of the imaginary part at
small doping. This behavior are very similar (i.e. it appears as
a smooth continuation) to the one displayed by the self-energy of
the normal state ED-CDMFT study published in ref.\cite{marce05}.
With respect to this latter study however, here we display the
{\it normal components of a superconducting solution} (while in
ref.\cite{marce05} no superconductivity is allowed) and the
energy resolution achieved ($\beta t=300$) is an order of
magnitude smaller than in ref.\cite{marce05} ($\beta t=32$). This
allows us to extract the very low-energy properties. In
particular, as expected by the fact that $\hat{\Sigma}$ has to be
negative in order to respect casuality, we can observe in Fig.
\ref{Eig-Sig-nor-cluster} that the slope $\alpha$ of the
Im$\Sigma\sim \alpha \omega_n$ is always negative. The fact then
the $\omega_n\to 0$ behavior of Im$\Sigma$ is linear indicates
that these eigenvalues components have Fermi liquid properties.
Following a standard Fermi liquid approach, it is instructive to
define cluster quasiparticle residua $Z_X= \left(
1-\partial\Sigma_X/\partial\omega_n\right)$ with $X= (0,0),
(0,\pi), (\pi,\pi)$ , even if these quantities have a real
physical meaning only in correspondence of a real Fermi surface
(i.e. $Z_X$ has to be interpreted here as renormalized cluster
quantity). In the standard picture of the Mott
transition\cite{revmodmft}, the quasiparticle residuum $Z\to 0$
as doping is reduced $\delta\to 0$. From Fig. \ref{Zcluster} we
see that only $Z_{(\pi,\pi)}$ appears to display this behavior,
while $Z_{(0,0)}$ and $Z_{(0,\pi)}$ clearly extrapolate to a
non-zero value. A last remark concerns once again the special
critical doping $\delta_c$, where remarkably the $Z_{(0,\pi)}$
shows a clear change in behavior. For $\delta> \delta_c$ it
closely follows $Z_{(\pi,\pi)}$, decreasing with doping as
expected in the standard Mott transition picture. At $\delta_c$,
however, $Z_{(0,\pi)}$ departs from $Z_{(\pi,\pi)}$, and shows a
behavior more similarly to $Z_{(0,0)}$. This once again indicates
a quick change in the physical properties of the system in
correspondence of $\delta_c$. We will come back later to
discussing the physical interpretation of these observations on
the cluster quasiparticle residua (see section \ref{sec4}).

%-------------------
\begin{figure}[!!t]
\begin{center}
\includegraphics[width=9cm,height=4.50cm,angle=-0]{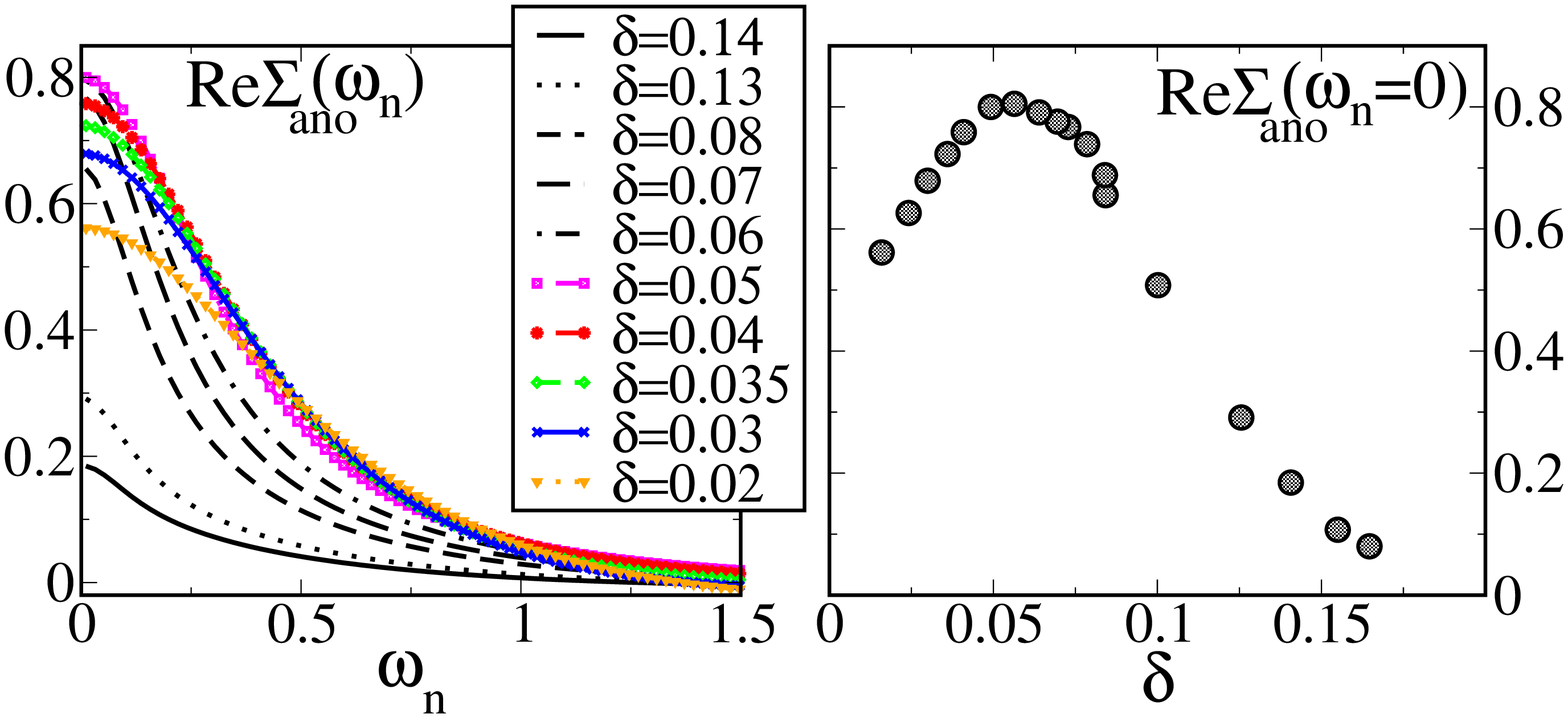}
\caption{(Color online). Left panel: real part of the anomalous
component of the cluster-self-energy $\Sigma_{ano}$ (the
imaginary part is negligible) vs Matsubara frequency $\omega_n$
for different doping $\delta$. Right panel:
Re$\Sigma_{ano}(\omega_n\to0)$ versus doping $\delta$.}
\label{Sig-ano-cluster}
\end{center}
\end{figure}
%-------------------

We now turn to the anomalous component of $\hat{\Sigma}_{c}$,
which we show in the left side of Fig. \ref{Sig-ano-cluster} on
the Matsubara frequency $\omega_n$. As resulting output of our
CDMFT solution, only the real part of the nearest-neighbor
component is appreciably non-zero on the Matsubara axis, and it
assumes a d-wave sign bond-alternating value on the
cluster-plaquette. Re$\Sigma_{ano}$ starts assuming appreciably
non-zero values in the over-doped side $\delta=0.14$, and grows
for decreasing doping, until $\delta_c \sim 0.08$ is reached. In
the under-doped side however at $\delta<\delta_p\sim 0.06$ the
curves change tendency, and they decrease by further decreasing
doping. The qualitative change in the behavior is better
enlightened by looking at the $\omega\to 0$ limit: in the
under-doped side the curves reach $\omega=0$ with a finite slope,
while in the under-doped side (after $\delta\leq 0.08$) the slope
has a smaller value. In right side of Fig. \ref{Sig-ano-cluster}
we show the $\omega_n \rightarrow 0$ extrapolated value of
Re$\Sigma_{ano}$ as a function of doping $\delta$. In first
approximation this value can be related to the superconducting
gap (as we will explain in detail in the section \ref{sec3}, see
eq. \ref{Ek}). The behavior of the Re$\Sigma_{ano}$ as a function
of doping shows therefore to be non-monotonic and roughly tracks
the behavior of the order parameter (see Fig. \ref{dOP}),
similarly to standard BCS theory. This result is fundamental and
it is in striking contrast with slave-boson resonating valence
bond theories\cite{anderson,liu,patrickrmp}, where the amplitude
associated to a particle-particle paring channel is a
monotonically decreasing function with doping and it has its
maximum close to the insulating transition. In section \ref{sec4}
we will better discuss the physical consequences of these results,
making connection with recent spectroscopy experiments on
cuprate-based materials.
%-------------------

With the ED-CDMFT procedure it is straightforward to analytically
continue on the real axis\cite{revmodmft} (differently from other
computational methods like e.g. Quantum Monte Carlo, which require
further approximate methods). The cluster Green's function
$\mathbf{G_c}(\omega_n)$ and the cluster self-energy
$\mathbf{\Sigma_c}(\omega_n)$ (via the Dyson's equation
\ref{selfconsistency}), are expressed in a pole expansion form
containing terms of the form $b/(\omega_n-a)$. To go from the
Matsubara to the real axis it is then enough to substitute $i
\omega_n \to \omega+ i \eta$, where $\eta$ is a small broadening
introduced to display poles (for more details see appendix
\ref{apxA}). The price to pay comes from the discreteness
introduced by truncating the bath in the impurity model (eq.
\ref{Ham-imp-full}) with a finite number $N_b$ of orbitals. This
reproduces continuous functions, like $\mathbf{G_c}(\omega_n)$
and $\mathbf{\Sigma_c}(\omega_n)$, through a finite number of
poles.

\begin{figure}[!!tb]
\begin{center}
\includegraphics[width=8cm,height=12.0cm,angle=-0] {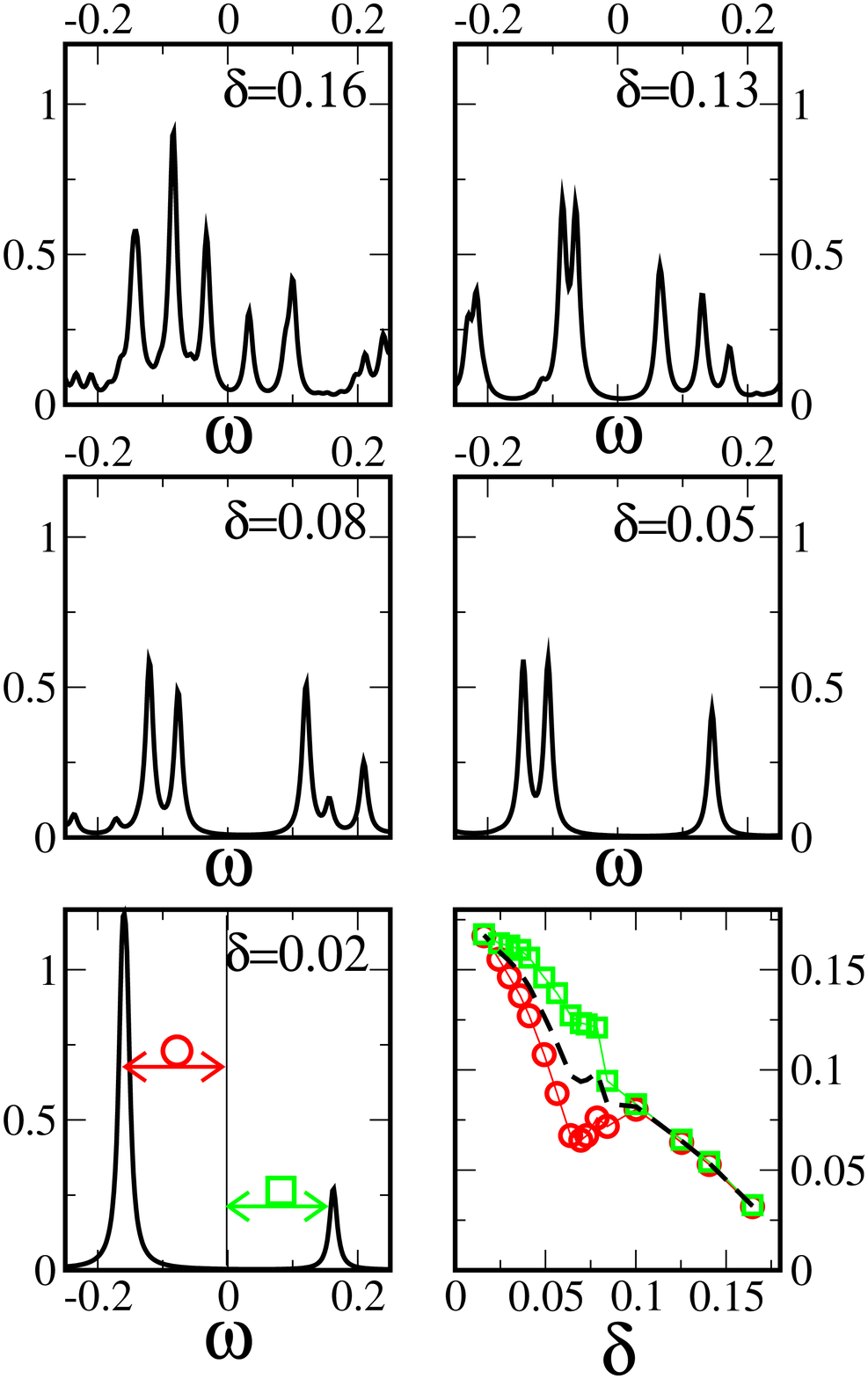}
\end{center}
\caption{(Color online). Local density of state $-\frac{1}{\pi}
G_{11}(\omega)$ obtained from the cluster-impurity solution (eq.
\ref{Ham-imp-full}) for various doping $\delta$ (from top to
bottom). The broadening used for display is $\eta=7\times
10^{-3}$. A measure of the energy scale associated with the
one-particle gap (always present in the superconducting state) is
given by the left (red circle) and right (green square) distances
of the ED poles from the Fermi level $\omega=0$. These are
displayed as a function of doping $\delta$ in the right bottom
panel.} \label{ImG11}
\end{figure}

In Fig. \ref{ImG11} we display on the real axes the local density
of states $N(\omega)= \frac{1}{\pi}\left[
G_c(\omega)\right]_{11}$, obtained from the cluster-impurity
output (see eq. \ref{selfconsistency}), for different doping
$\delta$ (from top to bottom), using a small imaginary broadening
i$\eta= 7\hbox{i}\times 10^{-3}$. As we will explain in detail in
the following section, in a d-wave superconducting state a linear
in $\omega$ "V" shaped density of states is expected for
$\omega\to 0$. This behavior cannot be captured by the
discreteness of our ED-CDMFT solution for small $\omega$, hence
we have rather a "U" shape for $\omega\to0$ given by the
broadening $\eta$. The shape of $N(\omega)$ and the energy scales
of the superconducting gap can however be estimated by the
location and intensities of the peaks. So, in looking at
$N(\omega)$ from the over-doped side (top row, left panel for
$\delta=0.16$) to the under-doped side (bottom row, left panel for
$\delta=0.02$) of the phase diagram, we can make interesting
observations:
\begin{enumerate}
  \item $N(\omega)$ is asymmetric function around
  $\omega=0$ at low doping ($\delta< \delta_c$). A rather symmetric
  shape is instead observed around optimal doping
  $\delta_c\sim 0.08> \delta> \delta\sim 0.06$
  (in  agreement with previous cluster DMFT results of ref.\cite{haule-ctqmc}) .
  \item the total superconducting gap, which can be evaluated by measuring
  the distance of the spectral peaks from $\omega=0$, is increasing
  by reducing doping $\delta$.
\end{enumerate}
%-------------------------------------------------------------
%-------------------
To better elucidate these observations, we have measured the
distance from the Fermi level of the left (red circle) and right
peaks (green square), and displayed them as a function of doping
$\delta$ (bottom row, right panel). In the over-doped side
($\delta> \delta_c$) the peaks are equally distant from
$\omega=0$, i.e. the superconducting d-wave gap is  symmetric at
low frequency, as expected in a standard d-wave BCS theory. At
optimal doping $\delta_p<\delta<\delta_c$, however, the asymmetry
in the gap strikes in, and it is present in all the under-doped
side. This is in agreement with the cluster DMFT results obtained
in ref.\cite{haule-ctqmc} and it is in nicely agreement with
Scanning Tunneling experiments (STM\cite{davis05}) on cuprate
materials. The fact that $N(\omega)$ is most symmetric at optimal
doping $\delta_p<\delta<\delta_c$ (as evident from the red circles
which display a change in the slope as a function of $\delta$) is
in line with the observation carried in Fig. \ref{dOP} on the
special nature of this region (consistently e.g with the
avoided-quantum-critical-point scenario proposed in the CDMFT
study of ref.\cite{haule06}). The black dashed line is the average
of the right and left gaps, and the fact that it is increasing by
reducing doping (similarly to the predictions of resonating
valence bond slave boson theories\cite{anderson,liu,patrickrmp})
shows the important result on the presence in our solution of
another energy scale, different from the energy scale marked by
the behavior of Re$\Sigma_{ano}(\omega_n\to0)$ (which is instead
decreasing at low doping ($\delta<\delta_p$), see right panel
Fig. \ref{Sig-ano-cluster}). The interpretation of these two
energy scales brings to interesting physical insights, which can
be put in relation with recent experiments on the cuprate
materials
\cite{tacon06,tanaka06,kondo07,gomes2007,deutscher99,millis06,cho06,huefner-2008-71}.
This has been presented in a previous short
publication\cite{marce08} and it will be re-proposed more in
detail in the following sections.

\section{Momentum dependent quantities: periodization procedures}
\label{sec3}

We want now to interpret the cluster results we presented in the
previous section in terms of physical observables, which could be
possible related to experiments. The relevant information is
typically embodied in the one-particle Green's function, which in
a superconducting state can be conveniently written in a
Nambu-matrix notation:
\begin{equation}
\begin{array} {l}
\mathbf{G}^{-1}_{\sigma}(k,\omega) =  \\
 \\
\, \left(
\begin{array} {cc}
\omega- \xi_{k}- \Sigma_{k}(\omega) &- \Sigma_{ano}(k,\omega) \\
-\Sigma_{ano}(k,\omega) &\omega+ \xi_{k}+
\Sigma^{\ast}_{k}(-\omega)
\end{array} \right)
\end{array}
\label{Gk-sup}
\end{equation}
Here $\xi_{k}= t_k- \mu$ is the free band dispersion of our model
(see equation \ref{hamiltonian}). In order to determine the
Green's function we need therefore to determine the momentum
dependent self-energy from the cluster solution, i.e. we need a
periodization scheme. In previous work\cite{bpk,marce05,bumsoo06}
various periodization schemes have been proposed. The idea
consists in determining the most local quantity $W_{r}$, which
can be captured within the dimension of the cluster-impurity, and
construct its truncated Fourier expansion\cite{biroli04}:
\begin{equation}
W_{\sigma}(k)= \,\frac{1}{N_{c}}\, \sum_{\mu\nu} e^{-\imath
k\mu}\, W_{\sigma}(|\mu-\nu|)\,e^{\imath k\nu} \label{Wk}
\end{equation}
Smaller the neglected Fourier coefficients are (for $|\mu-\nu|>
\sqrt{N_c}$), compared to the cluster $W_{\sigma}(|\mu-\nu|)$,
more the $k$-dependent quantity $W_{\sigma}(k)$ is well
approximated. In the following we present two possible
cluster-quantities that can be adopted to construct the
$k$-dependent self-energy, showing in which cases they can be
considered good local quantities.

\subsection{Self-energy $\Sigma$-periodization }

The cluster self-energy (eq. \ref{Sig-cluster}) is a natural
candidate. It is convenient, for the discussions in the following
sections, to recast formula (\ref{Wk}) in terms of the cluster
eigenvalues (eq. \ref{Sig-diag}):
\begin{equation}
\hat{\Sigma}_{\sigma}(k)= \sum_{X}
\hat{\Sigma}_{X\sigma}\,\gamma_{X}(k) \label{Sigmak}
\end{equation}
where $\hat{\Sigma}_{\sigma}(k)$ is a 2 $\times$ 2 matrix containing
normal and anomalous components:
\begin{eqnarray}
\hat{\Sigma}_{\sigma}(k) &= &\left(
\begin{array}{cc}
\Sigma^{nor}_{k\uparrow}(\omega)        &  \Sigma_{ano}(\omega) \\
\Sigma_{ano}(\omega)                    &  - \Sigma^{nor}_{k\downarrow}(-\omega)
\end{array}
\right) \label{Sigmak-matrix}
\end{eqnarray}
We notice that with this formula the cluster eigenvalues
$\hat{\Sigma}_X$ are directly related to the corner points in
first quadrant of the Brillouin Zone $X=$$(0,0)$, $(0,\pi)$,
$(\pi,0)$ and $(\pi,\pi)$ (and we justify the notation introduced
in eq. \ref{Sig-diag}). $\gamma_{X}(k)$ are positive functions,
such $\sum_{X} \gamma_{X}(k)=1 \, \forall k\cite{tudor,tudor06} $:
\begin{eqnarray}
\gamma_{00}&=&\, \frac{1}{4} \left( 1+ \cos k_{x}+ \cos k_{y}+ \cos k_{x}\cos k_{y}\right) \nonumber \\
\gamma_{\pi\pi}&=&\, \frac{1}{4} \left( 1- \cos k_{x}- \cos k_{y}+ \cos k_{x}\cos k_{y}\right) \nonumber \\
\gamma_{0\pi}&=&\, \frac{1}{4} \left( 1+ \cos k_{x}- \cos k_{y}- \cos k_{x}\cos k_{y}\right) \nonumber \\
\gamma_{\pi 0}&=&\, \frac{1}{4} \left( 1- \cos k_{x}+ \cos k_{y}- \cos k_{x}\cos k_{y}\right) \nonumber \\
\label{gamma-k}
\end{eqnarray}
We remark that, by construction, with this procedure we assume
that the system is a simple Fermi liquid. We have in fact shown in
Fig. \ref{Sig-nor-cluster} that the normal components of the
eigenvalues of the cluster self-energy have Fermi-liquid behavior
(Im$\Sigma_{X}\rightarrow 0$ for $\omega\rightarrow 0$). Their
simple linear combination extends this property to all the
$k$-space. The anomalous component of the lattice self-energy
$\Sigma_{ano}(k)$  turns out to have a d-wave shape
\begin{equation}
\Sigma_{ano}(k)= 2 \Sigma_{ano} \, \left( \cos k_{x}-\cos k_{y} \right)
\label{Skano}
\end{equation}
in agreement with the symmetry of the superconductive gap measured
in experiments on cuprates\cite{damascelli,campuzano}.

%--------------------------------------------------------------
\subsection{Cumulant $\mathcal{M}$-periodization}
%-------------------------------------------------------------
In a normal state study of the two-dimensional Hubbard Model of
ref.\cite{tudor,tudor06}, it has been shown that a more suitable
local quantity to describe the Mott transition is the two point
irreducible cumulant $\mathcal{M}$, which arises from the atomic
limit by perturbatively expanding the hopping term $t$ in
Hamiltonian (\ref{hamiltonian}). It is simply related to the
normal-component lattice self-energy $\Sigma^{nor}_{k}$ by:
\begin{equation}
\mathcal{M}^{nor}_{k}(\omega)= \frac{1}{\imath\omega +\mu-
\Sigma^{nor}_{k}} \label{MkSk}
\end{equation}
In the cluster-impurity we have $2 \, N_{c} \times N_{c}$
cumulant relations (2 is for the spin degeneracy), conveniently
represented by a $N_{c} \times N_{c}$ cumulant-cluster-matrix
$\hat{\mathcal{M}}_{\sigma c}$:
\begin{eqnarray}
\mathbf{\hat{\mathcal{M}}^{nor}_{\sigma c}}(\omega)  & =& \left[
(\imath\omega+\mu) \mathbf{1}- \mathbf{\hat{\Sigma}^{nor}_{\sigma
c}} \right]^{-1} \label{M-cluster}
\end{eqnarray}
where $\mathbf{1}$ is the $N_{c} \times N_{c}$ identity matrix.
The eigenvalue of the cumulant matrices are straightforwardly
related to the eigenvalues of the self-energy matrices (eq.
\ref{Sig-diag}):
\begin{eqnarray}
\mathcal{M}^{nor}_{X}= \, (\,\imath\omega +\mu- \Sigma^{nor}_{X}
\,)^{-1} \label{M-egn}
\end{eqnarray}
where the notation is again $X=$$(0,0)$, $(0,\pi)$, $(\pi,0)$ and
$(\pi,\pi)$. As in the eq. (\ref{Sigmak}), we obtain the lattice
cumulant $\mathcal{M}^{nor}_{k}$ by periodizing the eigenvalues
of the cluster cumulants:
\begin{equation}
\mathcal{M}^{nor}_{k}= \sum_{X}\,
\mathcal{M}^{nor}_{X}\,\gamma_{X}(k) \label{Mknor-egn}
\end{equation}
Inverting eq. (\ref{MkSk}), we finally obtained the
normal-component lattice self-energy $\Sigma^{nor}_{k}$ in the
$\mathcal{M}$-periodization.
%%-----------------------------------------------------------------
%--------------------------------------------------------------
\subsection{Nodal and Antinodal dichotomy:
$\Sigma$-vs $\mathcal{M}$-periodization}
%================================================================
%-------------------------------------------------------------
\begin{figure}[!!tb]
\begin{center}
\includegraphics[width=8cm,height=6cm,angle=-0] {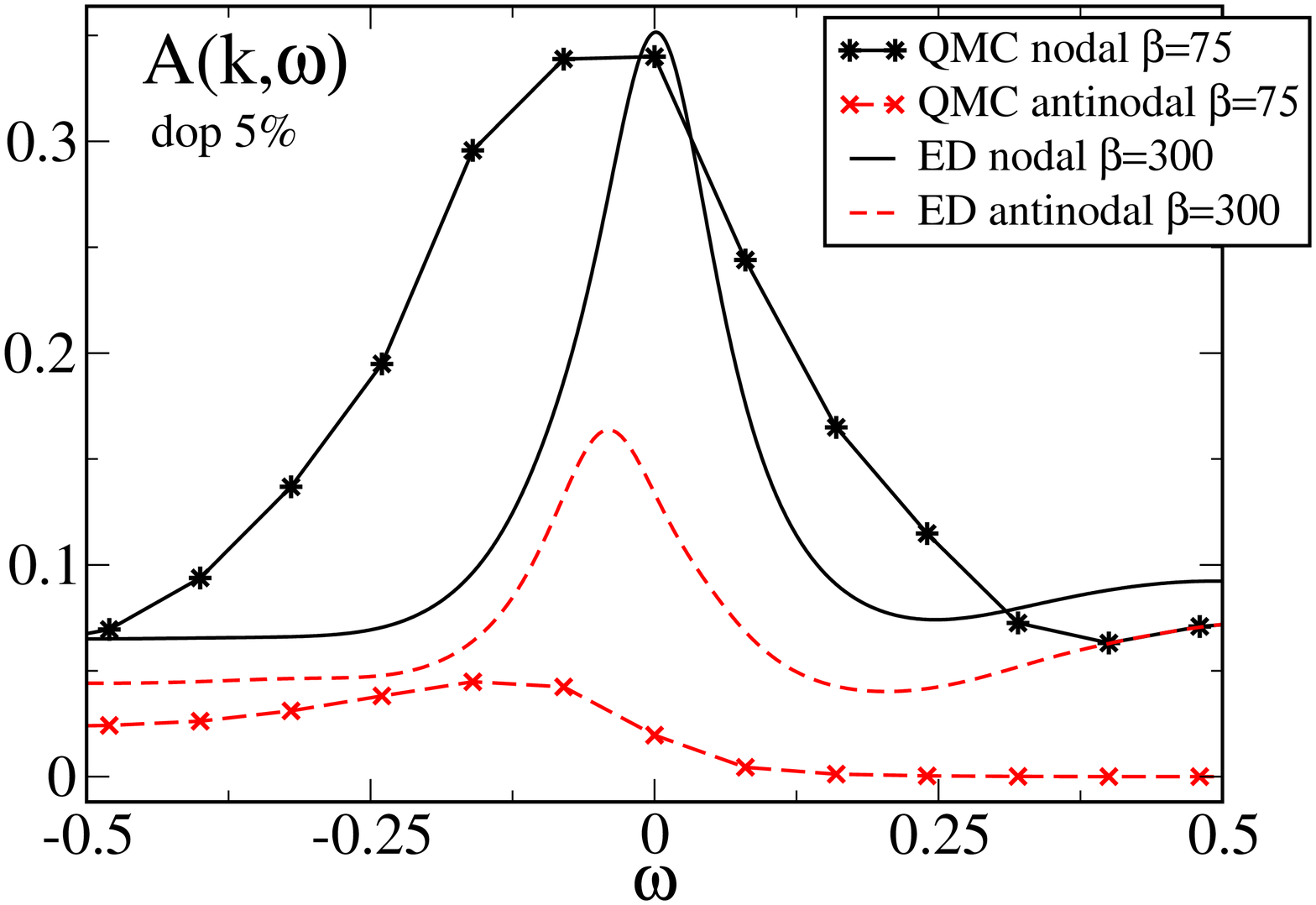}
\end{center}
\caption{(Color online). We compare the nodal and antinodal
quasiparticle peaks obtained from the normal component of the
superconductive ED-CDMFT-solution at zero temperature and the
QMC-CDMFT normal state solution at much higher temperature ($1/T=
75$). The functions on the real axis with QMC are obtained with
the maximum entropy method. To display the ED-CDMFT curves the
slightly $\omega$-dependent broadening
$\eta(\omega)=0.075+\omega^{2}$ was used (see discussion in
appendix \ref{apxA}).} \label{ED-QMC}
\end{figure}
%---------------------------------------------------------------

It is not trivial to decide whether it is better to periodize the
self-energy $\Sigma$ or the cumulant $\mathcal{M}$. The choice
could be strongly dependent on the physical properties of the
system, which are not {\it a priori} well known. And it is not
trivial either understanding {\it a priori} to which extend the
two approximated schemes could be able to describe such
properties. In this case, we rely on experimental results on
cuprate materials, like e.g. the already mentioned
ARPES\cite{damascelli,campuzano}, to fix a physically reasonable
starting hypothesis. It is a well established experimental fact
that approaching the Mott insulator the normal state Fermi surface
(measured at temperatures above T$_{C}$) breaks up, displaying
well defined quasiparticles in the regions (nodes) close to the
centers of the quadrants of the Brillouin Zone
$k\sim(\pm\frac{\pi}{2},\pm\frac{\pi}{2})$, while quasiparticle
disappear in the strong scattering regions (anti-nodes) close to
the corners of the quadrants $k\sim(0,\pm\pi)$ and $(\pm\pi,0)$.
Besides photo-emission experiments, measures of transport
properties show anomalous (non Fermi liquid) power-low exponents
in the temperature-dependence, especially in the under-doped
regime of cuprate materials. In this case, a series of
phenomenological approaches, which try to explain these
experimental observations in the framework of Boltzmann
theory\cite{PSK,Millis1998,Zheleznyak1998,Hlubina1995}, have been
based on dividing the momentum space in regions of high
quasiparticle scattering rate, {\it hot spots} around the
antinodal points, and regions of low quasiparticle scattering
rate, {\it cold spots} around the nodal point, where
quasiparticles have a much longer lifetime. The simple idea
underlying this choice is that the nodal region behaves as a
standard Fermi liquid, while the system in the antinodal region
is like an insulating state.

This {\it nodal/antinodal dichotomy} fits into the CDMFT frame as
the natural path taken by the system to approach the doping-driven
Mott transition \cite{marce05,bumsoo06,tudor,tudor06}. This is a
general property independent of the periodizing scheme adopted.
Some important differences however hold between the $\Sigma$ and
the cumulant $\mathcal{M}$ periodizations, which we will
illustrate in the following.

In Fig. \ref{ED-QMC} we show for example the spectral functions
$A(k,\omega)= \, -\frac{1}{\pi} \hbox{Im} G^{nor}_{k}(\omega)$ in
the nodal $k\sim (\frac{\pi}{2},\frac{\pi}{2})$ (black continuous
line) and antinodal $k\sim (0,\pi)$ (red dashed line) points of
momentum space, obtained via $\mathcal{M}$-periodization for the
case $U/t=12$, $t^{\prime}=-0.3t$ and $5\%$ doping. We confront
our zero temperature ED-CDMFT result with a
QMC-CDMFT\cite{bpk,olivier-note} at much higher temperature
($T=1/75 t$), where the system is in the normal state. In the ED
case we have extracted the normal part of the superconducting
solution (setting $\Sigma_{ano}=0$ in eq. \ref{Gk-sup}),
interpreting it as a low-temperature normal state parent of the
high-temperature QMC-solution. This is far from being a trivial
statement, as the normal component of a superconducting solution
is not generally a normal state solution. Our aim in this picture
is however to present a qualitative comparison between two very
different impurity-solving methods in two very different regimes
to show the generality and solidity of the nodal/antinodal
dichotomy concept. This figure serves also to compare results of
maxent analytic continuation of Fye-Hirsch QMC scheme with the ED
results at low temperatures. The qualitative agreement is
reasonably good, supporting the observation we made above on
experimental results and the dichotomy nature of the physical
properties in this system.

We observe in particular in Fig. \ref{ED-QMC} that the antinodal
quasiparticle peak is very broad in the high-temperature QMC
solution, denoting a short lifetime of quasiparticles, while it is
sharper in the low-temperature ED solution, much more than the
narrowing due to the different temperature (we have scaled the
heights of the peaks taking the high of the nodal quasiparticle
peak as reference to fix the scale between the ED and QMC
curves). This goes in the direction of ARPES experimental
observations, which show a sharpening of the antinodal
quasiparticle in going from the normal to the superconductive
state\cite{ding} by decreasing temperature. We finally point out
that with the $\mathcal{M}$-periodization used in Fig.
\ref{ED-QMC}, in the antinodal point the quasiparticle peak
shifts to negative energies, opening a pseudogap. This is also in
line with the above mentioned ARPES observations. On the
contrary, by using the $\Sigma$-periodization the antinodal peak
remains always (for all dopings) at the Fermi level, but with
reduced spectral weight compared to the nodal point\cite{marce05}.

All these observations, either from experimental facts and from
the output of our method, point towards a dichotomy of the nodal
and antinodal regions of momentum space, which show different
coherence energies. The nodal region has the highest coherence
scale and sharp quasiparticles. In the antinodal region a
pseudogap opens in the spectrum, and quasiparticles, if present
at the gap-edge, are more broad and incoherent. It is therefore
natural to assume that the Fermi-liquid nodal region is better
portrayed by the $\Sigma$-periodization (which as we mentioned
above describes a Fermi liquid by construction), while the
antinodal insulating region is better described by the
$\mathcal{M}$-periodization (as shown in Fig. \ref{ED-QMC}).
%================================================================
\begin{figure}[!!tbp]
\begin{center}
\includegraphics[width=8cm,height=6cm,angle=-0] {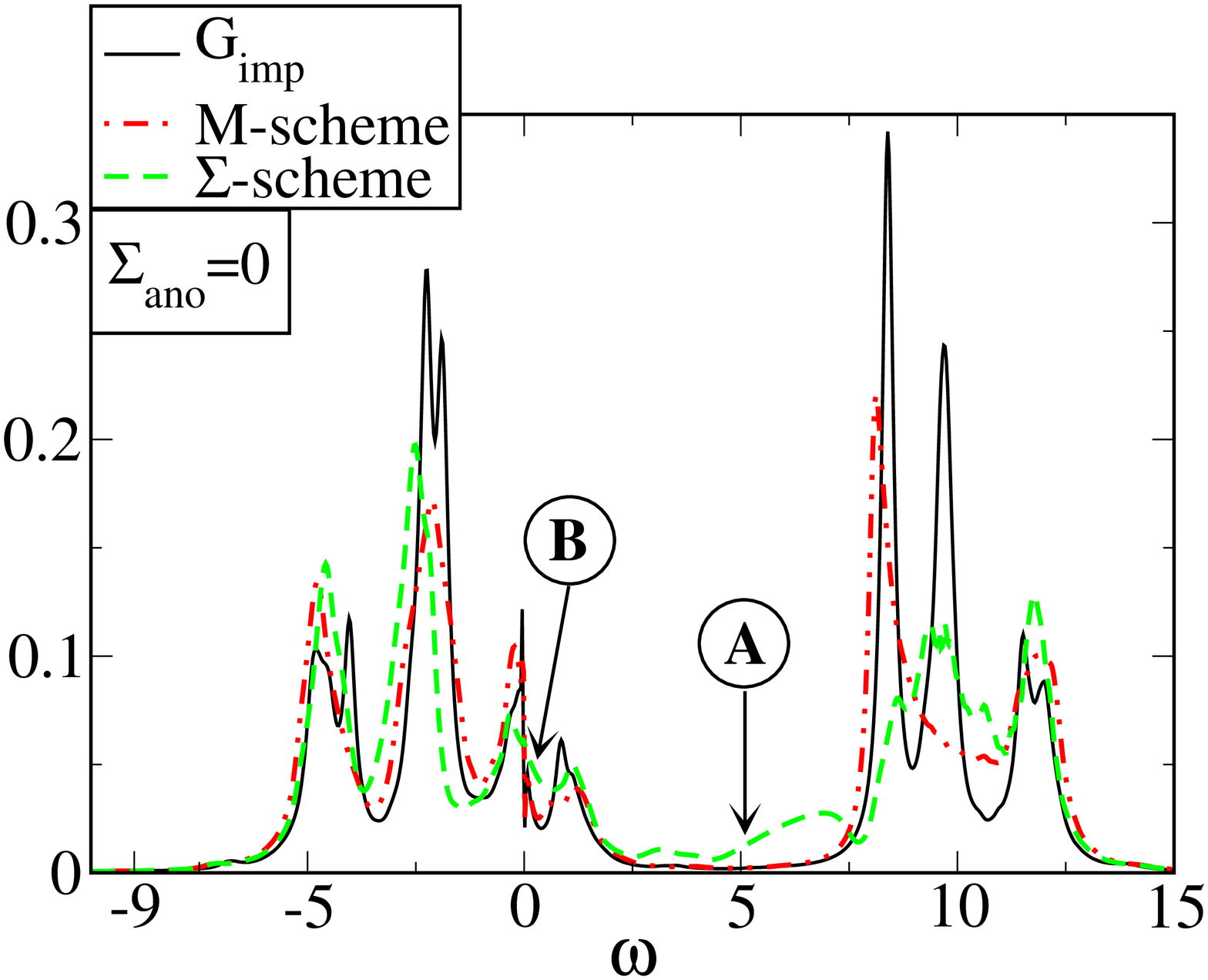}\\
\includegraphics[width=8cm,height=6cm,angle=-0] {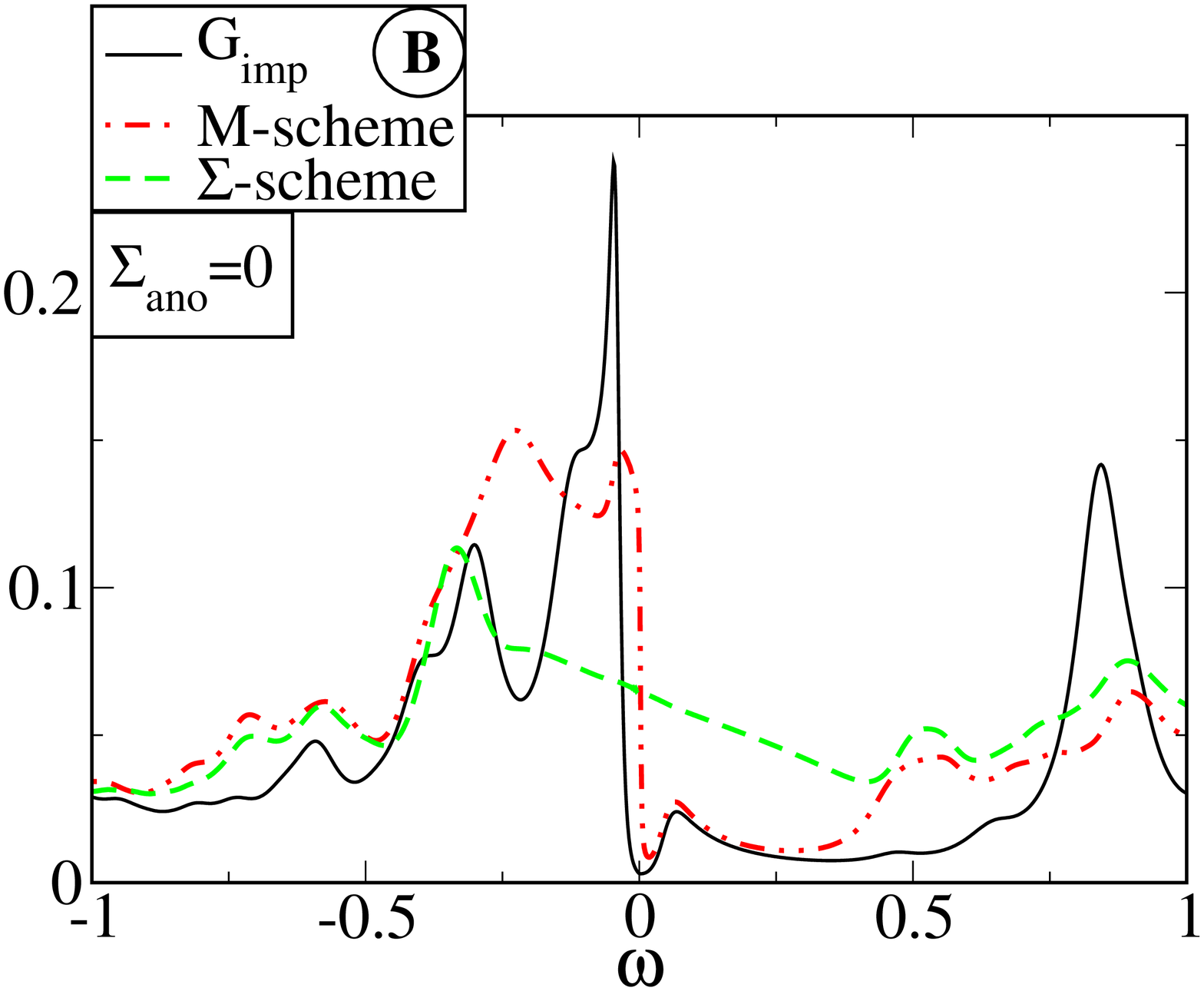}\\
\includegraphics[width=8cm,height=6cm,angle=-0] {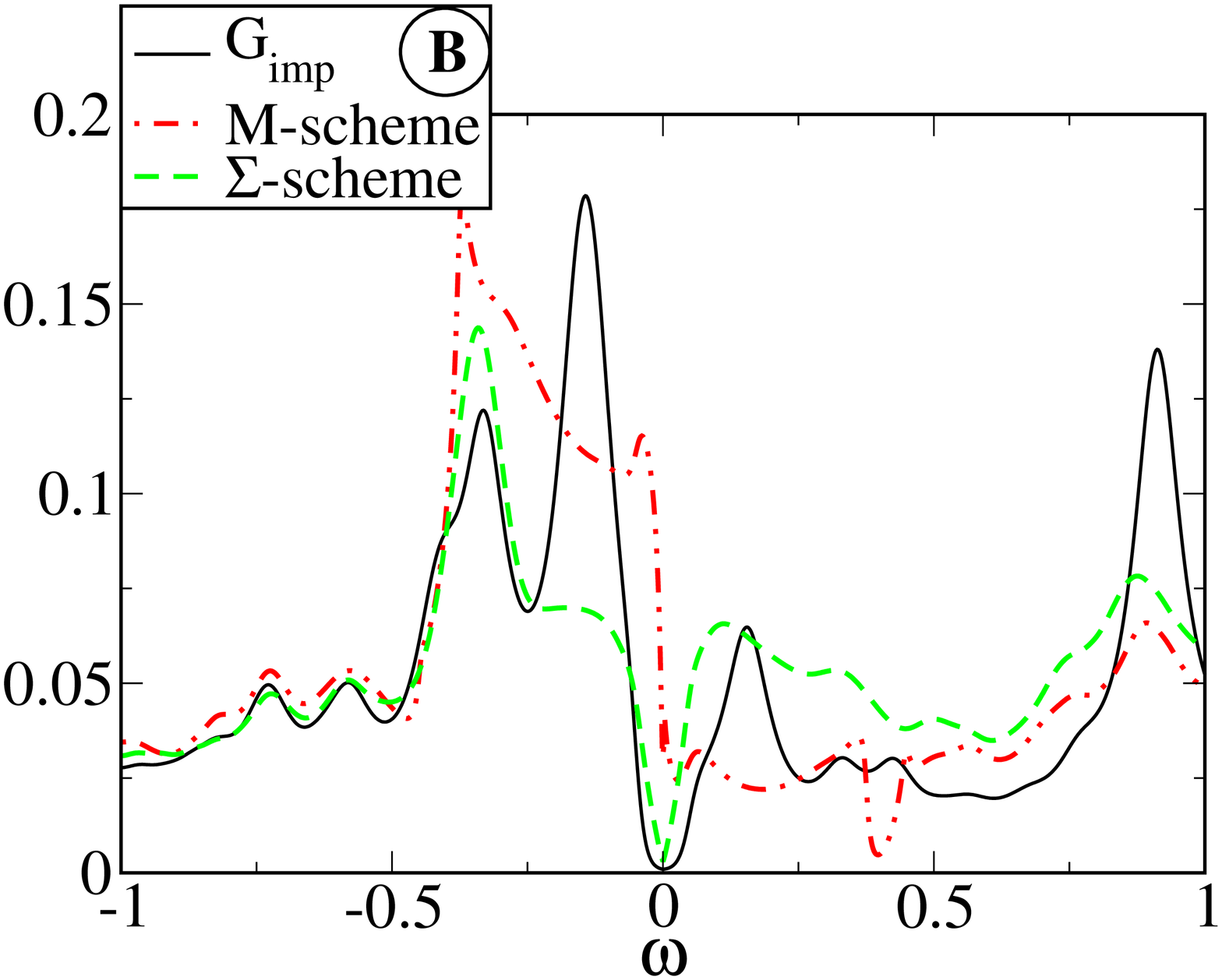}
\end{center}
\caption{(Color online). {\bf Top panel}: reconstructed local
density of states $-\frac{1}{\pi}\sum_{k} G^{nor}_{k}(\omega)$
for the normal component ($\Sigma_{ano}=0$) of the Green's
function at small doping ($\delta=0.05$). $G^{nor}_{k}(\omega)$
has been obtained by either the cumulant
$\mathcal{M}$-periodization (red point-dash line labeled
$M$-scheme) and by the $\Sigma$-periodization (green dash line
labeled $\Sigma$-scheme), and confronted with the local density
of states of the cluster-impurity solution (black continuous
line). {\bf Middle panel}: close-up at the Fermi level. {\bf
Bottom panel}: the local density of states of full
superconducting solution ($\Sigma_{ano}\neq 0$) selects in the
$k$-summation only the nodal points. A low-energy confront with
the cluster-impurity solution shows a different matching of the
cumulant and self-energy periodizations with the cluster-impurity
result. The display on the real axis of the Green's function has
been obtained by introducing a $\omega$-dependent broadening
i$\eta(\omega)$ (see appendix \ref{apxA} for details). }
\label{DOS-SM2}
\end{figure}
%---------------------------------------------------------------

Taking this last assumption as starting point, we compare the
results of the $\Sigma$ and $\mathcal{M}$-periodization,
stressing virtues and defaults in the frame of the physical
observations carried over above. A stringent test is given by
reconstructing the local density of states
$N(\omega)=\,-\frac{1}{\pi}\sum_{k} \hbox{Im}
G^{}_{k}(\omega)_{11}$ (where $G^{}_{k}(\omega)$ is obtained by
eq. \ref{Gk-sup}), which can be compared with the local Green's
function $G_{imp}= (\hat{G}_c)_{11}$, obtained directly in the
cluster-impurity solution. This test is presented in Fig.
\ref{DOS-SM2}. First, it is once again instructive to separately
study the normal component of the superconducting state by
setting $\Sigma_{ano}=0$ in eq. (\ref{Gk-sup}). We focus the
attention on a small-doping case $\delta=0.05$ close to the Mott
transition point. The top panel displays a full energy range
which includes the lower and upper Hubbard bands. The
cluster-impurity density of states $N(\omega)$ is represented by
the continuous black line, the $\mathcal{M}$-periodization result
by a red dot-dash line and the $\Sigma$-periodization result by a
green dash line. Already at small doping, in the region labeled
(A), the formation of a Mott gap is visible in the
cluster-solution. The $\mathcal{M}$-periodization is capable to
describe this part of the spectrum quite well (matching the
continuous black line of the impurity result). On the contrary,
the $\Sigma$-periodization creates artificial states in the Mott
gap. This is true also at low energy (region labeled B). A close
up on the Fermi level ($-t< \omega< t$) is displayed in the
middle panel (with again $\Sigma_{ano}=0$). The
$\mathcal{M}$-periodization reproduces the formation of a
low-energy pseudogap in the normal component of the spectra, as
already remarked in previous normal-state
studies\cite{tudor,tudor06}, while, once again, the
$\Sigma$-periodization introduces artificial states that fill the
pseudogap. This simple test therefore evidences the failure of the
$\Sigma$-periodization in well describing the high and low energy
normal-component of Green's function, in the regions of momentum
space where a pseudogap is present, like in the antinodal
$k$-point of Fig. \ref{ED-QMC}. The scenario is different if we
look at the low-energy {\it superconducting} density of states
$N(\omega)$ by restoring $\Sigma_{ano} \neq 0$ in eq.
(\ref{Gk-sup}), as presented in the bottom panel of Fig.
\ref{DOS-SM2}. Here it is the $\Sigma$-periodization line (green
dash line) which better portrays the cluster spectrum (black
continuous line). The $\mathcal{M}$-periodization this time
introduces spurious states close to the Fermi level ($\omega=0$).
This is not contradictory with respects to the result presented
above on the normal component of the system (top and middle
panels). By adding the d-wave superconducting gap in fact, we
selects at low energies ($\omega\rightarrow 0$) only the region
of momentum space close to the nodal points, where the gap is
zero. These points dominate in fact the sum
$-\frac{1}{\pi}\sum_{k} \hbox{Im} G^{}_{k}(\omega)_{11}$.
$N(\omega)$ for $\omega\to 0$ is therefore a direct probe of the
nodal point behavior only (while without superconducting gap, i.e.
$\Sigma_{ano}=0$, all $k$-points eventually contribute in the
summation at low energies).

We clarify this last statement. Let's assume as starting point
that at the nodes quasiparticles are well defined on the full
range of doping. We can extract from eq. (\ref{Gk-sup}) the low
energy ($\omega_{n}\rightarrow 0$) Green's function:
\begin{eqnarray}
\hbox{Re}\Sigma_{k}(\omega_{n}) &\sim&
\hbox{Re}\Sigma_{k}(0)\nonumber
\\
\hbox{Im}\Sigma_{k}(\omega_{n}) &\sim& (1-Z^{-1}_{k}) \omega_{n}
\nonumber \label{Sigma-w0}
\end{eqnarray}
and
\begin{eqnarray}
G_{}(k,\omega_{n})&\sim&  \frac{ \frac{Z_{k}}{2} ( 1-
\tilde{\xi}^{}_{k}/E_{k} ) } {\imath \omega_{n}+ E_{k}}+ \frac{
\frac{Z_{k}}{2} ( 1+ \tilde{\xi}^{}_{k}/E_{k} ) } {\imath
\omega_{n}- E_{k}} \label{Gk-sup-W0}
\end{eqnarray}
%-----------------------------------------------------------------
For convenience's sake, we have enlighten the quasiparticle
dispersion
\begin{eqnarray}
\tilde{\xi}_{k}&=& Z_{k} \, \left[ \xi_{k} + \hbox{Re} \Sigma_{k}(0) \right] \nonumber \\
E_{k}  &=&\,\sqrt{|\tilde{\xi}_{k}|^{2}+
|Z_{k}\,\Sigma_{ano}(k,0)|^{2}} \label{Ek}
\end{eqnarray}
The nodal point is the only one gapless at low energies and the
quasiparticle spectrum can be linearized (eq. \ref{Ek}):
\begin{eqnarray}
E_{k_{nod}}  &=& \,\sqrt{|\tilde{\xi}_{k_{nod}}|^{2}+
|Z_{k_{nod}}\,\Sigma_{ano}(k_{nod})|^{2}} \nonumber \\
 &=& \,\sqrt{ v_{nod}^{2} k^{2}_{\bot}+ v_{ano}^{2} k^{2}_{\|}}
\label{Ek-linear}
\end{eqnarray}
$v_{nod}= |\nabla_{k} \tilde{\xi}_{k}|$ is the quasiparticle
Fermi velocity perpendicular to the Fermi surface, $v_{ano}=
Z_{nod} | \nabla_{k} \Sigma_{ano}(k)|$ is parallel to the Fermi
surface.
%----------------------------------------------------------------
After analytic continuation $\omega_{n}\rightarrow \omega+\imath
\delta$, it is now easy to calculate the low energy behavior of
the one-particle density $N(\omega)= \frac{1}{\pi} \, \sum_{k}\,
\hbox{Im} G_{k}(\omega) $
\begin{equation}
N(\omega) \sim \frac{1}{\pi} \sum_{j}^{\tiny \hbox{nodes}}\,
\frac{Z_{nod_{j}}}{v_{nod_{j}} v_{ano_{j}} }  \, \omega
\label{DOS}
\end{equation}
i.e. $N(\omega)$ is linear in frequency close to the Fermi energy
and the slope is uniquely determined by the quasiparticle nodal
velocity $v_{nod}$, by the nodal derivative of the
superconducting gap $v_{ano}$ and by the nodal quasiparticle
residuum $Z_{k_{nod}}$. In other words, the low-energy density of
states in the d-wave superconductor is a direct measure of the
spectra at the nodes. This result explains the low-energy
spectrum of the cluster density of states (black-continuous line
in Fig. \ref{DOS-SM2}), which is roughly linear for $\omega\to 0$,
in the limit of the energy-resolution given by ED impurity solver.

The fact that here the $\Sigma$-periodization better portrays the
cluster result is consistent with the hypothesis of Fermi-liquid
behaviour at the nodal points of momentum space, well described
by the self-energy. The results of the normal component of the
system suggest instead that in other regions of momentum space
(at least close to the Mott transition point), a pseudogap opens
in the spectrum and quasiparticles die in the anti-nodes. In the
latter regions a cumulant $\mathcal{M}$-periodization is more
appropriate than the $\Sigma$-periodization.
%===============================================================
\section{Properties of the nodal and antinodal points}
\label{sec4}

In this section we follow the observations presented above on the
nodal/antinodal dichotomy and study the physical properties of
the system in the nodal and antinodal points, completing in
detail the work presented in ref.\cite{marce08}. In order to be
able to make contact with the experiments, we need to extract the
fully momentum dependent Green's function $\hat{G}_{k}(\omega)$
(see equation \ref{Gk-sup}), which can be generally related to
the response functions. We employ therefore a periodizing
procedure to extract the $k$-dependent normal and anomalous
components $\Sigma_{k}(\omega)$ and $\Sigma_{ano}(k,\omega)$ of
the $k$-dependent self-energy. As discussed above, it is
reasonable to start by assuming a d-wave shape of the
superconducting gap, which is naturally obtained in our scheme by
periodizing the anomalous component of the cluster self-energy
via eq. (\ref{Skano}). At the nodes, we periodize also the normal
component of the self-energy via eq. (\ref{Sigmak}). This
guarantees in particular Fermi liquid properties in the nodal
region of momentum space. At the anti-nodes instead the insulating
properties of the normal component of the system (like the
formation of a pseudogap at low energy and of the Mott gap at
higher energies, see Fig. \ref{DOS-SM2}) are better portrayed by
periodizing the cumulant $\mathcal{M}^{nor}$, according to eq.
(\ref{Mknor-egn}).
%------------------%
\begin{figure}[!!tb]
\begin{center}
\includegraphics[width=9cm,height=9.00cm,angle=-0] {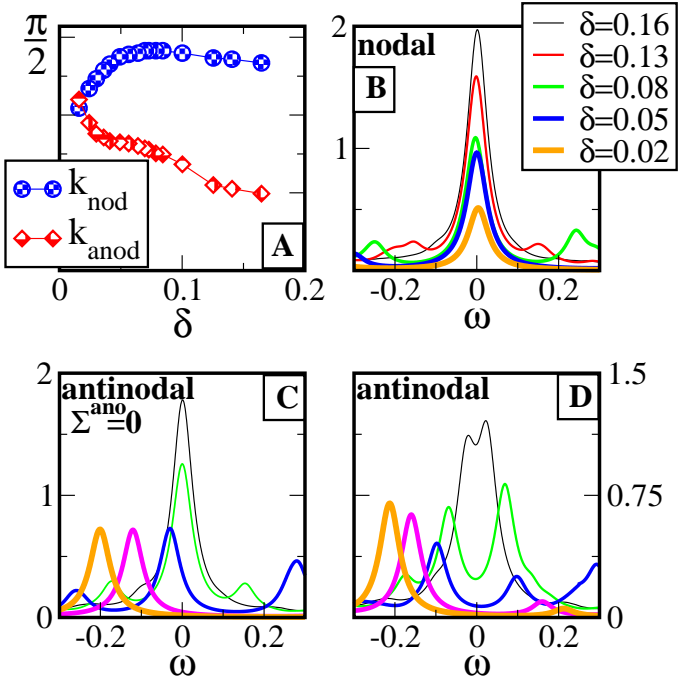}
\end{center}
\caption{(Color online). {\bf Panel A}: nodal $k_{nod}$ and
antinodal $k_{anod}$ positions of the quasiparticle peaks in the
first quadrant of the Brillouin Zone as a function of doping
$\delta$. {\bf Panel B}: Nodal peaks are found always at the
Fermi level and decrease with doping $\delta$ in approaching the
insulator. {\bf Panel C}: At high doping ($\delta> \delta_c\sim
0.08$) antinodal peaks are also found at the Fermi level in the
normal component of the system (set $\Sigma_{ano}=0$ in eq.
\ref{Gk-sup}). For ($\delta< \delta_p\sim 0.06$) however they
shifts to negative energy opening a pseudogap. {\bf Panel D}: The
actual antinodal spectra ($\Sigma_{ano}\neq 0$ in eq.
\ref{Gk-sup}) show to become asymmetric for $\delta> \delta_c$,
when the pseudogap starts opening in the normal component (see
panel C, and bottom right panel of Fig. \ref{ImG11} ).}
\label{Peaks}
\end{figure}
%---------------------------------------------------------------

By adopting this procedure we can first obtain quasi-particle
spectra in the {\it nodal and antinodal points of momentum
space}, as it is shown in Fig. \ref{Peaks}. In order to define the
$k$ vectors in the nodes and in the anti-nodes, we follow an
operative definition, similar to the one used in the ARPES
experiments of ref.\cite{tanaka06}. The nodal point $k_{nod}$
[antinodal point $k_{anod}$] is chosen as the one having the
sharpest quasiparticle peak in moving on the path
$(0,0)\to(\pi,\pi)$ [$(0,\pi)\to(\pi,\pi)$] of momentum space.
The vectors $k_{nod}$ and $k_{anod}$ as a function of doping
$\delta$ are shown in panel A of Fig. \ref{Peaks}. We notice that
$k_{anod}$ is a monotonic decreasing function of the $\delta$.
This is intuitively expected in a standard Fermi liquid, where
the approach to the Mott transition at $\delta=0$ is accompanied
by an increase of the volume enclosed by the Fermi surface in
momentum space, which is proportional to the density $n=1-\delta$
of the system (Luttinger theorem\cite{Luttinger60}). The
$k_{nod}$ vector on the contrary displays a non-monotonic
behavior, showing to decrease by reducing doping in
correspondence of the special doping $\delta_c\sim 0.08$, which
appears in our study as a critical point marking a change in the
physical properties of the system. In the following section we
will show that we can relate this behavior of the nodal and
antinodal $k$ points to a topology change in the Fermi surface.

In the remaining panels of Fig. \ref{Peaks}, we show the spectral
function Im$[\hat{G}_{k}(\omega)]_{11}$ (from eq. \ref{Gk-sup}).
As previously explained (see also appendix \ref{apxA}), within the
ED-CDMFT method it is possible to display Green's functions on
the real frequency axis $\omega$ by expressing them in a pole
expansion, displayed by adding in the denominator a small
imaginary part i$\eta$ (here we adopted $\eta=0.03t$). At the
nodal point (panel B) the d-wave superconducting gap is zero, and
a Fermi liquid quasi-particle peak is always found at the Fermi
level ($\omega=0$) for different doping $\delta$. The approach to
the Mott insulator ($\delta\to 0$) is marked by a progressive
reduction of the quasi-particle peaks. This behavior is
reminiscent of the Mott transition described in the standard
infinite dimensional Hubbard Model. At the antinodal point the
superconducting gap is maximal, a gap in the spectra is therefore
expected. Before looking at the full antinodal spectrum, however,
it is instructive to look at the contribution coming from the
normal component, which can provide information on the physical
properties of the liquid underlying the superconducting state.
The normal-component spectra can be simply obtained in our scheme
by zeroing the anomalous component of the self-energy
$\Sigma_{ano}(k,\omega)$ in eq. \ref{Gk-sup}. These spectra are
displayed in panel C of Fig. \ref{Peaks}. At doping $\delta>
\delta_c$, a Fermi liquid quasi-particle peak is also found at
the Fermi level ($\omega=0$). In this region of the phase
diagram, the normal properties of the system are therefore
Fermi-liquid-like (and also $k_{nod}$ and $k_{anod}$ are
monotonically decreasing with $\delta$, as described above). In
correspondence of the critical doping $\delta_c$ however, a
pseudogap opens and a quasi-particle peak is found at the gap
edge at negative energy ($\omega<0$). The pseudogap increases in
approaching the Mott transition ($\delta\to 0$), while,
differently from the nodal point, the peaks show a roughly
constant height. This behavior at the anti-nodes of the normal
component of this superconducting solution can be smoothly
connected to results previously obtained in CDMFT studies of the
normal state\cite{tudor,tudor06}. For $\delta<\delta_c$
therefore, the normal component of the system is not a Fermi
liquid in the strict sense, at least in the region of momentum
space close to the antinodal points (but a behavior unusual for a
Fermi liquid is also detected by the decreasing value of
$k_{nod}$ in panel A). This behavior appears also in the total
antinodal spectra (upon restoring the superconducting gap
$\Sigma_{ano}\neq 0$ in eq. \ref{Gk-sup}), which we show in panel
D. At doping $\delta> \delta_c$, the quasi-particle peaks present
in the normal component are parted into two bands by the opening
of a superconducting gap, resulting in the typical BCS symmetric
spectra. For $\delta< \delta_c$, however, the pseudogap already
present in the normal component super-impose to the
superconducting gap, resulting in asymmetric spectra. This
antinodal spectra nicely explain Fig. \ref{ImG11}, where the local
density of states $N(\omega)$, directly obtained from the
cluster-impurity solution, is displayed. The appearance of the
asymmetry for $\delta< \delta_c$ is therefore interpreted by our
$k$-momentum analysis as the appearance of the pseudogap phase,
which marks a departure from a Fermi liquid based BCS
superconductor, once again at the critical doping $\delta_c$. And
these observations must be directly linked to experimental
spectra, either in the anti-nodes of momentum space, obtained for
example with angle resolved
photo-emission\cite{damascelli,campuzano}, and locally in real
space, with for example scanning tunneling
spectroscopy\cite{davis05}.
%---------------------------------------------------------------
\begin{figure}[!!t]
\begin{center}
\includegraphics[width=8cm,height=6.0cm,angle=-0] {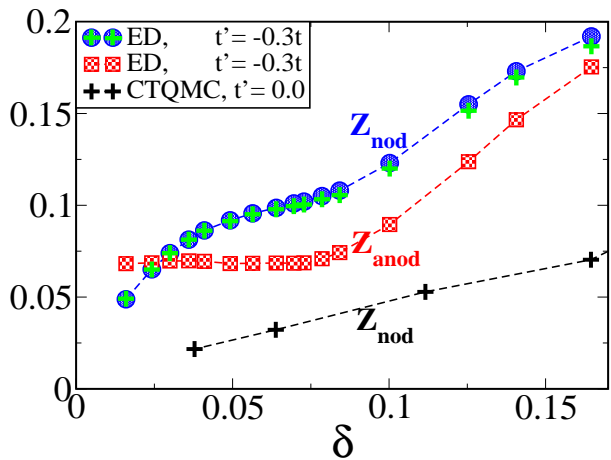}
\end{center}
\caption{(Color online). Quasiparticle residuum at the nodal point
$Z_{nod}$ and antinodal points $Z_{anod}$ as a function of the
doping $\delta$, evaluated as area of the quasiparticles peaks
shown in Fig. \ref{Peaks} (as a crosscheck, $Z_{nod}$ is also
evaluated as the slope of the imaginary part of the self-energy
on the Matsubara axis, green cross). While $Z_{nod}$
monotonically decrease with doping, as in the standard picture of
the Mott transition, the antinodal $Z_{anod}$ shows to stay
constant upon opening of the pseudogap ($\delta> \delta_c\sim
0.08$). For comparison's sake, we display the nodal $Z_{nod}$
obtained by CTQMC\cite{haule-ctqmc} on the two dimensional
Hubbard Model with similar parameters ($U=12t, t'=0.0,
\beta=200$). These quasiparticle residua should be compared with
their cluster counter-parts in Fig. \ref{Zcluster}.}
\label{Z-nodal}
\end{figure}
%===========================================================

In order to characterize the Mott transition, it is useful to
extract the quasi-particle residuum, which is defined as
\begin{equation}
Z_k= \left(1-\frac{\partial
\hbox{Re}\Sigma^{nor}(k,\omega)}{\partial\omega}\right)^{-1}_{\omega\to
0, k=k_{F}}\label{Zk}
\end{equation}
with $k_{F}=k_{nod}$ or $k_{anod}$ in our case. This quantity
corresponds to the area of the quasiparticle peaks (e.g. panel A
of Fig. \ref{Peaks}), it is unity in the non-interacting case,
less than unity in an interacting Fermi liquid. In the standard
description of the Mott transition (i.e. the infinite dimensional
Hubbard Model) $Z\to 0$ linearly as $\delta\to 0$ (see ref.
\cite{revmodmft}). It is interesting therefore to observe how this
quantity behave in the nodal and antinodal $k$-points by varying
doping $\delta$ , as we show in Fig. \ref{Z-nodal}. We can give
in this way a momentum space interpretation of the corresponding
cluster $Z_{X}$ that we have discussed in Fig. \ref{Zcluster}. As
we already said, at the nodes well defined quasiparticle peaks
are observed at every doping $\delta$ at the Fermi level. The
residuum $Z_{nod}$ is therefore well defined according to the
expression given above (which is strictly valid at $\omega=0$),
as in a typical Fermi liquid. This is confirmed by the good
numerical accord between the blue circles (calculated by
integrating the area of the peaks, which are displayed by
introducing the artificial broadening parameter $\eta=0.03t$) and
the green crosses (calculated more precise within our method by
using eq. \ref{Zk} on the Matsubara axis). For comparison's sake,
we present also the nodal quasi-particle residuum extracted in a
CDMFT-study implemented with a different impurity solver, the
CTQMC\cite{haule-ctqmc}, in the two dimensional Hubbard model
with Hamiltonian parameters $U=12t$ and $t'=0t$. The comparison
is only qualitative and it is aimed to get insight into the
physical trends. We indeed observe a monotonically decreasing
$Z_{nod}$ as a function of the doping $\delta$, similarly to the
standard infinite dimensional Hubbard Model. We cannot however
state within our numerical resolution if $Z_{nod}\to 0$ exactly
at the Mott point $\delta=0$ (as it seems also to suggest the
CTQMC result), or rather it extrapolates to a finite but very
small ($Z_{nod}< 0.02t$) value. In the antinodal point more
attention has to be paid in defining a quasi-particle residuum
$Z_{anod}$. As stressed above, for $\delta> \delta_c\sim 0.08$
quasiparticle peaks are present at the Fermi level (panel C of
Fig. \ref{Peaks}), and $Z_{anod}$ can be well defined by eq.
\ref{Zk}. For $\delta< \delta_c\sim 0.08$ however a pseudogap
opens. Even if a peak can be identified at the gap-edge, it is
not strictly speaking a Landau-Fermi liquid quasiparticle, as the
imaginary part of the self-energy is non-zero (even if small),
i.e. the quasiparticle has a finite lifetime. Formula \ref{Zk}
cannot be directly employed. However we can still calculate the
area of the peak, and display its behavior as a function of the
doping $\delta$. We find in our result that for
$\delta<\delta_c$, once the pseudogap opens and the peaks move to
negative frequency, the weight $Z_{anod}$ stays constant up to
the Mott transition point. The behavior of $Z_{nod}$ and
$Z_{anod}$ here presented has to be connected with the effective
cluster correspondents $Z_X$ (Fig. \ref{Zcluster}). According to
eq. \ref{gamma-k}, these are interpreted in momentum space as the
quasiparticle residua in the corner points of the first quadrant
of the Brillouin Zone (therefore far from the Fermi surface),
while $Z_{nod}$ and $Z_{anod}$, which are instead calculated on
the Fermi surface, have a real physical meaning. In spite of this
however, the cluster $Z$s already embody the physical properties
(a $Z$ going to zero and another non decreasing for $\delta\to
0$) characteristic of the Mott transition in this two-dimensional
system.

We stress that the description of the Mott transition we find in
this study of the two dimensional Hubbard Model is very different
from the standard Mott transition picture in infinite
dimension\cite{revmodmft}. In our system different regions of
momentum space behave very differently in approaching the
transition point. In this way we can go from a Fermi-liquid-based
superconductor (realized for $\delta> \delta_c$) into the Mott
insulator (at $\delta=0$) by passing through a phase
$0<\delta<\delta_p$, where the system is at the same time
insulating in the antinodal region and Fermi liquid in the nodal
region of momentum space. This latter appears to approach the
Mott point in the standard (infinite dimensional Hubbard Model)
way, with a quasiparticle residuum $Z_{nod}\to 0$ (at least
within our numerical precision). In the antinodal region instead
the quasiparticle peak (which underlies the superconducting gap)
stops reducing at $\delta=\delta_c$ and shifts to negative
energies opening a pseudogap. This behavior is reminiscent of the
{\it orbital selective Mott transition}, found e.g. in two band
Hubbard-like models
\cite{deleo-2008,ferrero-2005-72,medici-2005-72}, where the
spectral weight is not transferred from the low energy
($\omega=0$) to the Hubbard bands (located at a energy scale of
order $\sim U$), as in the standard Mott transition, but rather
onto a smaller energy scale of the order of an exchange coupling
$J\sim 1/U^{2}$, inside the Mott gap. In spite our model is a one
band one, different regions of momentum space appear to behave as
different bands. By decreasing doping $\delta$, a first "orbital
selective Mott transition" takes place at $\delta_c$ in the
antinodal regions, and a full Mott transition takes finally place
at $\delta= 0$.

%------------------------------------------------------------------------
\begin{figure}[!!t]
\begin{center}
\includegraphics[width=8cm,height=16.0cm,angle=-0] {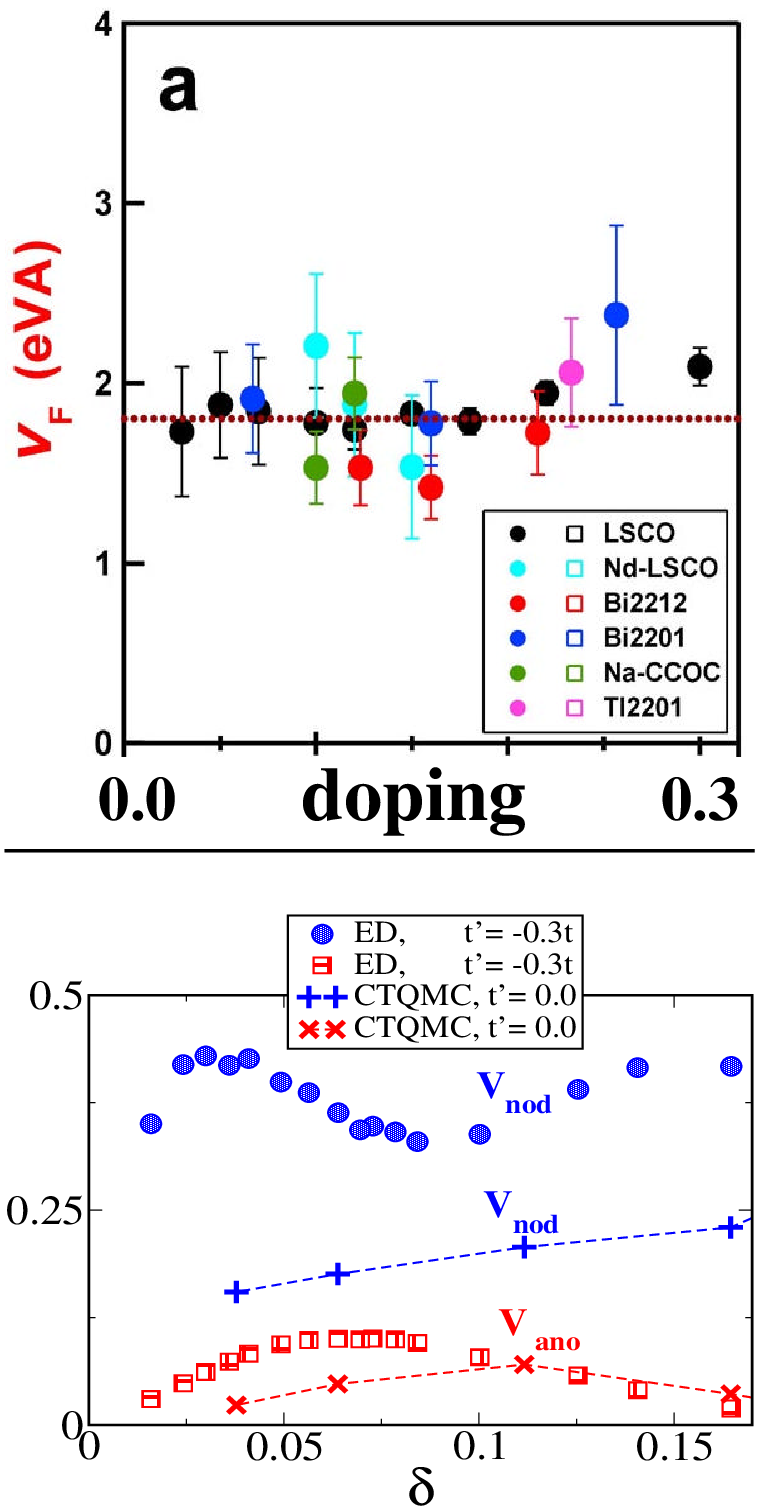}
\end{center}
\caption{(Color online). {\bf Top}: Nodal velocity as a function
of doping measured from the quasiparticle dispersion of different
materials (figure taken from ref.\cite{Shen-Nature03}). To be
confronted with the nodal component of the velocity $v_{nod}$
calculated in our work and displayed in the bottom. {\bf Bottom}:
Nodal anomalous velocity $v_{ano}= Z \Sigma_{ano}(k_{nod})$
(tangent component of the nodal velocity) and nodal quasiparticle
velocity $v_{nod}= Z|\nabla_{k} \xi_{k}|$ (component of the nodal
velocity perpendicular to the Fermi surface) as a function of
doping $\delta$ (units are $a_{o}t$, where $a_{o}$ is the square
lattice spacing). The trends are compared with
CTQMC\cite{haule-ctqmc} calculations on the two dimensional
Hubbard Model with similar parameters ($U=12t, t'=0.0,
\beta=200$). } \label{Z_Sigano-S}
\end{figure}
%==========================================================
Useful information on the nodal point can be extract by
performing a low energy expansion of the Green's function (see
eq. \ref{Gk-sup-W0} and \ref{Ek-linear}), taking advantage of the
Fermi liquid properties of the nodal point. In particular the
nodal velocity has two components, one coming from the normal part
$v_{nod}= |\nabla_{k} \tilde{\xi}_{k}|$ parallel to the Fermi
surface, and the other related to the superconducting gap
$v_{nod}= Z_{nod} | \nabla_{k} \Sigma_{ano}(k)|$, perpendicular
to the Fermi surface. $v_{nod}$ can be experimentally extracted,
e.g. from the ARPES quasiparticle dispersion at the
node\cite{Shen-Nature03}, while $v_{ano}$ can be determined e.g.
as the slope of the superconducting gap at the
node\cite{tanaka06}. In Fig. \ref{Z_Sigano-S} $v_{nod}$ and
$v_{ano}$ are displayed as a function of doping $\delta$. For
comparison's sake, in the top panel we show experimental nodal
velocity extracted from different materials (the figure has been
taken from the supplementary material of
ref.\cite{Shen-Nature03}). In the bottom panel we show our result,
and also insert for a qualitative comparison the CTQMC-CDMFT
result of ref.\cite{haule-ctqmc}. $v_{nod}$ shows to be greater
than $v_{ano}$, in agreement with experimental observation (see
e.g. ref.\cite{damascelli}). In ED-CDMFT it slightly oscillates
around a constant value from the over-doped to the under-doped
($\delta< \delta_c$) side of the phase diagram, while in the
CTQMC case it is slightly decreasing with decreasing doping. This
behavior is in good qualitative agreement with experimental
results reported in the top panel (a quantitative comparison
would roughly held, for a lattice spacing $a_{o}\sim 4 \AA$ and a
$t\sim 0.5$ eV, $v_{nod}\sim 1$ eV$\AA$, which is of the order of
magnitude of experiments). Remarkably, either with ED and with
CTQMC, $v_{ano}$ displays a dome-like shape, with a maximum
around optimal doping $\delta_p<\delta<\delta_c$ (notice that in
the ED case $t^{\prime}=-0.3t$ $\delta_c \sim 0.08$ while in the
CTQMC case $t^{\prime}=0$ $\delta_c \sim 0.12$). This is an
important result: as shown in formula \ref{Ek} and discussed in
ref.\cite{marce08}, the anomalous velocity $v_{ano}=
\sqrt{2}Z_{nod}\Sigma_{ano}(\omega=0)\sin k_{nod}$ can be
interpreted as a direct measure of the superconducting gap in the
nodal region, which reveals non-monotonic, in agreement with some
recent experimental spectroscopy results\cite{tanaka06,tacon06}.
The agreement of the trend of $v_{ano}$ between the ED and CTQMC
solutions shows that this result is solid from the theoretical
side too. We will come back to discuss the nodal and antinodal
gap more in detail at the end of this section.

\begin{figure}[!!t]
\begin{center}
\includegraphics[width=8cm,height=6.0cm,angle=-0] {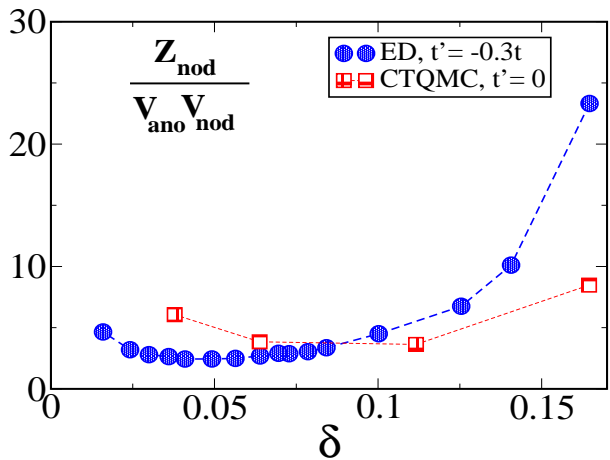}
\end{center}
\caption{(Color online). The slope of the low energy local density
of states $N(\omega)\sim \frac{Z_{nod}}{v_{\Delta} v_{nod}}
\omega$ is displayed as a function of doping $\delta$. Trends are
compared with CTQMC\cite{haule-ctqmc} calculations on the two
dimensional Hubbard Model with similar parameters ($U=12t, t'=0.0,
\beta=200$). } \label{gamma}
\end{figure}

At the nodal point, the combination of the quasiparticle residuum
$Z_{nod}$ and nodal velocities can give further information, which
can be confronted with experimental observable quantities and
which can further support the physical description drawn from our
CDMFT result. Basing on the Fermi liquid assumption at the nodes,
in eq. (\ref{DOS}) we have for example extracted the low energy
($\omega\to 0$) behavior of the local density of states
$N(\omega)\sim \frac{Z_{nod}}{v_{nod}v_{ano}}\, \omega$. This
value is displayed as a function of doping $\delta$ in Fig.
\ref{gamma}, comparing once again the ED-CDMFT results of this
work (with $t^{\prime}=-0.3t$) with the CTQMC-CDMFT results of
ref.\cite{haule-ctqmc} (with $t^{\prime}=0$). Once again it is
the trend we want to compare rather than the quantitative values.
Starting from the over-doped side, the slope of $N(\omega)$ is
decreasing monotonically by decreasing doping, until showing a
slight up-turn close to the Mott transition (observable both in
the ED and CTQMC cases). While the linearity of $N(\omega)$ it is
well established in scanning tunneling experiments\cite{davis05},
the behavior of the slope as a function of doping is at the
moment very difficult to extract (as it is not possible to obtain
absolute values for different densities). Our result (in
particular the up-turn tendency at small doping) has to be
therefore considered a theoretical prediction.

\begin{figure}[!!t]
\begin{center}
\includegraphics[width=8cm,height=16.0cm,angle=-0] {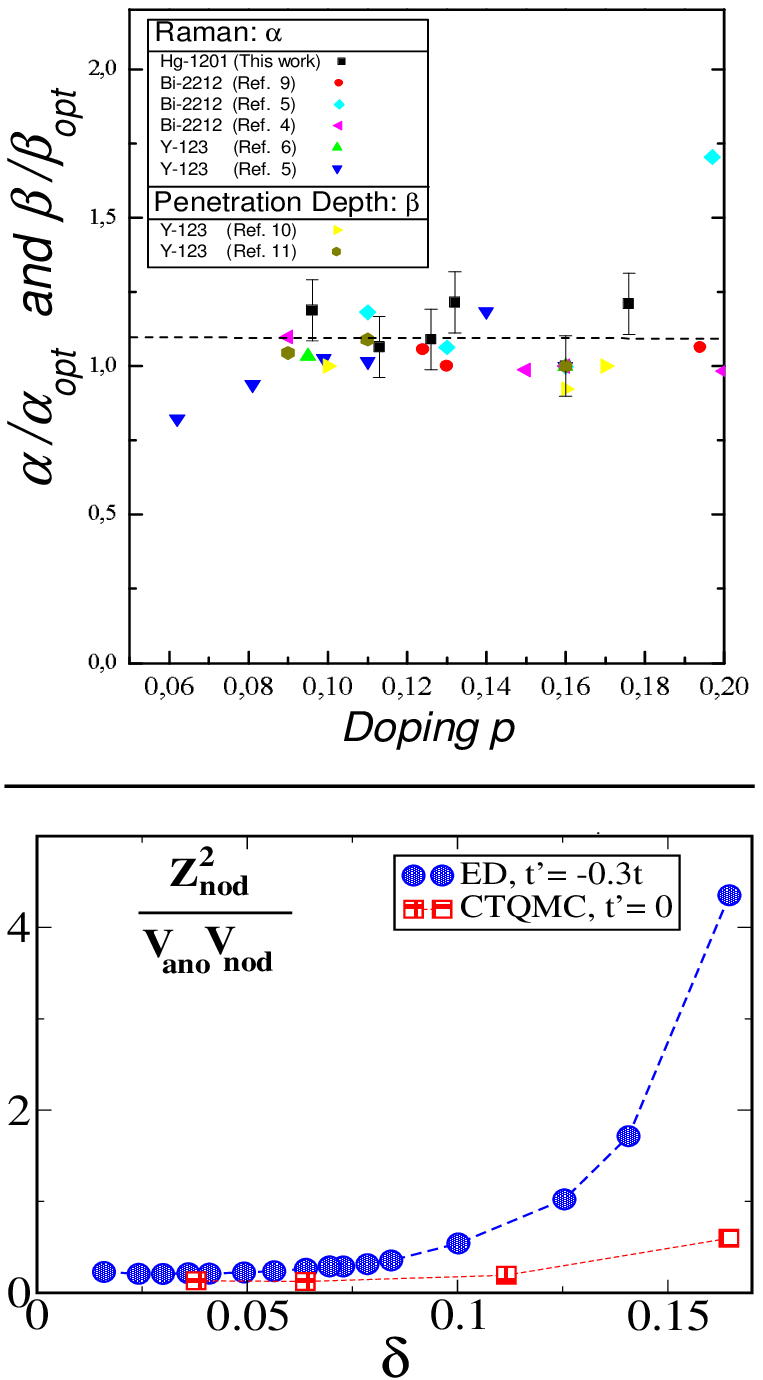}
\end{center}
\caption{(Color online). {\bf Top}: the linear coefficient of the
low energy Raman response $\chi_{B2g}(\omega)\sim \alpha \omega$
and the low temperature superfluid stiffness (from penetration
depth) $\rho_{s}(T)-\rho_{s}(0)\sim \beta T$, extracted from
various experiments (figure taken from ref.\cite{tacon06}).
Quantities are normalized at the optimal doping value. {\bf
Bottom}: The linear coefficient of the Raman response and
superfluid stiffness (see top panel) as extracted from our
calculation $\frac{Z_{nod}^{2}}{v_{\Delta} v_{nod}}$ and
displayed as a function of doping  $\delta$. Trends are compared
with CTQMC\cite{haule-ctqmc} calculations on the two dimensional
Hubbard Model with similar parameters ($U=12t, t'=0.0,
\beta=200$). } \label{alpha}
\end{figure}

A further ratio of experimental relevance can be connected at
first order to the low energy ($\omega\to 0$) linear behavior of
the $B_{2g}$ Raman response function  $\chi_{B2g}\sim
\frac{Z_{nod}^{2}}{v_{nod}v_{ano}}\, \omega$ (see
ref.\cite{tacon06}) and the low temperature ($T\to0$) behavior of
the superfluid stiffness $\rho_{s}(T)-\rho_{s}(0)\sim
\frac{Z_{nod}^{2}}{v_{nod}v_{ano}}\, \omega$, which can be
extracted from measures of the penetration depth
\cite{Bonn96,Panagopoulos98}. Our results (together with the
CTQMC results of Ref.\cite{haule-ctqmc}) are shown on the bottom
panel of Fig. \ref{alpha} and compared with the aforementioned
Raman and penetration depth data presented in the top panel (the
figure has been taken from ref.\cite{tacon06}). The remarkable
feature, found in experiments and supported in our calculation,
is the constant value displayed by this ratio in the under-doped
region ($\delta< \delta_c$). While in the experimental results a
sum rule is assumed, in order to being able to compare measures
from different doping/samples, our theoretical results are
derived from a bare strongly correlated electron model, where
other kind of assumptions and approximations (as explained in the
previous sections) are implied. The convergence of experimental
and theoretical results therefore strongly supports these
findings, presenting them as distinctive feature of the cuprate
superconductor nodal dispersion.

%----------------------------------------------------------------
%----------------------------------------------------------------

The behavior of the spectra  presented in Fig. \ref{Peaks} can be
directly connected to spectroscopy experiments. In particular,
recently a lively debate has risen on the momentum resolved
structure of the superconducting
gap\cite{deutscher99,kyung02,yang-2006-73,millis06,cho06,huefner-2008-71,valenzuela07,aichhorn-2007-99}.
We proceed as in ARPES experiments (see e.g. \cite{tanaka06}),
and extract the quasiparticle gap in the nodal and antinodal
points of momentum space, taking advantage of the periodizing
scheme we have introduced. To this purpose, it is convenient to
use the low energy ($\omega\to 0$) expansion carried out in eq.
\ref{Gk-sup-W0}. In particular we have seen that quasiparticles
peaks are always found in our result, even if not in a strict
sense (in the pseudogap region $\delta< \delta_c$ quasiparticle
peaks are located at the gap-edge and have finite lifetime), and
therefore we expect the expansion to be reasonably good at small
frequency (i.e. $\omega\leq$ the superconducting gap). We can in
this case write the total gap $\Delta_{tot}(k,\omega)$ as the
quadratic sum of two contributions (see eq. \ref{Ek}):
\begin{equation}
\Delta^{2}_{tot}(k,\omega)= \Delta^{2}_{nor}(k,\omega)+ \Delta^{2}_{sc}(k,\omega)
\label{gaps_kw}
\end{equation}
where $\Delta_{sc}(k,\omega)= Z_{k} \Sigma_{ano}(k,\omega)$ is the
usual d-wave superconducting gap (notice that it is directly
connected, except for constant factors, to the anomalous
component of the nodal velocity $v_{ano}\sim Z_{k_{nod}}
\Sigma_{ano}(\omega=0)$ discussed in Fig. \ref{Z_Sigano-S}),
$\Delta_{nor}(k,\omega)= \tilde{\xi}_{k}(\omega)= Z_{k} \, \left[
\xi_{k} + \hbox{Re} \Sigma_{k}(\omega) \right]$ is a normal
contribution to the gap which can arise only if, for some $k$ and
$\omega\to 0$, the normal component of the self-energy
$\Sigma_{k}(\omega)$ grows enough so that the band equation
$\tilde{\xi}_{k}=0$ cannot be satisfied, i.e. there is not Fermi
surface. Now, this does not take place at the nodes, where
quasiparticles are found for all dopings (panel B of Fig.
\ref{Peaks}), $\tilde{\xi}_{k}=0$, the system presents standard
Fermi liquid properties, and the total spectral gap coincides
with the superconducting gap $\Delta_{tot} \equiv \Delta_{sc}$.
(This already suggests that in looking for the ``real''
superconducting gap of the system one should look at the nodal
region gap. Recent theoretical\cite{haule-ctqmc} and experimental
Raman spectroscopy studies\cite{guyard-2008} point at this
direction.) In the antinodal region, instead, we have observed
that a pseudogap opens in the normal component for $\delta< 0.08$
(panel C of Fig. \ref{Peaks}), and this fact is associated in our
calculation with the appearance of lines in $k$-space close to
the antinodal region where the self-energy is diverging (see
ref.\cite{tudor,tudor06}). It is not possible therefore to
satisfy the equation $\tilde{\xi}_{k}=0$, and a Fermi surface
does not exists anymore. The normal contribution $\Delta_{nor}$
kicks in, and determines the properties of the total antinodal
gap $\Delta_{tot}$, originating the asymmetric spectra we already
described in panel D of Fig. \ref{Peaks}.
%----------------------------------------------------------------
\begin{figure}[!!t]
\begin{center}
\includegraphics[width=8cm,height=16.0cm,angle=-0] {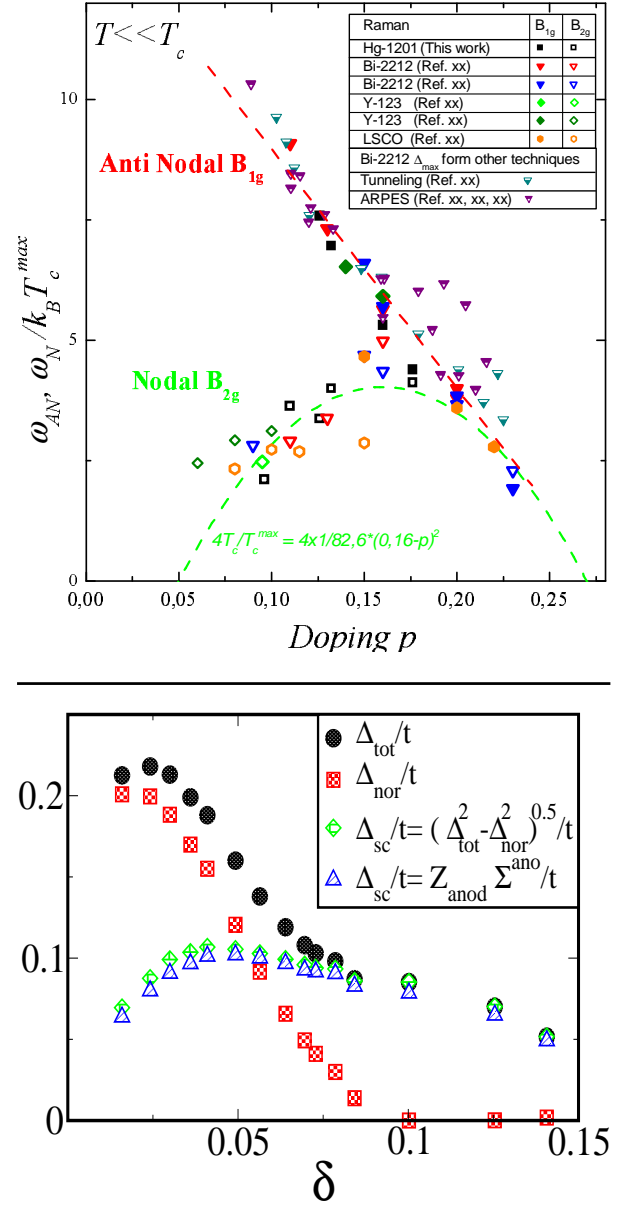}
\end{center}
\caption{(Color online). {\bf Top}: Nodal ($B_{1g}$) and antinodal
($B_{2g}$) quasiparticle gap extracted by different spectroscopy
experiments (in particular, the figure is taken from the Raman
spectroscopy results of ref.\cite{tacon06}) as a function of
doping. {\bf Bottom}: The antinodal gap $\Delta_{tot}$ and the
nodal $\Delta_{Sc}$ extracted from the spectra of Fig.
\ref{Peaks}, as a function of doping $\delta$. $\Delta_{nor}$ is
the pseudogap in the normal component extracted in panel C of
Fig. \ref{Peaks}.} \label{GAPS}
\end{figure}
%----------------------------------------------------------------

We can extract, like in ARPES experiments (e.g.
ref.\cite{tanaka06}), the antinodal gap $\Delta_{tot}$ from the
spectra of panel D of Fig. \ref{Peaks} by measuring the distance
of the quasiparticle peaks at the gap-edge from the Fermi level
$\omega=0$. In the same way, from panel C of Fig. \ref{Peaks}, we
can measure the normal contribution $\Delta_{nor}$, and display
them as a function of doping $\delta$. We can compare directly
with experimental results, which are shown in the top panel of
Fig. \ref{GAPS} (for convenience's sake we extract the picture
from the Raman results of ref.\cite{tacon06}, but ARPES points are
also displayed), while in the bottom panel we show the results of
our calculation. The antinodal gap $\Delta_{tot}$ is a monotonic
decreasing function of doping (curve labeled ``Anti Nodal
B$_{1g}$'' in the top panel of Fig. \ref{GAPS}), as it has been
known from experiments since a long time (see e.g.
ref.\cite{damascelli, campuzano}) and predicted in the most
popular theories of high temperature superconductivity ( e.g.
resonating valence bond theory\cite{anderson}, for a recent
general review see e.g.\cite{patrickrmp}). In our calculation we
show that at small doping an important contribution comes indeed
from the normal component $\Delta_{nor}$, which appears at
$\delta_c$ and it is also monotonic. From the experimental side,
the novelty comes from precise measures of the nodal gap,
recently obtained by Raman spectroscopy\cite{tacon06} and
ARPES\cite{tanaka06}, which show surprisingly that in this region
of momentum space the spectral gap is non-monotonic with doping
$\delta$, tracking instead the behavior of the critical
temperature $T_c$ (curve labeled ``Nodal B$_{2g}$'' in the top
panel of Fig. \ref{GAPS}). In our calculation the nodal gap
corresponds to the anomalous component of the nodal velocity
$v_{ano}$, which we have discussed in Fig. \ref{Z_Sigano-S}, and
which shows indeed a behavior strongly similar to these
experimental results. We clarify now how it is possible that the
nodal component of the gap tracks $T_c$ as a function of doping
$\delta$, while at the same time the antinodal component is
monotonic, by disentangling he superconducting contribution
$\Delta_{sc}= \sqrt{ \Delta^{2}_{tot}- \Delta_{nor}^{2}}$, which
we also display in the bottom panel of Fig. \ref{GAPS}. To check
the validity of our formula \ref{gaps_kw}, and making connection
with $v_{ano}\sim Z_{k_{nod}} \Sigma_{ano}$, which has been
evaluated at the nodal point, we also evaluate and display
$\Delta_{sc}= Z_{anod}
\Sigma_{ano}(k_{anod},\omega=\Delta_{nor})$, finding numerical
agreement. This shows indeed that the total antinodal gap
$\Delta_{tot}$ has indeed two distinct contributions, displaying
opposite trends with dopings. In the under-doped side
$\Delta_{nor}$ dominates in the antinodes, and creates a
monotonically increasing total gap. In the nodes, instead,
$\Delta_{nor}$ is zero and only the superconducting gap is
detectable. In our results we connect therefore the
experimentally observed two-gap phenomenon with the opening of
the pseudogap at the anti-nodes at a finite critical doping
$\delta_c$. And, according to our point of view, this is an
effect arising in a strongly correlated electron system that
approaches the doping-driven Mott transition in a two dimensional
lattice. The fact that $\Delta_{sc}$ tracks
$T_c$\cite{guyard-2008} is remarkably similar to the standard BCS
superconductivity. These results put strong constrains on the
theories of cuprate-based superconductivity, which have to
consider the presence of these two distinct components in the
spectral gap and its interplay with the dome-like shape of the
order parameter, the rising of the pseudogap phase and the
approaching to the Mott insulator. Our cluster results,
interpreted via a periodization procedure, well fit the
experimentally observations and give a simple interpretation in
terms of a combination of all these effects.

\section{A mixed-periodization scheme} \label{sec5}

The discussion presented in the previous subsections is valid
close to the nodal and antinodal points of momentum space, where
we have shown that periodizing the self-energy or the cumulant
its a reasonable approximation. Obtaining information in
intermediate region of momentum space, between the nodal and
antinodal ones, is beyond the limits of a pure $2 \times 2
$-plaquette study. In this region in fact the Fermi liquid
physical properties of the nodal points have to interlace in some
non-trivial way with the insulating-like properties of the
antinodal point. A detailed description of this phenomenon can be
taken into account by studying bigger cluster (i.e. obtaining a
better truncation of the Fourier expansion in eq. \ref{Wk}), at
present not possible with ED-CDMFT. Nevertheless we want to keep
in this section a low profile, and introduce a first order
description of the spectral properties in all momentum space,
which could be compared to experimental results (like e.g.
ARPES). We want therefore to introduce a periodizing scheme able
to describe
\begin{enumerate}
  \item a Fermi liquid quasiparticle in correspondence of the nodal point
  \item  the opening of a pseudogap in the antinodal points in the
          under-doped regime
  \item the formation of the Mott gap at high energies in approaching the Mott
   insulator (see top panel of Fig. \ref{DOS-SM2} region marked A).
\end{enumerate}
We can satisfy the first condition by using the self-energy
$\Sigma$-periodization. The second and third points are instead
obtained by using the cumulant $\mathcal{M}$-periodization, which
is able to describe the formation of the antinodal.
%================================================================
\begin{figure}[!!tb]
\begin{center}
\includegraphics[width=8.5cm,height=4.0cm,angle=-0] {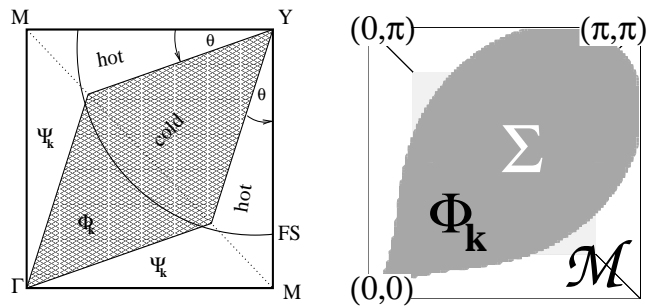}
\end{center}
\caption{Left: the shaded region is the patch $\Phi_{k}$ used in
the phenomenological Boltzmann approach of ref.\cite{PSK} in the
first quadrant of the Brillouin Zone. Right: the patch $\Phi_{k}$
used in this work (see formula \ref{Phi_k}).} \label{2patch}
\end{figure}
%================================================================
%---------------------------------------------------------------

In order to combine all these requests, we take advantage from
ideas introduced in phenomenological Fermi-liquid Boltzmann
approaches to the normal-state transport properties of cuprate
superconducting
materials\cite{PSK,Millis1998,Zheleznyak1998,Hlubina1995}. We
base in particular on the work of ref.\cite{PSK}, where the
division of momentum space in nodal Fermi-liquid-like regions,
{\it cold spots}, and antinodal insulator-like regions, {\it hot
spots}, was achieved by projecting the quasiparticle scattering
operator $C_{k k'}$ on a basis of patches $\{ \phi_{\alpha}\}$ in
momentum space with different temperature scattering-dependencies:
\begin{equation}
C_{k k'}=\, \sum_{\alpha\beta}\, \phi_{\alpha}(k) C_{\alpha\beta}(T) \phi_{\beta}(k)
\label{Ckk1}
\end{equation}
In this case, it is possible to solve exactly the Boltzmann
equation for the simple case of two patches (one marking the
nodal and the other the antinodal region). The shape and the
scattering properties of the cold patch (modeled by a small set of
parameters) was fixed by obtaining the best fit on few transport
quantities (resistivity, Hall coefficient), and a systematic
correspondence with others transport quantities
(magnetoresistance and termoelectric power) was then obtained.
For comparison, the nodal patch $\Phi_{k}$ used in ref.\cite{PSK}
is shown in the left hand side of Fig. \ref{2patch} (the
antinodal patch is simply defined as $1- \Phi_{k}$).

In the same spirit, we introduce here a mixed
periodization-scheme, by projecting the lattice self-energy
$\Sigma_{k}(\omega)$ on a nodal $\Phi(k,\omega)$ and an antinodal
$1-\Phi(k,\omega)$ patch in momentum space:
\begin{equation}
\Sigma_{k}(\omega)=    \, \Phi(k,\omega) \Sigma_{k}(\omega)
[\hat{\Sigma}]+ [1-\Phi(k,\omega)] \Sigma_{k}(\omega)
[\hat{\mathcal{M}}] \nonumber \label{Sigmak-mix}
\end{equation}
where $ \Sigma_{k}(\omega) [\hat{\Sigma}]$ is obtained by
periodization of the cluster $\Sigma$ and $\Sigma_{k}(\omega)
[\hat{\mathcal{M}}]$ by periodization of the cluster cumulant
$\mathcal{M}$ (eq. \ref{M-cluster}). The patch $\Phi(k,\omega)$
separates the $\Sigma$-periodized from the
$\mathcal{M}$-periodized regions (right hand side of Fig.
\ref{2patch}). For convenience's sake, we choose a form with $k$
and $\omega$ separable:
\begin{eqnarray}
\Phi(k,\omega)=\, \Psi(k) \, \Psi(\omega) \nonumber \\
%\hbox{where} \\
\Psi(\alpha)= \frac{1}{2} \,\left[ 1- \tanh( \beta_{\alpha}
r_{\alpha} ) \right]
\label{Phi_k}
\end{eqnarray}
where $r_{\alpha}$ is the distance from the a center in $k$ or
$\omega$ space:
\begin{eqnarray}
r_{k}&=&\, \sqrt{(k_{x}-k_{x_{c}})^{2}+(k_{y}-k_{y_{c}})^{2}}-r_{o}(\theta) \nonumber \\
r_{\omega}&=&\, |\omega|- \omega_{o}
\end{eqnarray}
Here $(k_{x_{c}},k_{y_{c}})\sim (\frac{\pi}{2},\frac{\pi}{2})$,
$r_{o}(\theta)= r_{o}\,
e^{-(\theta-\frac{\pi}{4})/\sigma_{o}^{2}}$ and $\theta=
\arctan(k_{y}/k_{x})$ can be chosen to properly shape the patch.
We fix through all the rest of the paper $\omega_{o}\sim 0.5 t$,
$\beta_{k}\sim 100$ and $\beta_{\omega}\sim 15$, which smooth the
boundaries of the patch. It remains to fix the parameters
$\mathbf{k}_{c}$, $r_{o}$ and $\sigma_{o}$, which have to be
chosen to mimic the phenomenological patch of Fig. \ref{2patch}.
There is of course a good degree of arbitrariness in its form. We
want however to have a first order qualitative description of the
physical properties in momentum space, with the aim to address
consideration that are only slightly dependent on the exact form
of the patch.

\subsection{Evolution of the Fermi Surface}

%===============================================================
\begin{figure}[!!tb]
\begin{center}
\includegraphics[width=8.5cm,height=5.6cm,angle=-0] {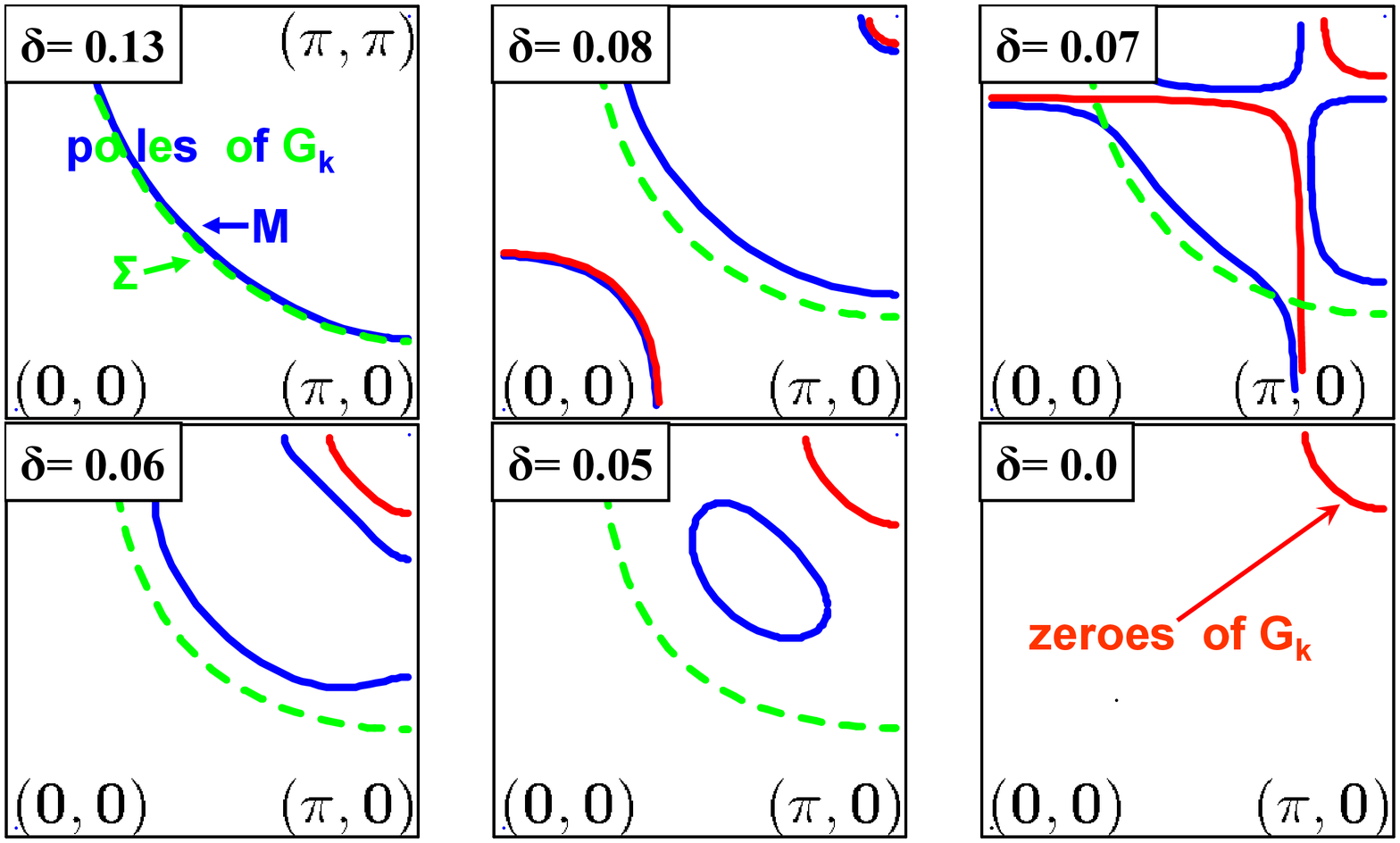}
\end{center}
\caption{(Color online). We show the evolution of the
normal-component Fermi surface (blue continuous line in the
$\mathcal{M}$-periodization scheme, green dashed line in the
$\Sigma$-periodization scheme), defined has $\xi_{k}+ \hbox{ Re
}\Sigma_k =0$ in eq. \ref{Gk-sup}. Panels span from the
over-doped region ($\delta=0.13$ top-left panel) to the Mott
insulator ($\delta=0.0$ bottom-right panel). {\bf By periodizing
the cumulant $\mathcal{M}$}, the Fermi surface disappears in the
insulating state and it is replaced by a line of zeroes of
$G_{k}$ (red lines). The way this takes place is via a
topological phase transition around $\delta_p\sim 0.06
<\delta<\delta_c\sim 0.08$, from a large hole-like Fermi surface
to a pocket-like Fermi surface. The low doping region is marked
by a phase where Fermi surface and lines of zeroes co-exists.
This gives origin to the pseudogap and the fragmentation of the
Fermi arcs in measures of spectra of the normal state system (see
e.g. Fig. \ref{Akw-SM}). {\bf By periodizing the self-energy
$\Sigma$} instead, by construction, the lines of zeroes cannot
appear. The pseudogap is realized only trough a modulation of the
spectral intensity along the Fermi surface line, which extends
always on all the Brillouin Zone (see also Fig. \ref{Akw-SM}).
 } \label{Poles-Zeroes}
\end{figure}
%=================================================================

In order to fix the dimension of the patch $\Psi(k)$ in $k$-space
it is useful to follow the evolution of the Fermi surface in
approaching the Mott insulator with the two periodization schemes
$\Sigma$ and $\mathcal{M}$, as shown in the panels of Fig.
\ref{Poles-Zeroes}. At high doping (in our case $\delta=0.13$)
all the system is well described by a Fermi liquid, the
self-energy is mostly local, the $\Sigma-$ (green dash line on
the left top panel of Fig. \ref{Poles-Zeroes}) and $\mathcal{M}-$
(blue continues line on the left top panel of Fig.
\ref{Poles-Zeroes}) periodizations give in practice the same
result. We can choose to describe the system with the
$\Sigma$-periodization and $\Psi(k)$ covering all the $k$-space
(for example see the left panel of Fig. \ref{Patch-mixed} for
$\delta=0.13$, where the patch is marked by the gray region
covering all the quadrant). By reducing doping, however, the
off-diagonal components of the cluster self-energy $\hat{\Sigma}$
are not negligible anymore, and the two periodizations produce
different results. In particular, as we stressed in the previous
sections, doping $\delta_c \sim 0.08$ and $\delta_p\sim 0.06$ are
special points. In the $\mathcal{M}-$periodization the Fermi
surface shows a striking topological phase transition, produced
by the appearance of lines of zeroes of the Green's function
$G_{k}(\omega \to 0)$ at the Fermi level (marked by a continuous
red line in Fig. \ref{Poles-Zeroes}). The effect of the
appearance of the lines of zeroes of the Green's function is at
the origin of the opening of a pseudogap in the spectral function
close to the $(0,\pi)-(\pi,\pi)$ and $(\pi,0)-(\pi,\pi)$ sides of
the first quadrant in the Brillouin zone (as shown for example in
panel C of Fig. \ref{Peaks}) . In the $\Sigma$-periodization
instead a more continuous evolution of the original high-doping
Fermi surface (green dashed line) takes place at all dopings up
to the Mott insulating state, where it disappears. We notice
however that the doping $\delta_c$ still marks a change in the
curvature of the Fermi surface (even if with the $\Sigma$
periodization the effect is more difficult to be noticed at naked
eye), as evidenced in the downturn of the $k_{nod}$ vector as a
function of doping in panel A of Fig. \ref{Peaks}. This effect
was first noticed in ref.\cite{marce05}, where we show that the
result of periodizing $\Sigma$ produces a Fermi surface which
enhances its hole-like curvature while reducing doping (and the
spectral weight reduces too with respect to the nodal point).
This goes in the direction of forming a hole pocket, which
however never arrives to be created within the
$\Sigma-$periodization. Within the $\mathcal{M}$ scheme instead,
the Fermi pocket forms at low doping and its progressive
reduction in approaching the Mott transition describes the way
the Fermi surface disappears. In the Mott insulating state Fermi
lines have of course totally disappeared, but a line of zeroes of
$G_{k}$ (i.e. a line of divergent self-energy
$\Sigma^{nor}_{k}(\omega=0)$) remains (in red) close to the
corner $k=(\pi,\pi)$ of the first quadrant of the Brillouin Zone.

It is very difficult to state how close to the real solution one
or the other of the two descriptions are. Particular intriguing is
the $\mathcal{M}$ periodization result. Not only it well portrays
the pseudogap in anti-nodes, it also produces, together with the
Fermi surface, lines of zeroes of $G_{k}$. In this way it
describes a continuity from the Fermi liquid at high doping (where
only the Fermi surface is present) to the Mott insulating state
(where only lines of zeroes are present). To this respect, it is
clear that the $\Sigma$ periodization scheme fails in describing
the Mott state, as it is unable by construction to build up lines
of zeroes. Within our analysis however, we cannot claim that in
the real physical system a Fermi pocket, together with lines of
zeros, is actually present. Recent experiments on cuprate
systems, where it has been possible to induce a low temperature
normal state by the application of an external magnetic field,
have actually observed de Haas-van Alphen oscillations compatible
with a Fermi pocket picture\cite{Doiron-Leyraud07}. Hall
resistivity measures, extracted at low temperature by
suppressing  superconductivity with the application of an external
magnetic field\cite{balakirev-2003,balakirev-2007,daou-2008}, are
also compatible with the scenario of a topological phase
transition of the Fermi surface. Our results are also in strong
resemblance with the theoretical study of ref.\cite{yang-2006-73},
where similar conclusions on the evolution of the Fermi surface
and the appearance of lines of zeroes have been drown starting
from an ad hoc model for the doping-driven Mott insulator
transition in two dimensions. Other non-perturbative microscopic
approaches have drawn conclusions in similar directions (see e.g.
ref.\cite{Plakida07,avella07}). There are however some caveats in
concerning the hole pocket which have to be considered.

In first place, according to the generalized Luttinger
theorem\cite{Luttinger60,Dzyaloshinskii03,essler-2003-90,konik-2005}
the volume enclosed between the Fermi surface and the line of
zeroes (if present) should be equal to the particle density
$n=1-\delta$. This theorem is quite respected (but not so much at
low doping) by the Fermi surface derived with the
$\Sigma$-periodization (green dashed line of Fig.
\ref{Poles-Zeroes}). It is not clearly obeyed instead at low
doping by $\mathcal{M}$-periodization, as evident e.g. by looking
at the area enclosed between the Fermi pocket (blue line) and the
lines of zeroes (red line) in the panel $\delta=0.05$ of Fig.
\ref{Poles-Zeroes} (close to half-filling the volume which gives
the correct density should be half of the quadrant). Far from
stating that the system is violating Luttinger theorem (which in
strongly correlated system is actually a
possibility\cite{rosch-2007-59,stanescu-2007-75}), this effect is
likely an artifact coming from the truncated Fourier expansion in
eq. \ref{Wk}, which can be improved only by increasing the
cluster size. A possible scenario is that in real systems the
line of zeroes is closer to the zone diagonal which goes form
$(0,\pi)\to (\pi,0)$, as actually we find at half-filling for a
chemical potential value in the middle of the Mott gap $\mu\sim
U/2$ (the panel displayed in Fig. \ref{Poles-Zeroes} has a
chemical potential close to the Mott gap edge $\mu\sim
U/2-\Delta_{M}/2$, where $\Delta_{M}$ is the total Mott gap).
This is in fact what was propose in ref.\cite{yang-2006-73} as
starting hypothesis. In this way, the side of the hole-pocket
facing the $(\pi,\pi)$ corner point is "cancelled" by the
proximity of the lines of zeroes (see for example the spectral
functions of Fig. \ref{Patch-mixed}), and the resulting picture
is a Fermi arc in the nodal region, which is replaced by a lines
of zeroes in the antinodal regions.

In second place, the discussion we have previously carried out
(see Fig. \ref{DOS-SM2} in section \ref{sec3} and the description
of nodal properties in section \ref{sec4}) shows that the low
energy nodal point is better portrayed by periodizing $\Sigma$,
i.e. at the nodal point we rather have a Fermi arc more than a
pocket. This is important if the nodal point properties
(presented in the previous sections \ref{sec3} and \ref{sec4})
have to be well portrayed. The real solution result at the nodes
is likely to lay in between the $\mathcal{M}$ and $\Sigma$
periodization schemes.

\subsection{The choice of the patch shape}
%================================================================
\begin{figure}[!!tb]
\begin{center}
\includegraphics[width=8.5cm,height=2.8cm,angle=-0] {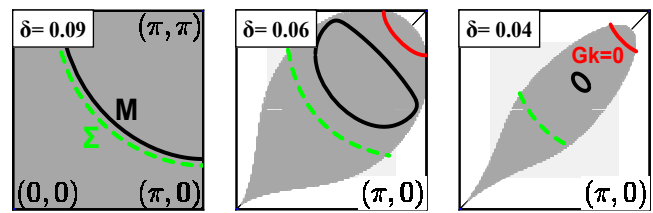}
\end{center}
\caption{(Color online). The dimension of the patch $\Psi_k$ in
momentum space (gray shaded region) is determined in such a way
that the $\mathcal{M}$-Fermi surface (black line) is cut away and
replaced by the $\Sigma$-Fermi surface (dashed green line). The
red line are where the self-energy $\Sigma_k$ is diverging and
$G_{k}=0$. } \label{Patch-mixed}
\end{figure}
%================================================================
\begin{figure}[!!tb]
\begin{center}
\includegraphics[width=8.5cm,height=2.8cm,angle=-0] {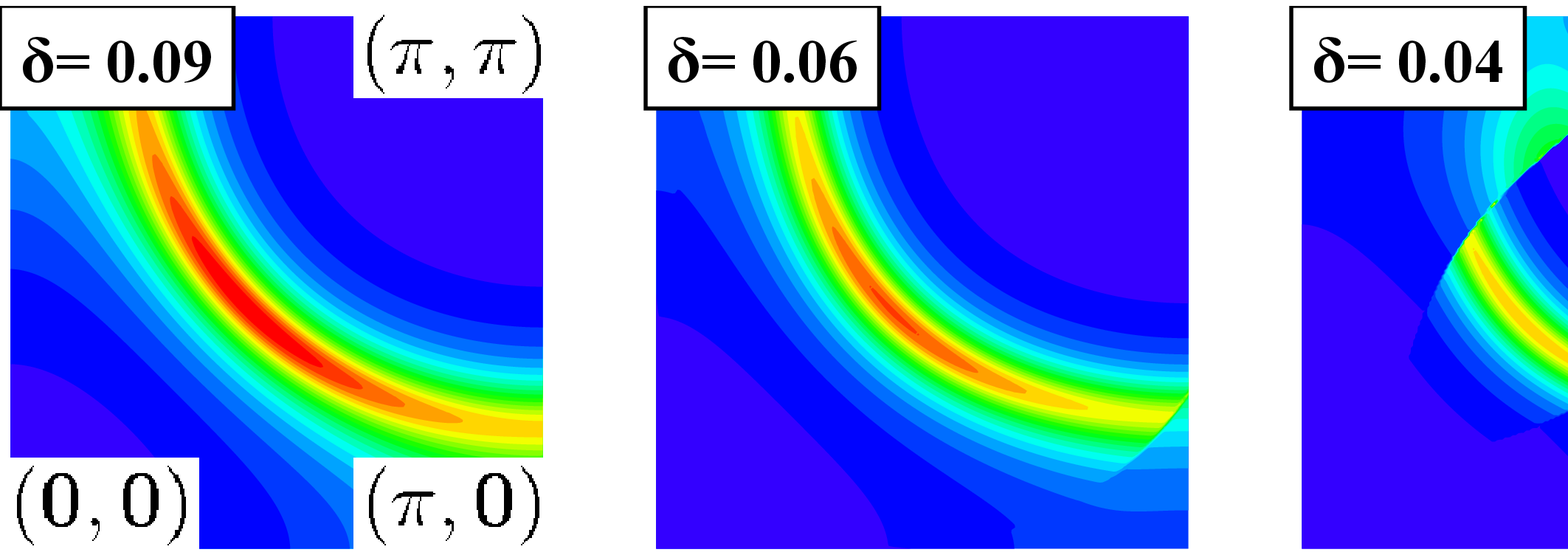}
\includegraphics[width=8.5cm,height=2.8cm,angle=-0] {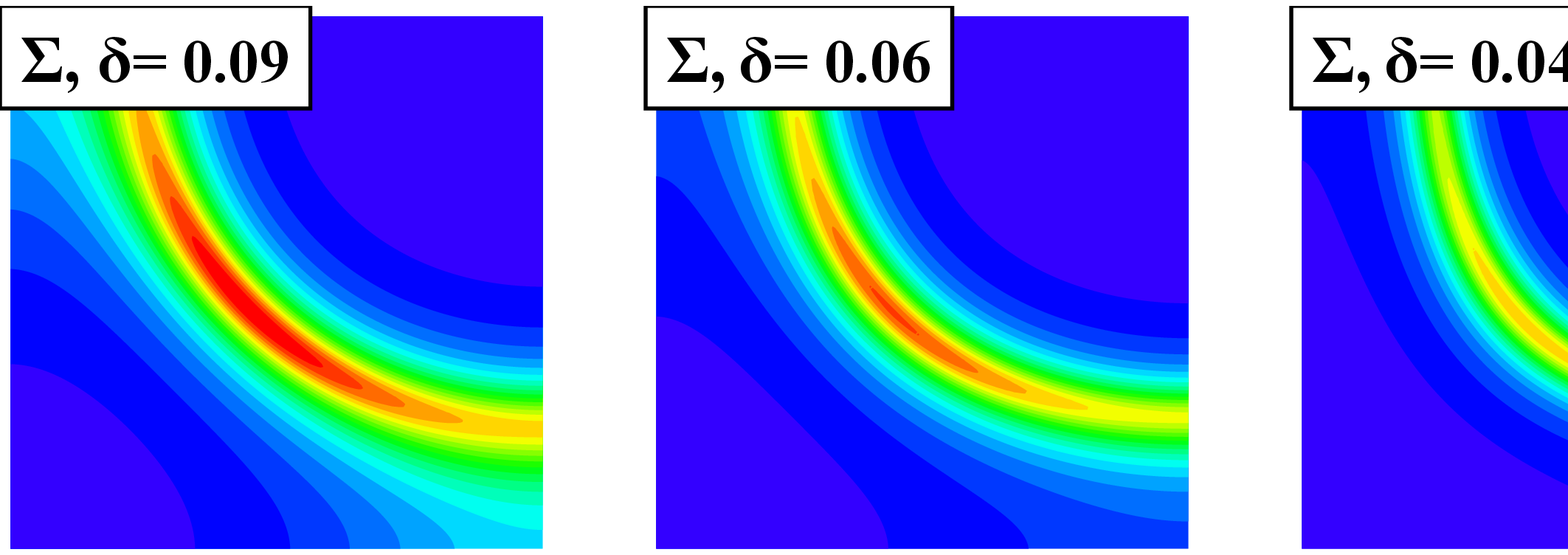}
\includegraphics[width=8.5cm,height=2.8cm,angle=-0] {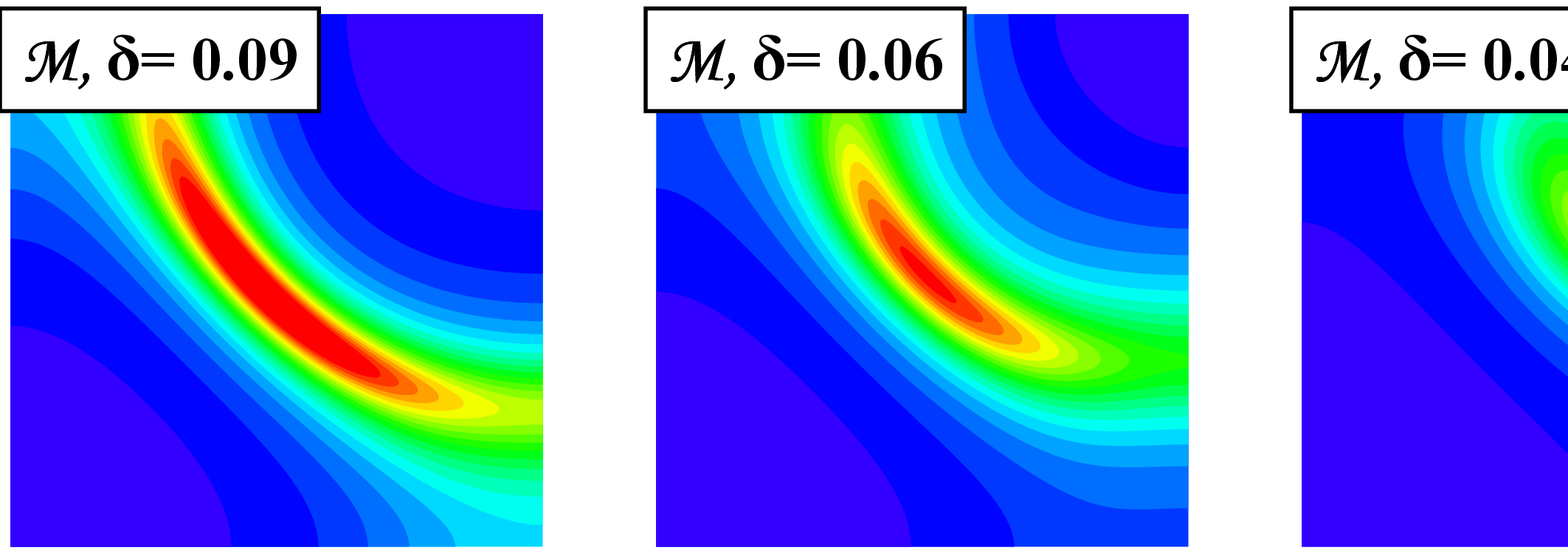}
\end{center}
\caption{(Color online). The spectral function $-\frac{1}{\pi}
\hbox{Im}G_{k}(\omega\to 0)$ in the first quadrant of the
Brillouin Zone obtained with the mixed periodization (top row) is
confronted with one obtained from the self-energy ($\Sigma$) and
the cumulant ($\mathcal{M}$) periodizations (second and third row
from the top). The broadening $\eta=0.052$, the color scale
$x=0.5$ (see appendix \ref{apxA} for details).} \label{Akw-SM}
\end{figure}
%================================================================
\begin{figure}[!!tb]
\begin{center}
\includegraphics[width=8cm,height=5.0cm,angle=-0]
{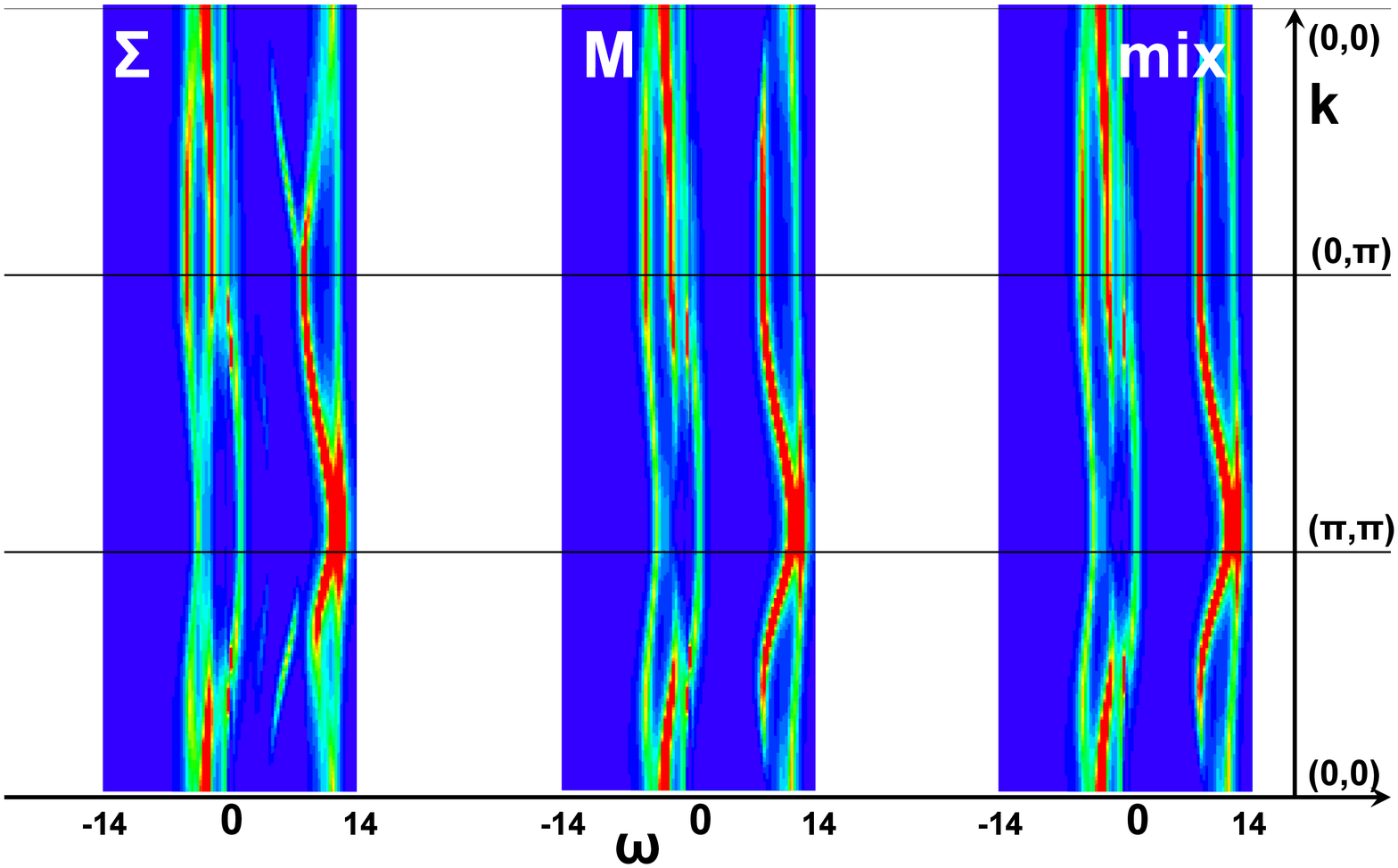}\\
\includegraphics[width=8cm,height=5.0cm,angle=-0]
{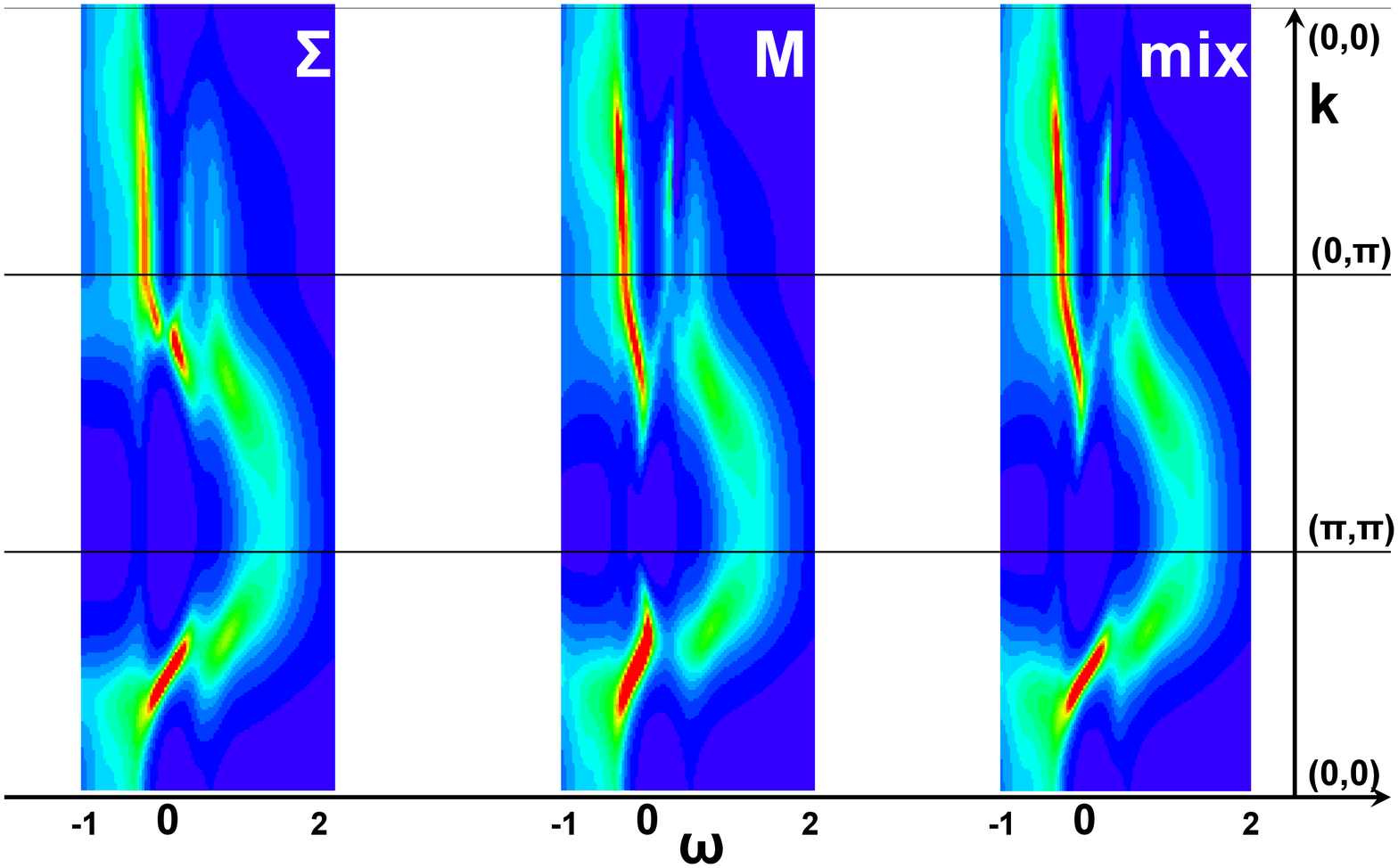}\\
\end{center}
\caption{(Color online). Comparison between the
$\Sigma$-periodization scheme, the cumulant
$\mathcal{M}$-periodization scheme and the mixed-periodization
scheme for a low doped system $\delta=0.05$. On the top row we
show the full energy range $-14<\omega<14$ covering the lower
Hubbard and upper Hubbard bands. The gross features are very
similar. Notice the $\Sigma$-scheme introduces fake density of
states in the Mott Hubbard gap. In the bottom row we show a close
up at low energy $-1<\omega<2$, where the methods are most
different. The mixed scheme is constructed to retain a
$\Sigma$-periodization character close to the nodal region $k\sim
(\frac{\pi}{2},\frac{\pi}{2})$ and a $\mathcal{M}$-periodization
character in the antinodal region $k\sim (0,\pi)$. Spectral
functions are displayed by introducing a $\omega$ dependent
broadening i$\eta(\omega)$ and the maximal color scale value are
$x=0.20 [0.25]$ for the top [bottom] panel (see appendix
\ref{apxA}). } \label{SMmix}
\end{figure}
%================================================================

In line with the discussion above, we choose therefore to assign
to the nodal region the path $\Psi_{k}$, which uses the $\Sigma$
periodization. The rest of the $k$-space (covered by $1-\Psi_{k}$)
is described with the $\mathcal{M}$ periodization. By following
this criterion we establish the evolution of $\Psi(k)$ with
doping, as shown in Fig. \ref{Patch-mixed}. The size and shape of
the patch $\Psi_{k}$ is determined so that the $\mathcal{M}$
Fermi surface is cut away, and the only piece of Fermi surface
inside $\Psi_k$ is the one produced by periodizing $\Sigma$ (green
dashed line). By reducing doping $\delta$ the size of the patch
$\Psi_{k}$ progressively reduces, until possibly disappearing in
the Mott insulating state ($\delta=0$) where only the
$\mathcal{M}$-scheme reproduces well the gapped spectra. We have
fixed the center of $\Psi_{k}$ by choosing $(k_{x_{c}},
k_{y_{c}})= (k_{x_{nod}}, k_{y_{nod}})$, the momentum coordinate
of the nodal point described in panel A of Fig. \ref{Peaks}. A
complete description of the patch parameters $r_{o}$ and
$\sigma_{o}$ (see formula \ref{Phi_k}) as a function of doping is
given in the following table:
\begin{center}\label{patch-para}
\begin{tabular}{||c||c|c|c|c|c|c|c||}
    \hline
$\delta$     & $>$0.08   & 0.08    & 0.06      & 0.055   & 0.05    & 0.04     & 0.03   \\
    \hline \hline
$r_{o}$      & 3.00    & 2.20      & 2.20      & 2.20    & 2.20    & 2.20     & 2.20    \\
\hline
$\sigma_{o}$ & $\pi/2$ & $\pi/2.4$ & $\pi/2.8$ & $\pi/5$ & $\pi/8$ & $\pi/12$ & $\pi/16$ \\
    \hline \hline
\end{tabular}
\end{center}

We notice that with the choice of this patch most of the lines of
zeroes disappear from the quadrant the Fermi level. As mentioned
above, we are not able to make definitive statement about the
actual position of these lines of zeroes. If it were closer to the
quadrant diagonal (as conjecture in work\cite{yang-2006-73}),
they would re-appear in the $1-\Psi_k$ region of momentum space,
which is described by the cumulant. As stressed in the previous
subsection, this would be important if one want to respect the
Luttinger theorem on the particle counting. In this mixed
periodization used for the $2\times 2$ plaquette result however
this does not take place. We remark however that in the
$1-\Psi_k$ region scattering rate (i.e. self-energy) is very high
(similarly to the results in ref.\cite{avella07}), and this fact
is ultimately the reason for the appearance of a pseudogap,
whether or not the lines zeroes are effectively present. The
division of the $k$-space via the patch $\Psi_k$ is surely
artificial, and it has not pretext of describing in detail the
real properties of the system. It has however capable of
capturing in a unique scheme either the virtues of the
$\Sigma$-periodization (above all in the nodes) and of the
$\mathcal{M}$-periodization (above all in the anti-nodes), which
portray with some good degree of confidence the physical
properties in different regions of momentum space (as we
discussed in the previous sections).

A confront between the $\Sigma$, $\mathcal{M}$ and mixed
periodizations is presented in Fig. \ref{Akw-SM} and Fig.
\ref{SMmix}. In Fig. \ref{Akw-SM} we show the spectral density
$A(k,\omega)= \, \frac{1}{\pi}$Im$G^{nor}_{k}(\omega\to 0)$ in
the first quadrant of the Brillouin Zone for decreasing doping
$\delta$ (from left to right). The mixed-periodization scheme (top
row) is confronted with the $\Sigma$ and $\mathcal{M}$
periodizations (bottom rows), showing how the patch $\Psi_k$ is
interlacing them. The well known phenomenon of the Fermi arc
breakup\cite{damascelli,campuzano} is reasonably well described
by all methods, showing this is a solid result of CDMFT.
Moreover, the similarity with spectra calculated in previous
CDMFT work on the normal state\cite{tudor,tudor06} shows the
smooth continuity between the normal component spectra of this
superconducting state solution with the spectra of a normal state
result. In Fig. \ref{SMmix} we show the band spectrum
$A(k,\omega)$ plotted as a function of the energy $\omega$ along
the path $(0,0)\rightarrow (\pi,\pi)\rightarrow
(0,\pi)\rightarrow (0,0)$ in the first quadrant of the Brillouin
Zone. In the top panel we confront the full energy range $-14<
\omega< 14$ covering upper Hubbard band and lower Hubbard band.
At this energy resolution the three schemes are qualitatively
very similar. We just stress that the $\Sigma$-periodization
artificially introduces spectral weight in the Mott gap (as
discussed in Fig. \ref{DOS-SM2} and evident in the figure around
$\omega\sim 3-7 t$). This justify the choice of cutting the patch
$\Psi(\omega)$ at low energies $\omega \leq 0.5 t$, using the
$\mathcal{M}$-periodization in the remaining of the energy range.
In the bottom panel of Fig.(\ref{Poles-Zeroes}) we show a close
up at low energy ($-1< \omega< 2$). In this energy range the
methods most differ, however the qualitative results are still
very similar. In particular the mixed scheme has been designed to
well describe a Fermi liquid linear dispersion at the node, and
the right description of the Mott gap, especially in the
antinodal region. In the analysis of the following sections we
will therefore apply the mixed scheme introduced here through
formula \ref{Sigmak-mix} to periodize the normal component of the
self-energy  (while the anomalous component
$\Sigma_{ano}(\omega)$ is always obtained through formula
\ref{Skano}, implying by construction a d-wave shape of the
superconducting gap).

\subsection{Local density of state with the mixed scheme}
%---------------------------------------------------------------
\begin{figure}[!!tb]
\begin{center}
\includegraphics[width=9cm,height=7.0cm,angle=-0] {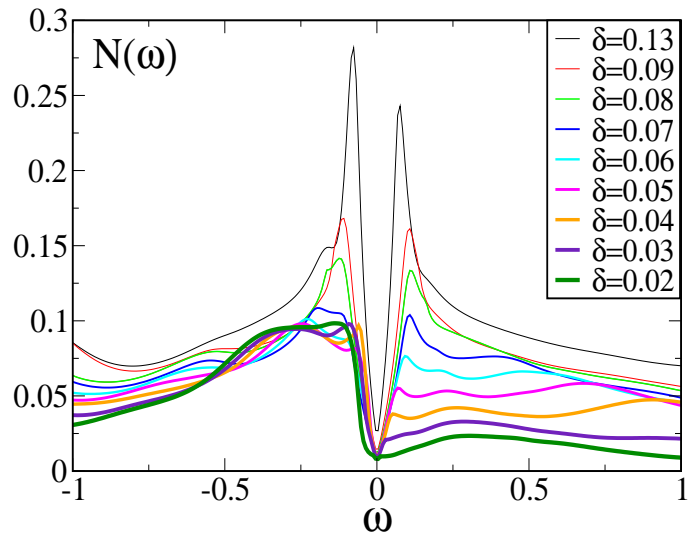}
\end{center}
\caption{(Color online). Local density of states $N(\omega)$ as a
function of the doping obtained with the mixed-periodization
scheme. The V-shape at $\omega\rightarrow 0$ is expected from a
d-wave superconductor with nodal Fermi liquid quasiparticles. For
doping $\delta< \delta_p\sim 0.6$ spectra develop a marked
asymmetry. This result interprets the cluster result of Fig.
\ref{ImG11} and it is in good agreement with scanning tunneling
experiments (see e.g. ref.\cite{davis05}). The display on the
real axis of the Green's function has been obtained by
introducing a $\omega$-dependent broadening i$\eta(\omega)$ (see
appendix \ref{apxA} for details).} \label{DOS-V2}
\end{figure}
%------------------------------------------------------------

We use now the mixed periodization to re-calculate the local
density of states $N(\omega)= -\frac{1}{\pi}\sum_{k} \,
\hbox{Im}G_{k}(\omega)$ at low energy, beyond the cluster energy
resolution (which is due to the finite dimension of the truncated
Anderson impurity model used to implement CDMFT, see Fig.
\ref{ImG11}). As widely explained above, this relies on the
implicit assumption that the superconducting gap has a d-wave
form (see eq. \ref{Skano}) and at the node we have a well defined
Fermi liquid arc. The result is shown in Fig. \ref{DOS-V2}. A
"V-shaped" $N(\omega)$ is observed either in the over-doped and
under-doped regions. Coming from the over-doped side towards the
under-doped side, the slope is always decreasing, until reaching a
saturating value at the small doping (until eventually showing a
small up-turn for the smaller doping, see Fig. \ref{gamma}). The
slopes well fit the analytical value extracted at low energy in
formula \ref{DOS}. On a wide range of energy ($-1< \omega< 1$),
the V-shape is quite symmetric in the over-doped region
$\delta\geq \delta_c\sim 0.08$ (as already remarked in the
CTQMC-CDMFT study of ref.\cite{haule-ctqmc}). Spectra become
strongly asymmetric in the under-doped region($\delta\leq
\delta_p\sim 0.06$), when the pseudogap in the normal part of the
system opens and super-impose to the superconducting gap (as
commented in panel D of Fig. \ref{Peaks}). The qualitative
behavior of these curves should be compared with the raw cluster
result of Fig. \ref{ImG11}. The periodizing mixed scheme we
introduced should be considered as the best fit we could achieve
to our cluster DMFT results, basing on few reasonable solid
assumptions on the physical properties of the system (like the
d-wave superconducting gap, the pseudogap formation, the Fermi
liquid properties at the node). This allows us to recover a
momentum dependent Green's function and access physical
quantities comparable to experimental results on cuprates. The
qualitative behaviour of $N(\omega)$ we determined well portrays
in fact results of scanning tunneling experimental\cite{davis05},
and it supports a comparison in momentum space.

%%%%%%%%%%%%%%%%%%%%%%%%%%%%%%%%%%%%%%%%%%%%%%%%%%%%%%%%%%%%%%%%%%%%%
%------------------------------------------------------------
\section{Quasiparticle spectra from the superconductor to the insulator.}
\label{sec6}
%-------------------------------------------------------------

In this section we use the mixed-periodization scheme and derive a
detailed description of the connection between the spectra of the
Mott insulator and the superconductor by varying doping.

\subsection{Doping a Mott insulator}
%-------------------------------------------------------------
\begin{figure}[!!tb]
\begin{center}
\includegraphics[width=8cm,height=5cm,angle=-0] {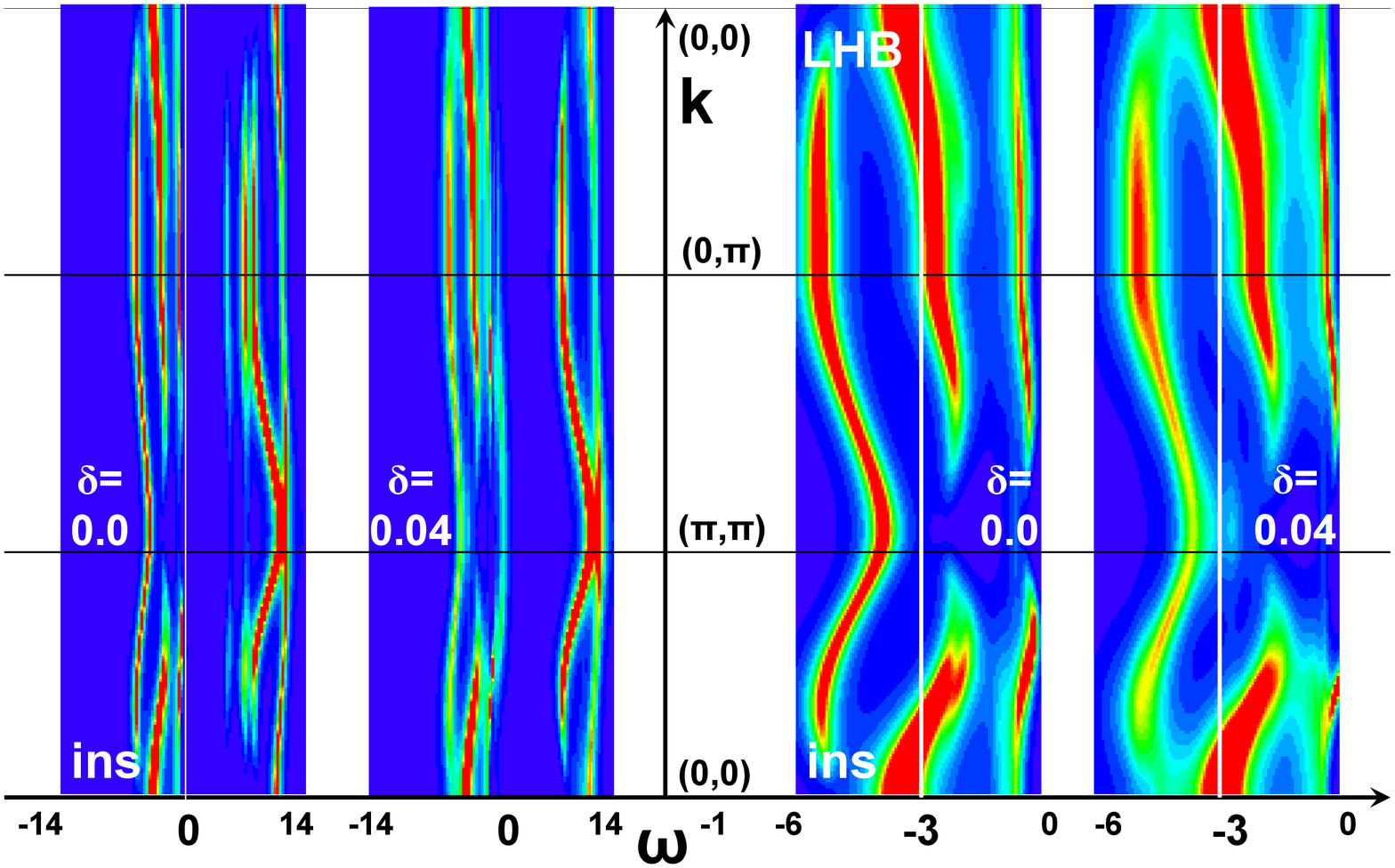}
\includegraphics[width=8cm,height=5cm,angle=-0] {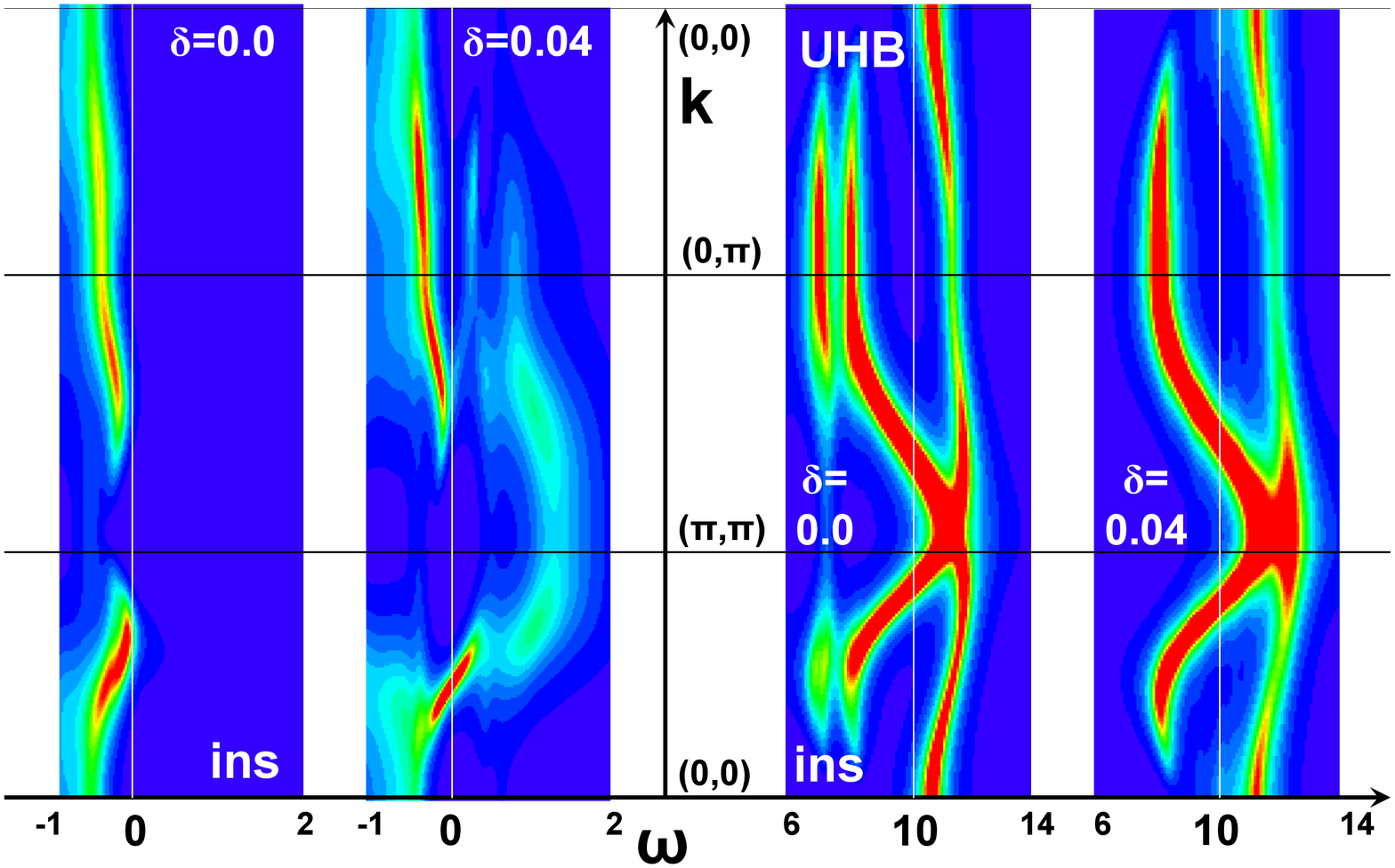}
\end{center}
\caption{(Color online). Band dispersion $A(k,\omega)$ along the
path $(0,0)\rightarrow (\pi,\pi)\rightarrow (0,\pi)\rightarrow
(0,0)$ in the first quadrant of the Brillouin Zone. The
Mott-insulating state ($\delta=0$) is confronted with a small
doping ($\delta= 0.04$) state. {\bf Top-left} corner: the full
range of energies ($-15< \omega< 15$) covered by the one-particle
spectrum is displayed. The gross band-structure, displaying a
lower Hubbard band (LHB) ($-6< \omega< 0$) and a upper band (UHB)
($10< \omega< 15$) separated by a Mott gap ($0< \omega< 10$)
remains un-changed in going from the insulator into the metal.
{\bf Top-right} corner: detail of the LHB, which rigidly shifts
in going from the insulating to the metallic state. {\bf
Bottom-left} corner: detail of the low energy ($-1< \omega< 2$)
feature. In going into the metallic state, the shift of the bands
is not rigid and spectral weight appears at positive energy as
soon as doping is added into the system. {\bf Bottom-right}
corner: detail of the UHB. Upon doping, a rigid shift in the
metallic state is accompanied by a transfer of spectral weight to
low energies. The color scale maximum value is always $x=0.20$,
except for the lower left panel, where it is set $x=0.25$. We
refer to appendix \ref{apxA} for the choice of the broadening
i$\eta(\omega)$.} \label{Ek-w-ins-met}
\end{figure}
%---------------------------------------------------------------
To this purpose, we analyze in Fig. \ref{Ek-w-ins-met} the
spectral functions, $A(k,\omega)= \, -\frac{1}{\pi}\hbox{Im}
G(k,\omega)$ in the $\omega-k$ space, along the $k$-path
$(0,0)\rightarrow (\pi,\pi)\rightarrow (0,\pi)\rightarrow (0,0)$
in the first quadrant of the Brillouin zone. We compare the
insulating state ($\delta=0$) with a slightly doped state
($\delta=0.04$). In the left top panel we display a wide
energy-range $-14<\omega<14$, which covers the lower (LHB) and
upper (UHB)  Hubbard bands. We notice that, upon adding a small
doping, the gross structure of the lower Hubbard band ($-6<
\omega< 0$) and the upper band ($10< \omega< 15$), which are
separated by a Mott gap ($0< \omega< 10$), remains substantially
un-changed. A detail of the lower Hubbard band, a closeup of the
Fermi level ($-1< \omega< 2$) and a detail of the upper Hubbard
band are presented in the right top, left bottom, right bottom
panels respectively. The LHB (top right) rigidly shifts in going
from the insulating to the metallic state, loosing spectral
intensity, which goes to build up quasiparticles at the Fermi
level $\omega=0$ (bottom left). The novel outcome from the CDMFT
calculation, as compared with the single-site DMFT
\cite{revmodmft}, consists in the anisotropic fashion
quasiparticles first occupy the Fermi level in momentum space.
This is better seen in a closeup of the band dispersion around the
Fermi level (bottom left). The region where firstly quasiparticles
appear is close to $k=(\frac{\pi}{2},\frac{\pi}{2})$. In fact,
already in the insulating state ($\delta=0$), we observe that
around $k=(\frac{\pi}{2},\frac{\pi}{2})$ a heavy-particle
hook-shaped band is closest to the $\omega=0$ level, while in the
proximity of $k=(0,\pi)$ there is a "pseudogap" and the band
disperses at negative energies. This kind of band-structure
survives upon doping, as we can observe in the $\delta=0.04$
panel. Contrary to the Hubbard bands, however, the low-energy
band does not shift rigidly as doping is added to the insulator,
rather it stays pinned at the Fermi level, and the shifts in
frequency is only a small fraction of the changing in chemical
potential. We also observe the appearance of spectral weight at
positive energy, coming from both the LHB and UHB, which starts
building up a full Fermi-liquid-like band, as we show more in
detail in the following. In the bottom right corner of Fig.
\ref{Ek-w-ins-met} we finally show the UHB. In this case a rigid
shift in the metallic state is accompanied by a strong reduction
of spectral weight. In fact, according to our result, the UHB
narrows with respect to the insulating state.
%-------------------

\subsection{Approaching the Mott insulator from the over-doped side}

\begin{figure}[!!tbh]
\begin{center}
\includegraphics[width=8cm,height=6cm,angle=-0] {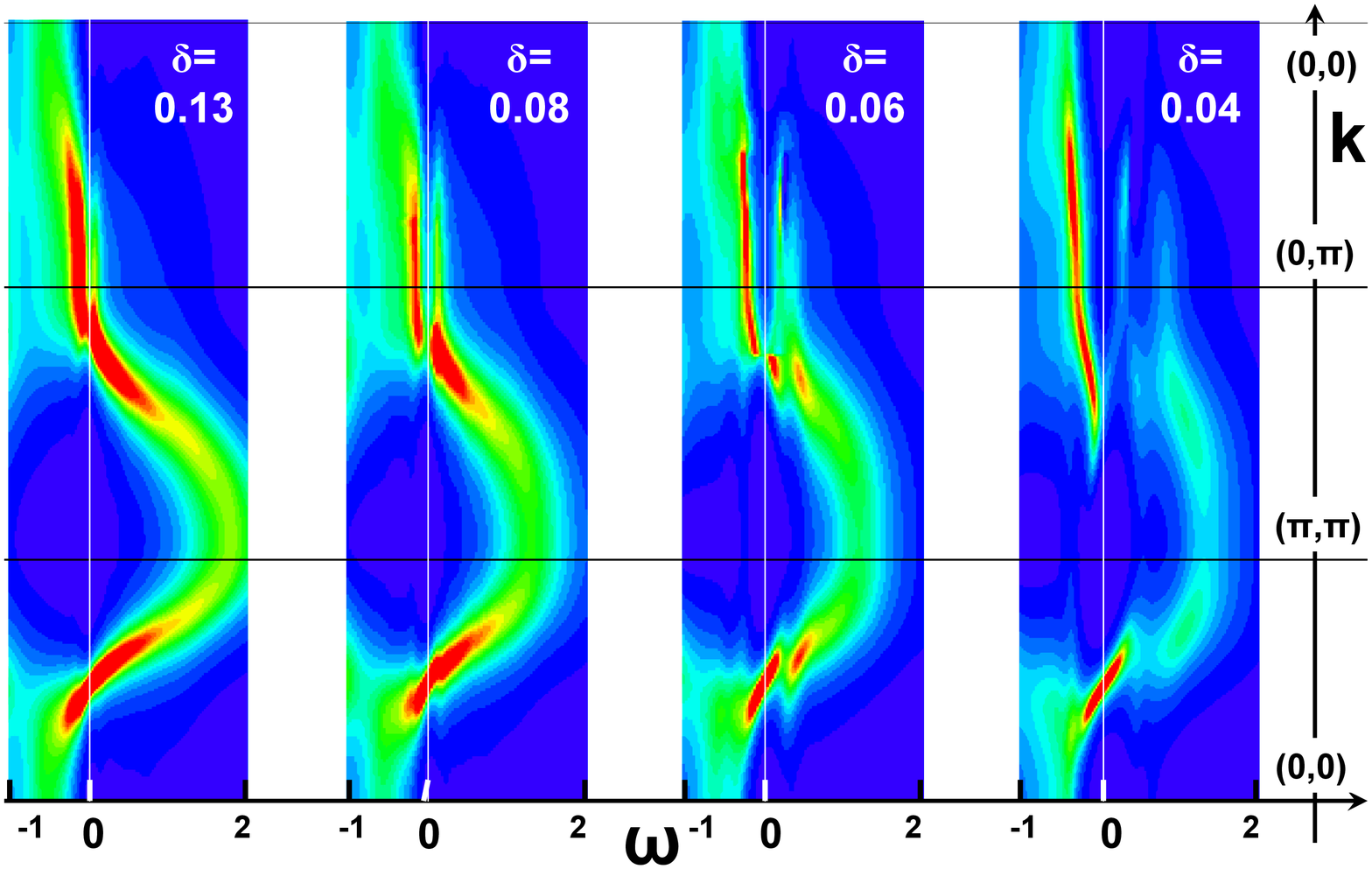}\\
\includegraphics[width=8cm,height=6cm,angle=-0] {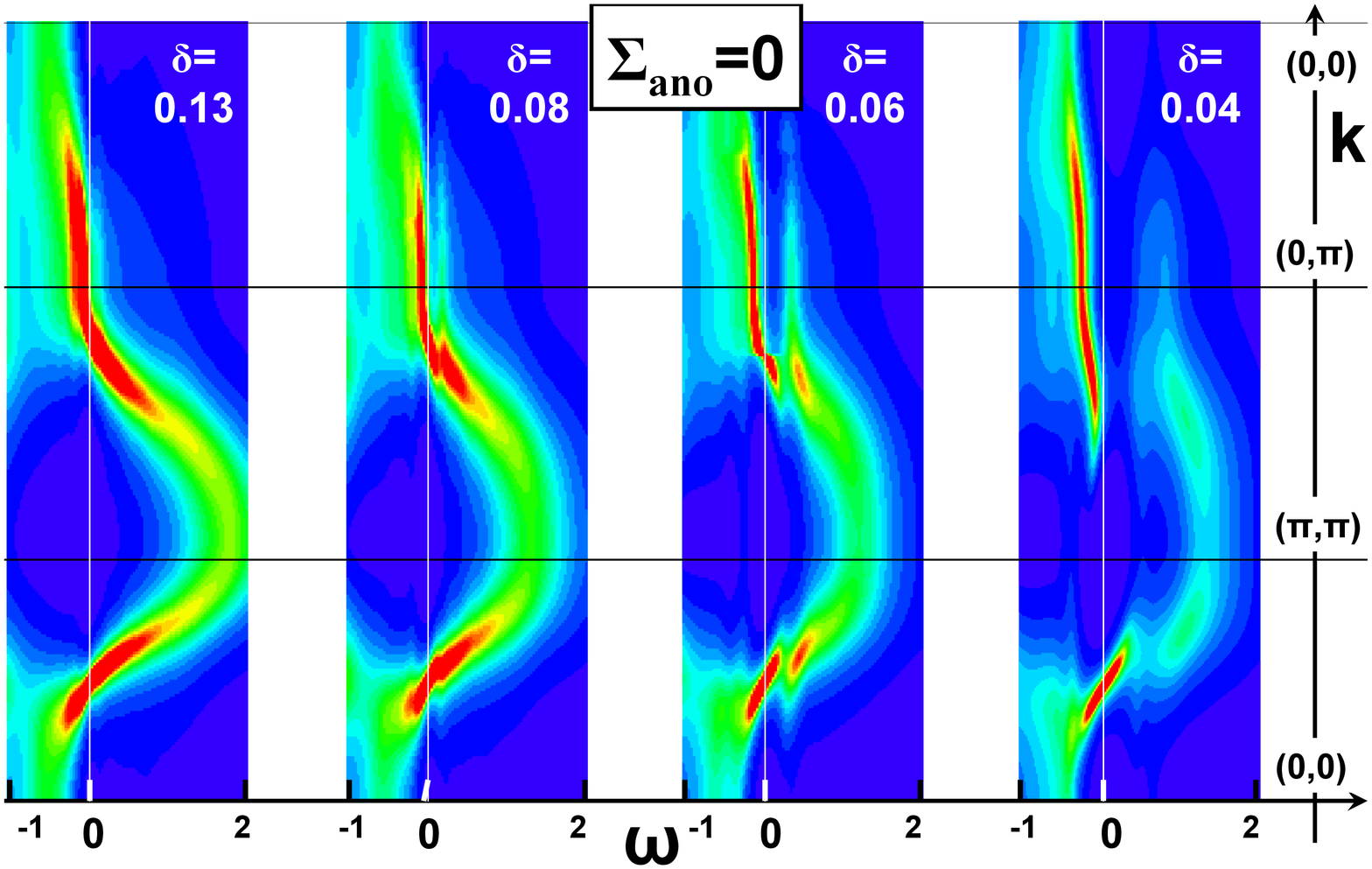}
\end{center}
\caption{(Color online). {\bf Top panel}: we show the evolution of
the low-energy one-particle spectral band from the over-doped
(left side, $\delta=0.13$) to the under-doped (right side,
$\delta=0.04$) state. A Fermi-liquid-like band at high doping
(left side, $\delta=0.13$) is progressively destroyed by
decreasing doping. Around the region
$k=(\frac{\pi}{2},\frac{\pi}{2})$ the dispersion presents a Fermi
liquid quasiparticle crossing the Fermi Level. Around the region
$k=(0,\pi)$, the d-wave superconducting-state opens a gap by
removing spectral weight from the Fermi level to positive and
negative energies. {\bf Bottom panel}: we show the contribution
to the low energy band coming from the normal component of the
system (we set the anomalous self-energy $\Sigma_{ano}=0$). In
particular we observe that in the under-doped side ($\delta\leq
0.08$) a gap (the pseudogap) opens in the $k=(0,\pi)$ region even
if the superconductive term is absent. For convenience's sake,
the color scale maximum value is set $x=0.30$ for $\delta=0.13$
and $x=0.25$ for all the other doping values. We refer to appendix
\ref{apxA} for the choice of the broadening i$\eta(\omega)$.}
\label{Ek-w-low-energy}
\end{figure}

We start now from the viewpoint of the highly doped system and
observe how the approach to the Mott insulator affects the
electronic structure of the superconducting state. The high-doping
system offers the advantage of having more standard
Fermi-liquid-like properties (the patch $\Psi(k)$ introduced in
the previous section covers all the momentum space).
%-------------------
\begin{figure}[!!bt]
\begin{center}
\includegraphics[width=9cm,height=6.0cm,angle=-0] {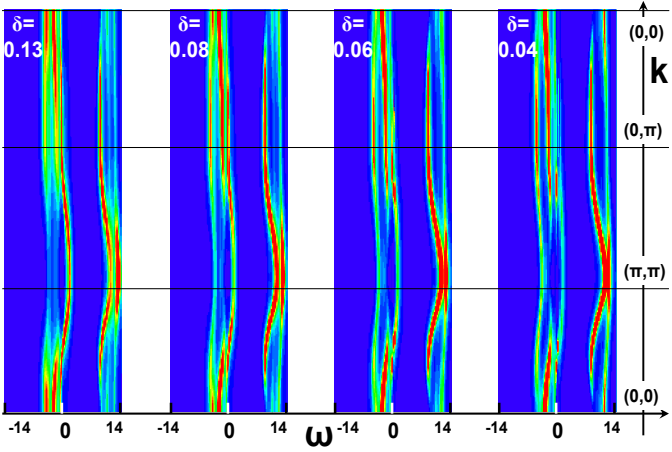}
\end{center}
\caption{(Color online). Evolution of the one-particle spectrum
from the over-doped (left side, $\delta=0.13$) to the under-doped
(right side, $\delta=0.04$) state. The typical Fermi-liquid band,
well visible in a narrow energy range ($-1<\omega<1$) around the
Fermi level at high doping $\delta=0.13$, is progressively
destroyed by reducing doping, while the upper and lower Hubbard
bands build up acquiring spectral weight . These features are
shown in detail in Figures \ref{Ek-w-met-UHB-LHB} and
\ref{Ek-w-low-energy}. The color scale maximum value is $x=0.20$
and we refer to appendix \ref{apxA} for the choice of the
broadening i$\eta(\omega)$.} \label{Ek-w-met-full}
\end{figure}
%------------------------------------------------------------------
In Fig. \ref{Ek-w-met-full} we show the electronic band
$A(k,\omega)$ in the $k-\omega$ space, once again in the path
$(0,0)\rightarrow (\pi,\pi)\rightarrow (0,\pi)\rightarrow (0,0)$
of the first quadrant of the Brillouin zone, on the full
energy-range $10< \omega< 15$ covering the LHB and UHB. Here we
want to display the evolution from high doping (left) to small
doping (right). At high a doping $\delta=0.13$ a Mott Hubbard gap
which separates a LHB structure from the UHB is already visible.
A narrow but Fermi Liquid-like band is however present, and it
crosses the Fermi level either in the region of momentum-space
around $k=(\frac{\pi}{2},\frac{\pi}{2})$ and $k=(0,\pi)$ (see
also Fig. \ref{Ek-w-low-energy}). Reducing the doping $\delta$, we
see that this narrow band loses intensity more and more (follow
the horizontal line at $k=(\pi,\pi)$ in $\omega\sim 0$) to the
advantage of the Hubbard bands, which instead gain spectral
weight in approaching the Mott insulator. To have a glance on how
this is taking place we look in the following figures at the
different energy-regions of the band in further detail.
%------------------------------------------------------------------
\begin{figure}[!!tb]
\begin{center}
\includegraphics[width=8cm,height=6cm,angle=-0] {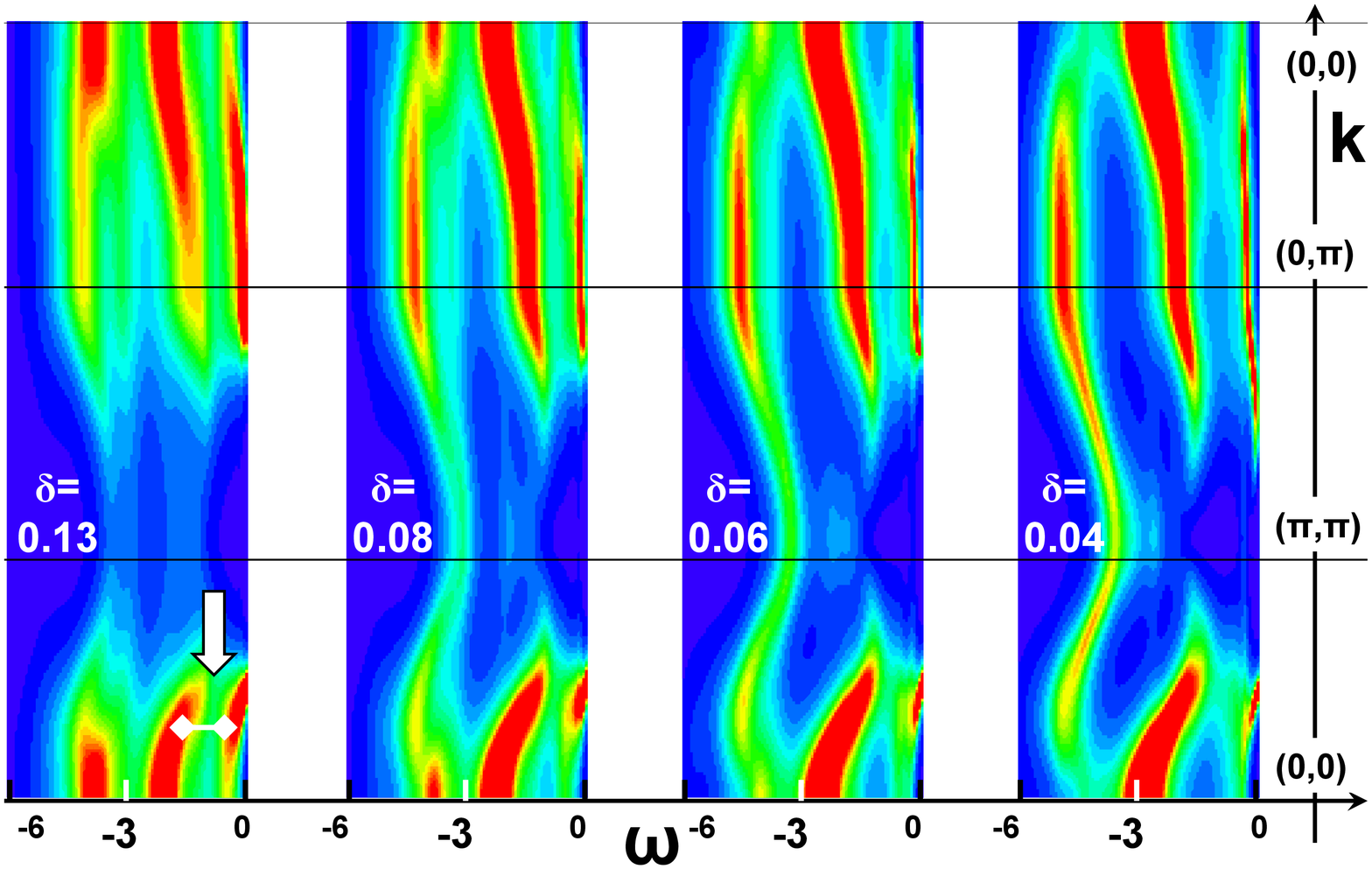}\\
\includegraphics[width=8cm,height=6cm,angle=-0] {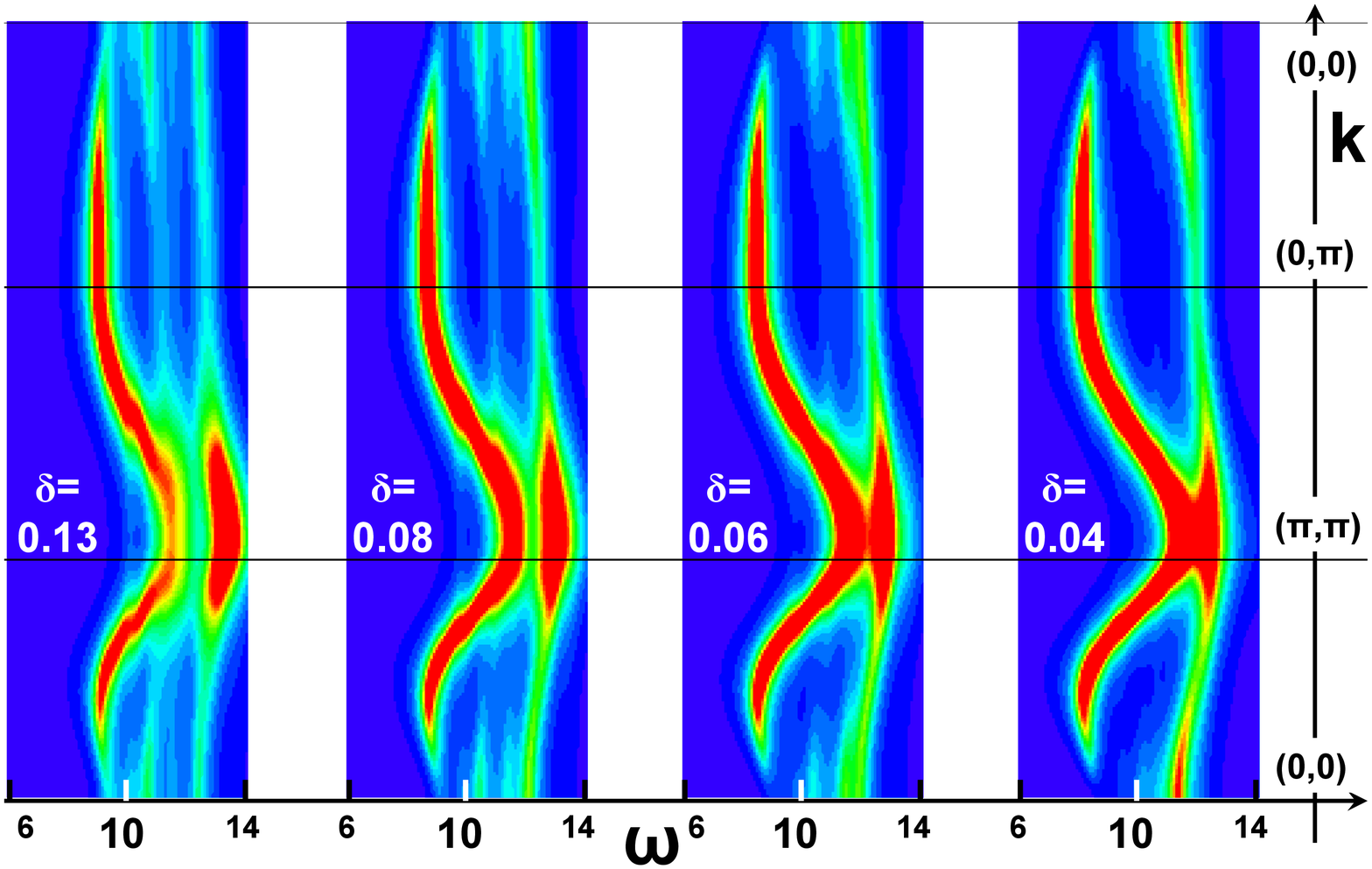}
\end{center}
\caption{(Color online). {\bf Top panel:} Evolution of the LHB
(top) from the over-doped (left side, $\delta=0.13$) to the
under-doped (right side, $\delta=0.04$) state. The LHB rigidly
shift according to the change in chemical potential by reducing
doping, acquiring structure (follow the horizontal line at
$k\sim(0,\pi)$. {\bf Bottom panel:} Evolution of the UHB from the
over-doped (left side, $\delta=0.13$) to the under-doped (right
side, $\delta=0.04$) state. The UHB rigidly shift according to the
change in chemical potential by reducing doping and, as evident
in the color intensity, it gains spectral weight. The color scale
maximum value is $x=0.20$ and we refer to appendix \ref{apxA} for
the choice of the broadening i$\eta(\omega)$.}
\label{Ek-w-met-UHB-LHB}
\end{figure}
%------------------------------------------------------------------
%------------------------------------------------------------------

Fig. \ref{Ek-w-met-UHB-LHB} shows the UHB and the LHB in detail.
The statements made above are confirmed: both Hubbard bands gain
spectral weight in decreasing doping, mainly in the region of
momentum space close to the anti-nodes $k\sim(0,\pi)$ (follow once
again the horizontal line), while rigidly shifting with respect to
the change of chemical potential $\Delta \mu$ (one can actually
show that the shift of the bands in energy is equal to $\Delta
\mu$). The behavior of the Hubbard bands is therefore in
agreement with a picture describing the approach to the Mott
insulator as a rigidly moving bands, which transfer part of their
weight to low energy. As mentioned in the previous subsection,
the novelty of our CDMFT result in finite dimension, with respect
to the standard vision of the Mott transition given by single-site
DMFT in infinite dimension, is that this transfer of spectral
weight takes place in a very anisotropic fashion, with the
antinodal regions $k\sim (0,\pi)$ getting insulating before the
nodal ones.
%-------------------
%-------------------

A closer look to the behavior at low energy, Fig.
\ref{Ek-w-low-energy}, reveals much richer phenomena taking place.
In the top panel we show the spectra resulting from the
one-particle Green's function $G_{11}(k,\omega)$ (eq.
\ref{Gk-sup}) in the superconducting state. First of all,
contrary to the behaviour of the Hubbard bands, the low energy
part of the band does not shift proportionally to $\Delta\mu$, in
agreement with the observation already made in Fig.
\ref{Ek-w-ins-met}. The Fermi-liquid-like band at high doping
(left side, $\delta=0.13$) is progressively destroyed by
decreasing doping (with progressive reduction of spectral weight
in the arc at $0< \omega< 2$). In the region close to $k=(0,\pi)$
instead, the d-wave superconducting-state opens a gap by removing
spectral weight from positive and negative energies (a Bogoliubov
band is formed at $\omega>0$). As widely discussed in the
previous sections however, the presence of lines of zeroes in the
Green's function kicks in the under-doped region ($\delta<0.08$),
opening a pseudogap in the normal component of $G_{11}(k,\omega)$
around $k=(0,\pi)$. This is shown in the bottom row of Fig.
\ref{Ek-w-low-energy}, where the same panels of the top row are
reproduced by imposing the anomalous self-energy identically zero
$\Sigma_{ano}=0$ in eq. \ref{Gk-sup}. Switching off
superconductivity has in general little effect on the
quasiparticle bands, except indeed close to the antinodal point
$k\sim(0,\pi)$ and close to the Fermi level $\omega\sim 0$ (the
nodal points $k\sim (\frac{\pi}{2},\frac{\pi}{2})$) are
practically unaffected). The differences are most evident in the
over-doped region (see e.g. $\delta\sim 0.13-0.08$ close to $k\sim
(0,\pi)$), where the superconducting gap has disappeared and the
quasiparticle band is reconstructed. The weight in the Bogoliubov
branch at $\omega>0$ also disappears. For $\delta>\delta_c\sim
0.8$ differences are much less evident (above all $\delta=0.04$).
A gap is also present in the normal component solution, the
structure of the quasiparticle dispersion around $k\sim (0,\pi)$
is very reminiscent of the superconducting solution in the
corresponding top row. It seems that in the under-doped region the
superconducting gap appears to complete the structure already
present in the normal component solution. This indicates that the
pseudo-gap appearing in the normal component of the system (and
present at temperatures above T$_C$) and the d-wave
superconducting gap are possible answers of the system to the
same instability, and they co-exist in the under-doped region
(while in the over-doped only the superconducting gap is
present). This instability is, in our view, connected to the
approach to the Mott transition, and it reflects the anisotropic
way chosen by the Hubbard system in two dimension to approach the
insulating state: the regions around the anti-nodes $k\sim
(0,\pi)$ become insulting (at $\delta=\delta_c\sim 0.08$) before
the regions close to the nodes $k\sim
(\frac{\pi}{2},\frac{\pi}{2})$ (which become finally insulating
$\delta=0$).

\section{Low-energy kink in the quasiparticle spectra }
\label{sec7}
%=======================================================
\begin{figure}[!!t]
\begin{center}
\includegraphics[width=9cm,height=6.0cm,angle=-0] {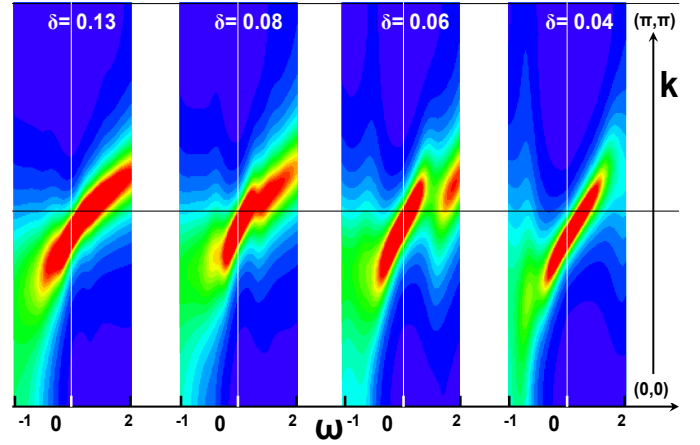}
\end{center}
\caption{(Color online). The the low energy kink-feature in the
spectral intensity at the nodal point for various values of
doping. For convenience's sake, the color scale maximum value is
set $x=0.35$ for $\delta=0.13$ and $x=0.30$ for all the other
doping values. We refer to appendix \ref{apxA} for the choice of
the broadening i$\eta(\omega)$. } \label{kink0}
\end{figure}
%-------------------%

We can now compare the quasiparticle spectra presented in the
previous section with the spectra measured, e.g. by
photo-emission, in cuprate-based systems. Unlike the linear
dispersion predicted by simple band calculations, in recent years
a series of experiments on the electronic structure of many H-TC
superconductor compounds has shown sharp breaks in the dispersion
of spectra , "kinks", at binding energies of the order of 50-80
meV\cite{Lanzara01,Z-XShen02,he-2001-86,Hwang04,cuk-2004-94}.
Sudden changes in the quasiparticle velocity were reported by a
factor two or more. This break in the dispersion is evident at
and away from the $d$-wave node line, and its magnitude shows
doping and temperature dependence. Kinks may provide useful
information on the nature of the coupling between electrons and
possible single-particle or many-body excitations, which are at
the origin of strongly-correlated many-body properties of the
system. In H-TC superconductors these feature have been
associated either with phonons\cite{Lanzara01,Z-XShen02} or
spin-fluctuation based\cite{he-2001-86,Hwang04} pairing
mechanisms.

In Fig. \ref{kink0} the dispersion we observe in the
quasiparticle band of our result around
$k=(\frac{\pi}{2},\frac{\pi}{2})$ is similar to the
experimentally observed kink (see e.g. ref.\cite{kaminski}). A
neat linearly dispersing quasiparticle crosses the Fermi level
($\omega=0$) in the $k$-$\omega$ plot, but around $\omega\sim
-0.2t$, the dispersion suddenly changes in slope and gets more
incoherent, as evidenced by the broadening spectra. If we set
$t\sim 300$meV as order of magnitude, we have that the kink
appears at $\omega\sim 60$meV, in good agreement with the
observed experimental energy range. The kink present in a wide
range of doping, from the under-doped to the over-doped regime,
and its slope is increasing with decreasing doping.

\begin{figure}[!!tbh]
\begin{center}
\includegraphics[width=8cm,height=17.5cm,angle=-0] {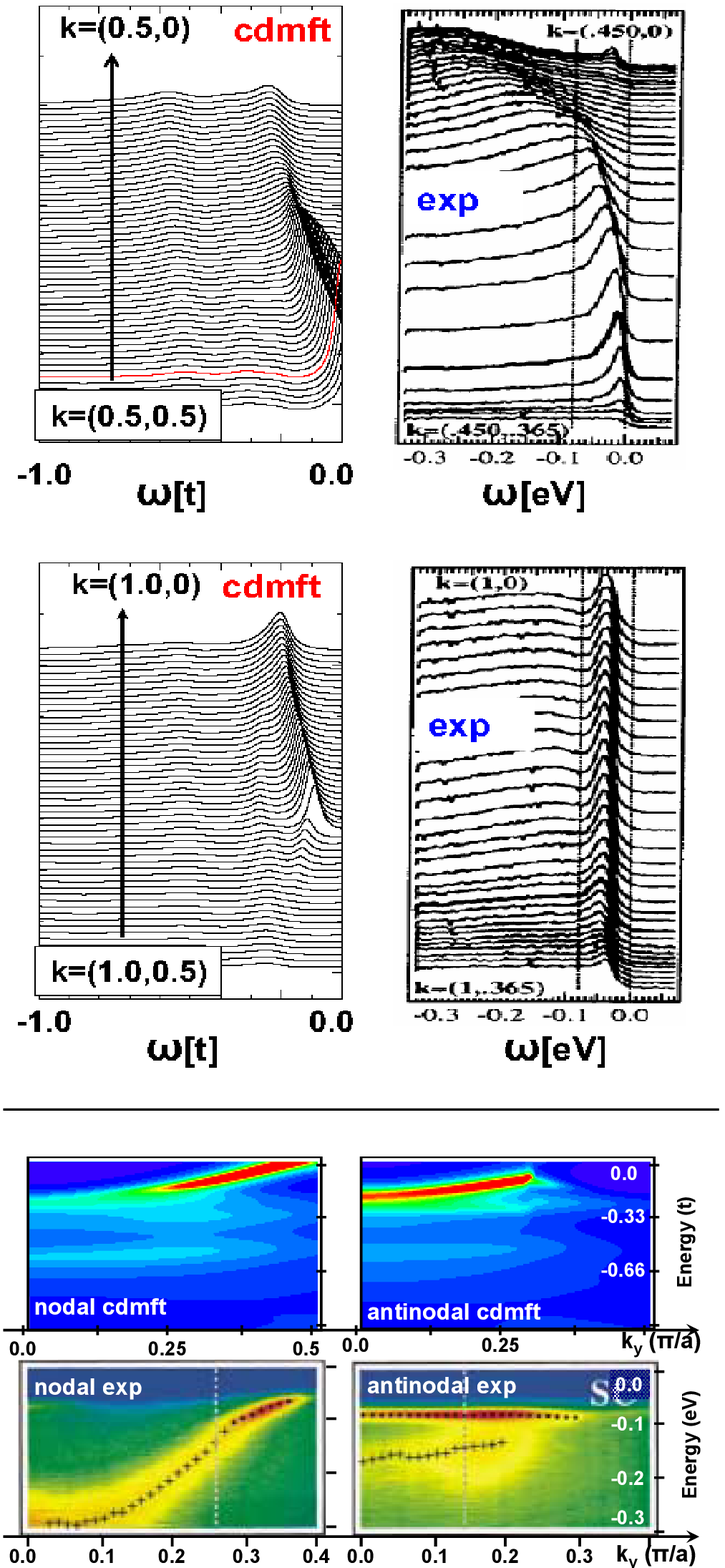}
\end{center}
\caption{(Color online). {\bf Top panel}: The low energy spectral
intensity $A(k,\omega)$ vs $\omega$ calculated within CDMFT
(left) is confronted with the experimental result of
ref.\cite{kaminski} in the nodal and antinodal regions. Vertical
cuts are traced in $k$-space close to the node and the antinode,
as traced on the figures. $\mathbf{k}$ is expressed in units of
$(\pi,\pi)$. {\bf Bottom panel}: Color-intensity plots of the
spectral intensity $A(k,\omega)$ in the $k-\omega$ space in the
nodal (left) and antinodal (right) cuts of momentum space (the
same used in the top panel). The color scale is set $x=0.30$ (see
appendix \ref{apxA}). In the bottom row the experimental result of
ref.\cite{kaminski} are displayed for comparison. We refer to
appendix \ref{apxA} for the choice of the broadening
i$\eta(\omega)$.} \label{kink1}
\end{figure}

This new energy scale in H-TC superconductors arises the debate on
the possible nature of the electron-electron coupling. In order
for superconductivity to take place in metals, it is necessary
that electrons bind into pairs, which condense in a
phase-coherent quantum state. In standard BCS superconductivity,
coupling between electrons and phonons (lattice vibrations)
drives the formation of the pairs. The existence of the kink
low-energy scale, not explicable in band theory calculation, may
give an hint on the low-energy nature of the interaction between
electrons, and hence help revealing the pairing mechanism of the
unconventional H-TC superconductivity. The single-band Hubbard
Model studied in this paper does not take into account phonons by
construction. Therefore the presence of the kink supports the
idea that the origin of these features are indeed purely
electronic, in agreement with the DMFT and CDMFT studies of
ref.\cite{byczuk-2007-3,chakraborty-2007}. In Fig. \ref{kink1} we
present the spectral function $A(k,\omega)= \,
-\frac{1}{\pi}$Im$G(k,\omega)$ calculated in our theory,
confronted with experimental ARPES data taken from
ref.\cite{kaminski}. Fig. \ref{kink1} shows the spectral function
$A(k,\omega)$ as a function of the energy $\omega$, along two
vertical cuts in the first quadrant of the Brillouin Zone in
correspondence of the nodal and antinodal region $(k_{nod},
0)\sim (\frac{\pi}{2},0) \rightarrow  (k_{nod}, k_{nod}) \sim
(\frac{\pi}{2},\frac{\pi}{2})$ and $(k_{anod}, 0)\sim (\pi,0)
\rightarrow  (k_{anod}, k_{anod})\sim (\pi,\frac{\pi}{2})$. The
system is close to optimal doping $\delta_c \sim 0.08$. In the
left column we display the CDMFT calculation, while the right
column the experimental data, taken from ref.\cite{kaminski}. The
nodal quasiparticle clearly show a dispersion, that from the
Fermi level propagates at negative frequencies until $\omega\sim
-0.2t$, where it has a sudden broadening, indicating a strong
incoherence. The antinodal quasiparticle shows instead a much
flatter dispersion at energy corresponding to the superconductive
gap, and the antinodal quasiparticle (which shows to have less
weight than the nodal quasiparticle) does not lose much
coherence. This plots have a good resemblance with the
experimental data on the right column. The comparison between the
energy scale of our results in the left $-t< \omega<0$ and of the
experimental data $-300$eV$< \omega< 0$ shows also that the
esteem we have used $t\sim 300$eV is a reasonable order of
magnitude.
%-------------------
%-------------------%

The same comparison is also presented in the intensity color plot
at the bottom of Fig. \ref{kink1}, where the nodal (left side)
and the antinodal (right side) dispersion calculated with CDMFT
are displayed in the top row, while experimental plots of
ref.\cite{kaminski} are in the bottom row. From these diagrams,
it is more evident the quasiparticle-like dispersion in the nodal
point, which compares very well to the data from photo-emission.
In the photo-emission data however at $\omega\sim -0.075$eV the
quasiparticle appears loosing coherence (effect marked by the loss
of red color that gets yellow) and at the same time the slope of
the dispersion changes substantially. In the CDMFT calculation
instead, as soon as the quasiparticle loses coherence (around
$\omega\sim -0.2t$), the dispersion stops (i.e. is
$\omega(k)\sim$ constant), and no spectral weight is present in
the region at smaller energy ($\omega< -0.2t$). The kink appears
in our result as a "gap" in the $\omega$ vs $k$ spectra. A
similar phenomenon takes place at higher energies, as marked by an
arrow in the top-left panel of Fig. \ref{Ek-w-met-UHB-LHB}, where
an evident gap is present in between the low energy band and the
lower Hubbard band. This could be possibly associated with the
"water-fall" dispersion features recently observed in many
cuprate materials (see e.g. ref.\cite{graf-2007-98,valla-2007-98})
at energies $\sim 350-600$ meV, where the same $k$-vector is
marking the dispersion $\omega(k)$ in a rather extensive range of
energy. In our calculation this phenomenon is described as a gap
in the dispersion. The antinodal point, where the superconducting
gap is present, is non-dispersing at low energy (right-bottom
side in Fig. \ref{kink1}), and the resemblance of our result with
the photo-emission data is rather good.
%-------------------%

\section{The Hall resistivity to detect the topological phase transition of
the Fermi Surface}\label{sec8}

We have shown in section \ref{sec5} that in the under-doped region
($\delta< \delta_p\sim 0.06$) a different regime sets in our
solution, marked by a topological transition from a large to a
small Fermi surface (which reduces to an arc or pocket). The
strong reduction of the Fermi surface area should correspond to a
strong reduction of the carrier density too.
%=============================================================
\begin{figure}[!!t]
\begin{center}
\includegraphics[width=8cm,height=6.0cm=angle=-0] {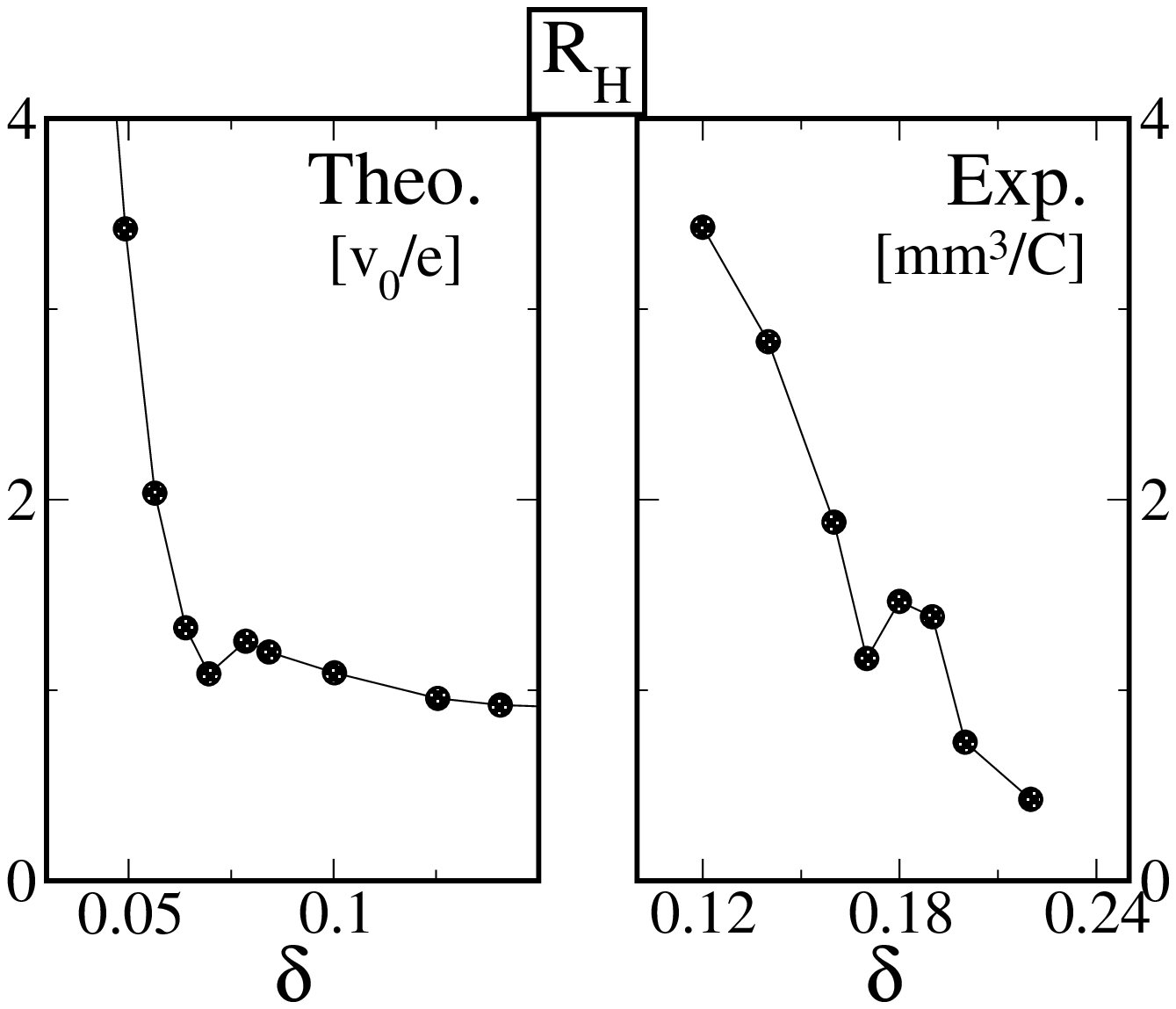}
\end{center}
\caption{Hall resistivity $R_{H}$ as a function of doping
$\delta$. The topological phase transition of the Fermi Surface is
efficiently detected by a rapid change of the behavior in the Hall
resistivity R$_{H}$, which indicates a change of the carrier
density. {\bf Left:} our theoretical result (units are in the
unitary cell volume $v_{o}$ over electron charge $e$). {\bf
Right:} the experimental result for $LSCO$ extracted from
ref.\cite{balakirev-2007}.} \label{RH-fig}
\end{figure}
%--------------------------------------------------------------%
This could be in principle experimentally detected by measuring
the Hall resistivity $R_{H}= -\frac{1}{n_c e}$, which is directly
related to the carrier density $n_c$, provided superconductivity
can be suppressed and the underlying normal liquid extracted. This
latter is far from being a trivial procedure, which has many
caveats. A standard method is for example to apply a magnetic
field, which suppresses superconductivity. Recent publications
have addressed this problem in $BSLCO$\cite{balakirev-2003}
($Bi_{2}Sr_{1.51}La_{0.49}CuO_{6+\delta}$) and
$LSCO$\cite{balakirev-2007} ($La_{2-\delta}Sr_{\delta}CuO_{4}$),
where magnetic field up to 65 Tesla was applied, and in
$Nd-LSCO$\cite{daou-2008}(La$_{1.6-x}$Nd$_{0.4}$Sr$_{x}$CuO$_{4}$),
where more moderate magnetic fields (up to 35 Tesla) can totally
suppress superconductivity. The results of these experiences on
the Hall resistivity show in fact to be compatible with a
reduction of the carrier density at low temperature in an
under-doped sample (as we will discuss more in detail in the
following).

In our study we can easily estimate the Hall resistivity by using
a Boltzmann approach, similarly for example to the
phenomenological study of ref.\cite{PSK}. The Hall resistivity
$R_{H}$ can be expressed in terms of the in-plane conductivity
(in the direction of an applied voltage) $\sigma_{xx}$ and the
transverse conductivity $\sigma_{xy}$ (perpendicular to the
applied voltage):
\begin{equation}
R_{H}=
-\frac{\sigma_{xy}}{\sigma_{xx}^{2}-\sigma_{xy}^{2}}\frac{1}{H}
\label{RH}
\end{equation}
where $H$ is the applied field. We can use the low-energy limit
(similarly to eq. \ref{Ek}) and extract the normal component of
the system (by setting $\Sigma_{ano}=0$ in eq. \ref{Gk-sup}). The
Green's function is conveniently written on the Matsubara axis as:
\begin{equation}
G(k,\omega_n)=\, \frac{1}{\hbox{i}\frac{(2n-1)\pi}{Z_k
\beta}-\zeta_k} \label{Gkw-RH}
\end{equation}
where a re-normalized dispersion $\zeta_{k}= \left[ \xi_{k} +
\hbox{Re} \Sigma_{k}(0) \right]$ and a inverse temperature
$\beta^{-1}$, re-normalized by the quasiparticle residuum $Z_k$,
are introduced. We consider this liquid as the base on which to
start applying a Boltzmann theory. The conductivities can then be
expressed in the first order in $H$ (see e.g. ref.\cite{PSK}):
\begin{eqnarray}
\sigma_{xx}&=& -2 e^{2} \tau\, \sum_{k} (v_{k}^{x})^{2} \,
\frac{\partial f_{(Z_k \zeta_{k})}}{\partial \zeta_{k}} \\
\sigma_{xy}&=& \frac{2 e^{3} H \tau^{2}}{\hbar c}\, \sum_{k}
v_{k}^{x} \, \left[ \, v_{k}^{y} \partial_{k_x} v_{k}^{y}- \,
v_{k}^{x} \partial_{k_y} v_{k}^{y} \, \right]\,
\frac{\partial f_{(Z_k \zeta_{k})}}{\partial \zeta_{k}}\nonumber
\end{eqnarray}
Here, $v_{k}^{\nu}= \partial \zeta_{k}/ \partial k_{\nu}$ is
$\nu=x,y$ component of the normal velocity, $f(x)= 1/(e^{\beta
x}+1)$ is the Fermi function, $\tau$ is the scattering time, which
at $T=0$ is, for example, given by the impurities, and that, for
convenience's sake, in a first order approach, we can hypothesize
$k$-independent. In the pure model, like the Hubbard Model, this
term is absent, so we have to add it in order to simulate the
finite $T=0$ resistivity of a real material. As the Hall
resistivity $R_{H}$ is anyway independent of a constant $\tau$,
we do not need to make any special further assumption about it.

The Hall resistivity resulting from our calculation implemented
via a mixed periodization and plug into the Boltzmann expression
is displayed in the left hand side of Fig. \ref{RH-fig} as a
function of doping $\delta$. The unit used to display $R_H$ is
the ratio between the unitary cell volume $v_o$ and the electron
charge $e$. To give an idea of the order of magnitude, taking for
$LSCO$ (see e.g. ref.\cite{cieplak06}) an average lattice spacing
$a_o\sim 4 \AA= 4\times 10^{-7} mm$ in the $Cu-O_2$ planes, and
$a_1 \sim 3 a_o$ in the $c$-axis $\frac{v_o}{e}\sim 1.2 \,
mm^{3}/C$, which well compares with the experimental results
extracted from ref.\cite{balakirev-2007} (but similar results
hold for $BSLCO$\cite{balakirev-2003}), and presented on the
right hand side. In the following, in drawing a parallel between
our theoretical result and the experimental data, one should keep
in mind that in our study the "optimal doping" (which we can only
identify with some degree of uncertainty as a maximum in the
order parameter $\delta_c\sim 0.08 <\delta<\delta_p\sim 0.06$,
see also the discussion in the conclusions) is situated at a
smaller doping than the experimental value $\delta_{opt}\sim
0.17$.

$R_{H}$ is positive, as expected by the hole-like Fermi surface
in the hole-doped system. As, in approaching the Mott insulator,
the localization of particles reduces free carriers, $R_{H}$ is
generally expected to monotonically decrease with doping. Two
regimes are clearly separable in our result (left side of Fig.
\ref{RH-fig}). At $\delta \sim \delta_p\sim 0.06$ we observe a
discontinuity in the behavior of $R_H$ (which slightly decreases
instead of increasing), and for $\delta<\delta_p$ a sudden
increase of $R_H$, related to the topological phase transition of
the Fermi surface presented in section \ref{sec5}, marks the
reduction in the carriers in approaching the Mott state. Such a
discontinuity is present and more evident in the experimental
data (right side of Fig. \ref{RH-fig}). The presence of a local
minimum at optimal doping $\delta_{opt}\sim 0.17$ represents a
crucial point in the results presented in
references\cite{balakirev-2003,balakirev-2007}. This behavior is
interpreted in terms of a quantum critical point, associated with
a change in the topology from a large hole-like Fermi surface
(realized for $\delta> \delta_{opt}$) to a small (pocket-like?)
Fermi surface, realized for $\delta< \delta_{opt}$). According to
their point of view, the force driving the superconductivity is
related to the fluctuations around a critical point. This
attractive force could overcome the mutual Coulomb repulsion of
electrons, delocalizing and freeing carriers right in proximity
of the quantum critical point. This fact would originate the
local minimum in the Hall resistivity $R_H$ (which is inversely
proportional to the number of free carriers). The critical point
would correspond then to the optimal doping $\delta_{opt}$, where
the critical temperature $T_C$ is the highest.

Our result shows qualitatively very similar trends, with a minimum
(even if milder) just at doping $\delta_p$, where the topological
transition of the Fermi surface takes place. This strongly
supports the analysis of our results and the conclusion we have
derived, in comparison with the experimental evidence on the Hall
resistivity.

In our 2$\times$2 plaquette study at $T=0$, however, we are not
unfortunately able to state if $\delta_p$ is or not the optimal
doping (which is defined at $T=T_C$). This point is as a matter
of facts close to the maximum of the d-wave order parameter (as
evident in Fig. \ref{dOP}). The study of ref.\cite{haule-ctqmc}
has pointed out that the exact determination of the optimal doping
may depend on temperature. Also, we cannot demonstrate that
$\delta_p\sim 0.06$ is a quantum critical point, as we have not
clearly identified an order parameter or the divergence, to some
order, in the free energy. Our study however finds good agreement
with the experimental results of
ref.\cite{balakirev-2003,balakirev-2007,daou-2008}, and it is not
in contradiction with their conclusions. We can in fact associate
the discontinuity of $R_H$ with a dramatic re-arrangement of the
electronic structure in the system, corresponding to the
topological transformation of the Fermi surface. This happening
is marked by a fast crossover region (between $\delta_c\sim
0.08<\delta<\delta_p\sim 0.06$), where the maximum of order
parameter is located, the pseudogap appears (at $\delta\sim
\delta_c$), the superconducting gap start decreasing (at
$\delta\sim \delta_p$). All these facts have striking
consequences on spectra and transport properties, which we have
presented throughout the paper.

More developments beyond the 2$\times$2 plaquette CDMFT are
needed, in order to be able to reveal if in reality
$\delta_c\equiv \delta_p$, or rather they represent two different
"transition" points. It would be then important to clarify if any
of these points are a quantum critical point, and their exact
connection with the optimal doping $\delta_{opt}$ of the system.
All these are important open questions left for the future
studies.

%-------------------------------------------------------------------------

\section{Conclusion}
\label{sec9}

\subsection{Comparison with the resonating valence bond mean field theory}

CDMFT can be viewed as a generalization of the earlier slave
boson resonating valence bond mean field
theory\cite{bza,ruckestein87}. It is therefore useful to put our
results in this context. RVB mean field theories had numerous
early successes, like the prediction of d-wave superconductivity
and a pseudogap phase having the same symmetry as the
superconducting state\cite{liu}. Both RVB mean field theories and
our CDMFT study are formulated in terms of  variables defined in a
plaquette. The main differences stem from the fact that in CDMFT
these variables are frequency dependent, and thus they are able
to properly describe coherence to incoherence crossover in the
momentum space. In the slave boson RVB mean field theory there
are two very important notions. The slave boson order parameter,
which  measures "Fermi liquid coherence", namely the emergence of
a well defined quasiparticle peak in the spectral function, and
the "spinon condensation" order parameters, both in the
particle-particle and particle-hole channel. The position of the
quasiparticle is shifted by a Lagrange multiplier which, adding to
the bare chemical potential, forms a quasiparticle chemical
potential. The dynamical mean field picture of CDMFT develops
these ideas, allowing them to acquire non trivial dependence in
momentum space.

To begin with, CDMFT describes naturally a high temperature state
showing poor coherence. As discussed in reference\cite{marce05}
and Fig. \ref{ED-QMC}, the peaks in the spectral functions in the
normal state are broad and incoherent. It is only in the
superconducting state that we can, for the first time, extract
the Fermi liquid parameters discussed in this work. Coherence
appears at low temperature in the nodal region more than in the
antinodal region, as it is clearly seen for example Fig.
\ref{ED-QMC}. Furthermore the weight of the quasiparticle is
different (and larger in most of the phase diagram) in the nodal
than in the antinodal region, as seen in Fig. \ref{Z-nodal}. At
small doping quasiparticles in the antinodal region cannot be
defined in a strict Landau-Fermi liquid sense, as a quasiparticle
peak cannot be identified at the Fermi level anymore. As a matter
of fact a pseudogap opens in the antinodal quasiparticle spectra.
In the nodal region instead, quasiparticles are always well
defined and their weight dramatically reduces as the doping is
reduced. The slave boson RVB picture describes the pseudogap
phase in terms of the formation of spin-singlets, parameterized
by two order parameters, one describing correlations in the
particle-particle channel and the second in the particle-hole
channel. A similar but more complete description is achieved in
terms of the dynamical anomalous and normal self-energies. Notice
the remarkable fact that the anomalous self-energy at low
temperatures and at low frequencies has a similar order of
magnitude as the normal self-energy (Fig. \ref{Sig-nor-cluster}
and Fig. \ref{Sig-ano-cluster}), indicating that, as in the RVB
mean field theory, they might both have a similar singlet-pairing
origin deriving from the approach to the Mott insulator. On the
other hand, the normal state self-energy has a more complicated
angular dependence than the simple harmonic (d-wave)
$k$-dependence of the anomalous self-energy. In this sense, the
normal and the anomalous gaps can also be viewed as competing for
the same electrons, suggesting that further refinements of CDMFT
along the lines of ref.\cite{stanescu04} are worth being pursued.

\subsection{The scenario presented by our CDMFT results}

While the dynamical frequency dependence introduced by DMFT well
describes the low-to-high energy crossover, as well known from
the infinite dimensional case\cite{revmodmft}, the momentum
dependence introduced by a cluster DMFT uncovers a wider spectrum
of $k$-dependent physical phenomena, which turn out fundamental
in describing the approach to the Mott transition in two
dimensions.

In the first part of the paper (section \ref{sec2}) we have
presented raw cluster quantities, which are direct output of the
CDMFT procedure, but which can be only partially interpreted in
physical terms. Nevertheless we have shown that the doping-driven
approach to the Mott transition takes place via an intermediate
regime (rising after an "optimal doping region" $\delta_c\sim
0.08> \delta> \delta_p\sim 0.06$), where physical properties
depart from the standard picture of a BCS superconductor. In
particular we were able to identify two distinct energy-scales,
one associated with the anomalous component of the self-energy
(see e.g. Fig. \ref{Sig-ano-cluster}), the other better
enlightened in the local density of states (see Fig.
\ref{ImG11}), showing different doping dependence. In the low
doping regime the local spectra also show a strong asymmetry in
$\omega$, as observed in experiments\cite{davis05} and contrary
to the expectations from a BCS superconductor.

In order to physically interpret the genuine cluster results, we
have restored the lattice translational invariance (broken in the
CDMFT procedure) by introducing a periodization scheme. In section
\ref{sec3} we have justified and compared two possible methods,
based on periodization of the cluster self-energy and the cluster
cumulant respectively. To this purpose, we have discussed the
physical properties of our system, taking advantage from either
the cluster results (supported also by the comparison with CDMFT
results obtained with QMC impurity-solver
methods\cite{bpk,olivier-note,haule-ctqmc}) and experimental
observation. We have performed a robust test on our approach by
reconstructing the local quantities from the momentum-dependent
Green's function $G(k,\omega)$ (see Fig. \ref{DOS-V2}), which
show to be not far from the corresponding cluster quantities
obtained directly in the impurity model.

By introducing a periodization procedure in momentum-space we
were able to analyze our result in terms of experimentally
observable quantities in the nodal and antinodal points of
momentum space (section \ref{sec4}), making contact with recent
spectroscopy experiments\cite{tacon06,tanaka06}. In particular,
we were able to interpret the two energy scale in terms of a pure
superconducting gap (dominant in the nodes) co-existing with a
normal component gap (related to the pseudogap of the normal
state and dominant in the anti-nodal region of momentum space at
low doping). We complete in this way the work presented in the
short publication of ref.\cite{marce08}.

We have then extended our procedure to describe in first
approximation physical properties in all momentum space (section
\ref{sec5}). The scenario which results present an under-doped
state where electronic structure undergoes a dramatic
re-arrangement. We could associate this fact with the appearance
of a topological phase transition of the Fermi surface (see e.g.
Fig. \ref{Poles-Zeroes}), driven by the appearance in momentum
space of lines of diverging self-energy, fingerprints of the Mott
physics. While these results may be quantitatively different for
different periodizing methods, in terms of a large enhancement of
the real part of the self-energies they indicate the same
qualitative trends. Enhancement of self-energy $\Sigma_{k}$ is
most relevant in the cumulant scheme. The formation of a
pseudogap in the antinodal region, however, only requires a large
value of the self-energy and not a strict divergence (which in
any case can only occur at $T=0$). Based on this physical idea,
at finite doping the antinodal region is closer to the Mott
insulating phase, while the nodal region is closer to the Fermi
liquid state (see the band structure of Fig. \ref{Ek-w-low-energy}
and the quasiparticle peaks Fig. \ref{Peaks}). We achieve in this
way a consistent picture of the evolution of the electronic
structure with doping within our formalism.

We were able to explain the strong asymmetry in the local density
of states, observed e.g. in scanning tunneling
experiments\cite{davis05}. We were able to give a complete
description of the evolution of electronic spectra with varying
doping, comparing the doped state with the Mott insulating one
(see section \ref{sec6}). In particular, we have shown that the
transfer of spectral weight in approaching the Mott transition
takes place in a strongly anisotropic fashion. We have described
the behavior of the low energy band close to the Fermi level,
which is strongly renormalized by the interaction and
progressively destroys by reducing doping, opening first a
pseudogap in the antinodal region. This latter appears to be
present already in the parent Mott insulator. We have stressed
how this phenomenon is connected to the aforementioned lines of
diverging self-energy. We have described how, from this underlying
structure of the normal component, a d-wave superconducting gap
rises. We have also shown (in section \ref{sec7}) how the
combination of these effects results in spectra which show "kink"
features similar to the ones observed in the electronic
dispersion of many photo-emission experiments. Finally we have
shown, using a simplified Boltzmann approach, how the rising of
the under-doped regime, associated in our study to a topological
phase-transition of the Fermi surface, is experimentally
detectable from the doping-dependent behavior of transport
quantities, like for instance a singularity in the  Hall
resistivity. All these properties are comparable with experiments
on cuprate materials, and the good agreement we have found
supports our study of the evolution with doping of the
superconducting state of the Hubbard Model, offering a
self-consistent scenario of the approach to the Mott transition
in two dimension. This is most relevant in connection with
cuprate H-TC superconductors, in which the evolution of the
electronic structure with doping (and its relation with H-TC
superconductivity) remains a fundamental open question. In
particular, in recent times, the issue of two (nodal and
antinodal) gaps \cite{tacon06,tanaka06} or one pure d-wave
gap\cite{campuzano07,campuzano07b,campuzano08} has been at the
center of the experimental and theoretical
debate\cite{millis06,cho06,huefner-2008-71}. In this contest, our
CDMFT results on the simplest electronic model for H-TC materials
show a good agreement with the two gaps scenario and, at the same
time, with many other electronic properties of cuprates.

An important question left open in our study is to determine the
exact nature of the small "transition region" between
$\delta_p\sim0.06$ and $\delta_c\sim 0.08$, where fundamental
changes take place in the physical properties of the system, in
going from the Fermi-liquid of the over-doped side into the
anomalous liquid of the under-doped side of phase diagram. In
particular, we have shown that lines of diverging self-energy
appear at the Fermi level, a pseudogap opens in the one-particle
spectrum and a topological phase transition of the Fermi surface
occurs (see Fig. \ref{Poles-Zeroes}). This small region of doping
locates, within the numerical precision of our result, the
"optimal doping", which we identify as a maximum in the d-wave
order parameter (Fig. \ref{dOP}). All these observations point
towards indicating a tight connection between the physical
happening taking place in this optimal doping region and the H-TC
mechanism. We are not able to state within our study if behind
lays a quantum critical point (which could provide the binding
force for a high critical temperature). This scenario is actually
supported by many
theories\cite{sachdev92,perali96,kivelson98,varma99} and
experimental studies\cite{ando95,boebinger96,balakirev-2003}, and
it could be that further developments of CDMFT (i.e. increasing
cluster size) could reveal that the two points $\delta_p$ and
$\delta_c$ actually coincide. Or two distinct points could be
actually present, and one (or both) could have the characteristic
of a quantum critical point. To reveal this, the right divergence
in any order of the free energy and an order parameter should be
clearly identified.

At present it is not possible to demonstrate that our solution
will survive in the thermodynamic limit (i.e. in the infinite
cluster size limit), in the sense that the real ground-state of
the Hubbard Model in two dimensions could be another one of the
possible competing instabilities, like, e.g. stripe ordering or
antiferromagnetism. The latter, in particular, is expected to be
the ground-state close to the Mott insulating state. Within a
(dynamical) mean-field approach, however, we can study the pure
paramagnetic phase, showing that it is a relevant phase even if
it is not the true ground-state of the system. We leave open the
physical question of what terms need eventually to be added in
the Hamiltonian in order to make this state a real ground-state.
In order to make our first order picture more rigorous, further
developments are needed. These should involve the cumulants
directly in the self-consistency condition, as proposed for
example in ref.\cite{stanescu04,tudor}, and should exploit the
flexibility of CDMFT, which can be formulated in terms of a set
of adaptive patches in momentum space (on this line see the recent
work\cite{ferrero-2008}). These extensions, as well as the use of
more powerful solvers, which could allow going beyond a 2$\times$2
cluster, are worth pursuing and are left for future studies.

\appendix

\section{Displaying function on the real axis with ED-CDMFT}
\label{apxA}

Once the ground-state $|gs>$ of the associate Anderson Impurity
Model (eq. \ref{Ham-imp-full}) has been determined via the Lanczos
procedure, it is possible to determined the zero-temperature
Green's function via a second Lanczos step (see e.g. the
review\cite{revmodmft}). To this purpose, one has to take as
initial vector $c^{\dagger}_{\mu} |gs>$ ($\mu$ denoting the
generic cluster-site index), and write the ground-state Green's
function in a continued-fraction expansion, describing the
"particle" and "hole" excitations:
\begin{equation}
G(\omega_{n})= G^{>}(\omega_{n})+ G^{>}(\omega_{n})
\end{equation}
with
\begin{equation}
G^{\alpha=>,<}(\omega_{n})= \frac{<gs|
c_{\mu}c^{\dagger}_{\mu}|gs>} {\hbox{i}\omega_{n}-a_{0}^{\alpha}-
\frac{b_{1}^{\alpha 2}}{\hbox{i}\omega_{n}-a_{1}^{\alpha}-
\frac{b_{2}^{\alpha 2}}{\hbox{i}\omega_{n}-a_{2}^{\alpha}- ...}}}
\label{fraction-expansion}
\end{equation}
The parameters $a_{j}^{\alpha}$ and $b_{j}^{\alpha}$ ($j=1...$
number of Lanczos steps, $\alpha=>, <$) output directly from the
second Lanczos step\cite{haydock1975}. In the ED-CDMFT procedure
the cluster Green's function and the self-consistency equation
\ref{selfconsistency} are evaluated on the Matsubara axis. This
implies introducing a parameter $\beta$ which determines the grid
of Matsubara points $\omega_{n}= (2n-1)\pi/\beta$, and which
plays the role of a fictitious temperature (our solution is
however at zero temperature). We can easily analytically continue
the Green's functions by replacing in the continued-fraction
expansion (eq. \ref{fraction-expansion})
\begin{equation}
\hbox{i} \omega \to \omega+ \hbox{i}\eta \label{anlytic-cont}
\end{equation}
where $\omega$ is the real axis frequency and $\eta$ a small
parameter used to display the poles. It is difficult to know {\it
a priori} the smallest value we can assign to $\eta$. It depends
on the physical problem considered, on the size of the associated
Anderson Impurity Problem used in the Lanczos procedure and on
the energy resolution imposed in satisfying the self-consistency
condition on the Matsubara axis (i.e. the parameter $\beta$). A
reasonable guess for a lower bound value would be for example
$\eta \approx \pi/\beta$. Moreover we expect $\eta$ to be
frequency dependent too. The Lanczos better determines the ground
state of the system, and portrays better low-energy properties.
The uncertainty in the ground-state vector $|gs>$ propagates in
the determination of the $a_{j}^{\alpha}$ and $b_{j}^{\alpha}$
coefficients (which is a further Lanczos step), and further
propagates in the periodization procedures (section \ref{sec3}).
The error in the second Lanczos step turns out much bigger at
higher frequency, and, if $\eta$ is chosen too small, this can
create enormous errors, even breaking casuality (which is instead
by construction always satisfied in the impurity solver output,
i.e. in the cluster Green's function of eq. \ref{Gloc}). A too big
value of $\eta$ results however in a poor resolution, which may
hide important features, above all at small energy.
%-------------------
\begin{figure}[!!h]
\begin{center}
\includegraphics[width=8cm,height=3.0cm,angle=-0] {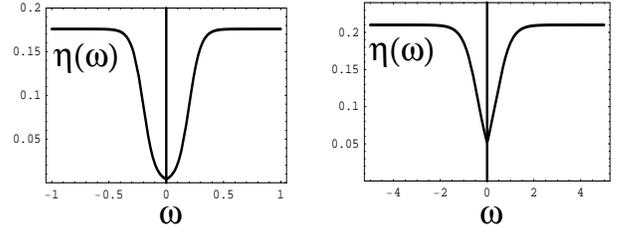}
\end{center}
\caption{Example of the broadening $\eta(\omega)$ used to display
spectral function on the real axis. {\bf Left:} the low energy
range $-1<\omega<1$ example of Fig. \ref{DOS-V2}. {\bf Right:}
the wide energy range $-14<\omega<14$ example of Fig.
\ref{SMmix}. For a complete list of $\eta(\omega)$ see table
\ref{eta-wt} } \label{deltaw}
\end{figure}
%---
In order to be able to display  at the same time high and low
energy features, we have introduced for convenience's sake in
sections \ref{sec3} (with the only exception of Fig.
\ref{ED-QMC}, see the corresponding caption), \ref{sec5},
\ref{sec6} and \ref{sec7} a $\omega$ dependent $\eta$,
arbitrarily choosing a function-shape which quickly separates a
low from a high energy range (see Fig. \ref{deltaw}):
\begin{equation}
\eta(\omega)= \eta_{0}+
\frac{\eta_{1}}{e^{-|\omega-\omega_{0}|/T_{w}}+1} \label{eta-w}
\end{equation}
The branching between the low and high energy regimes takes place
in a Fermi-function-like step at $\omega=\omega_{0}$ with
rapidity $1/T_{w}$. In Fig. \ref{deltaw} we report as example the
$\eta(\omega)$ adopted to display the local density of states of
Fig. \ref{DOS-V2} (on the left), where the low-energy range is
$-1<\omega<1$ , and (on the right) the $\eta(\omega)$ of the
density plots of the Fig. \ref{SMmix} and \ref{Ek-w-ins-met},
where the full energy scale $-14<\omega<14$, including the lower
and upper Hubbard bands, is considered. In the following we
report exactly the values of the parameters determining
$\eta(\omega)$ for all the figures involved:

\begin{center}\label{eta-wt}
\begin{tabular}{||l|c|c|c|c||}
    \hline
Figures& $\eta_{0}$ &  $\eta_{1}$    &   $\omega_{0}$  & $T_{w}$  \\
    \hline \hline
\ref{DOS-SM2}, top panel                & 0.01  & 0.15 & 0.05 & 0.01 \\
\ref{DOS-SM2}, middle and bottom panels & 0.001 & 0.05 & 0.05 & 0.01 \\
\hline
\ref{SMmix},\ref{Ek-w-ins-met},\ref{Ek-w-met-full},\ref{Ek-w-met-UHB-LHB},
\ref{Ek-w-low-energy},\ref{kink0}       & 0.01  & 0.2  & 0.4 & 0.3 \\
\hline
\ref{DOS-V2}                            & 0.001 & 0.175& 0.2 & 0.05 \\
\hline
\ref{kink1}                             & 0.025 & 0.04 & 0.2 & 0.025 \\
    \hline \hline
\end{tabular}
\end{center}
It is clear that spectral peaks displayed with different
$\eta(\omega)$ at different frequency present uncomparable heights
and widths. Our task is not however to make this kind of
comparison, rather to focalize on the position of the spectral
peaks and compare heights and widths at different doping (but at
the same frequency). This has to be kept in mind when analyzing
these figures.
%-------------------
\begin{figure}[!!h]
\begin{center}
\includegraphics[width=5cm,height=0.30cm,angle=-0] {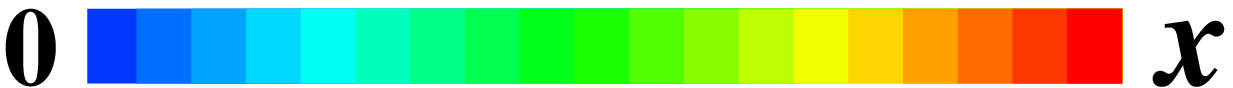}
\end{center}
\caption{Color scale code adopted in displaying the color-density
plots. $x$ is the maximum value of the scale.} \label{color-scale}
\end{figure}
%---

Finally, the color-code we adopted to display the density-plots
(see e.g. figures \ref{SMmix}, \ref{Ek-w-ins-met},
\ref{Ek-w-met-full}, \ref{Ek-w-met-UHB-LHB},
\ref{Ek-w-low-energy}, \ref{kink0}) is shown in Fig.
\ref{color-scale}. $x$ is the maximum value of the scale, that we
have chosen according to the picture (see the corresponding
caption). If the value of the function displayed is bigger than
$x$, the color remains red.

\section{Superconducting Bath-parameterization within CDMFT}
\label{apxB}

The general form of the associated cluster-Anderson impurity
Hamiltonian $\mathcal{H}_{\rm imp}$ can be written:
\begin{eqnarray}
\mathcal{H}_{\rm imp}&=&\sum_{\mu \nu \sigma}E_{\mu \nu
\sigma}c_{\mu \sigma }^{\dagger }c_{\nu \sigma}+  U \sum_{\mu}
n_{\mu\uparrow}n_{\mu\downarrow} + \nonumber \\
&&+\sum_{k \sigma}\epsilon _{k \sigma} a_{k\sigma
}^{\dagger}a_{k\sigma }  +\sum_{k\mu \sigma} V_{k\mu
\sigma}a_{k\sigma }^{\dagger}
c_{\mu \sigma} + {\rm h.c.}+ \nonumber \\
&&+\sum_{k\mu \sigma} V^{\rm sup}_{k\mu \sigma}a_{k\sigma }^{\dagger} c^{\dagger}_{\mu \bar{\sigma}} +
 \sum_{k\mu \sigma} V^{\dagger \rm sup}_{k\mu \sigma} c_{\mu \bar{\sigma}} a_{k\sigma }
\label{Ham-imp-apx}
\end{eqnarray}
Here $\mu,\nu=1, ..., N_c$ label the sites in the cluster and
$E_{\mu \nu \sigma}$ represents the hopping and the chemical
potential within the cluster. $\epsilon _{k \sigma}$ is the
energy level of the $(k,\sigma)$ orbital in the free electron
bath, $V$s represent the hybridization hopping amplitude either
for a particle-desctution/construction and for the
singlet-destruction/destruction (construction/construction)
between the impurity-cluster and the bath. In order to drive the
solutions towards physically interesting regions of the
bath-parameters space, we have introduced a reduced
bath-parametrization\cite{venky05} which allows to exploit the
symmetries in the square lattice to gain a better physical
insight of the Green's function symmetries. Moreover using fewer
parameters the work required by the minimization procedure is
faster and the result simpler to interpret:
%%%%%%%%%%%%%%%%%
\begin{eqnarray}
\mathcal{H}^{R}_{\rm imp}&=&\sum_{\mu \nu \sigma}E_{\mu \nu
\sigma}c_{\mu \sigma }^{\dagger }c_{\nu \sigma}+ U \sum_{\mu}
n_{\mu\uparrow}n_{\mu\downarrow} \nonumber \\
&& \sum_{m m' \sigma}\epsilon _{m m' \sigma}^{\alpha}a_{m'\sigma
}^{\dagger\alpha
}a_{m\sigma }^{\alpha} +\sum_{m\mu \sigma} V_{m\mu \sigma}^{\alpha}a_{m\sigma }^{\dagger\alpha } c_{\mu \sigma} + {\rm h.c.} +  \nonumber \\
&& + \sum_{\alpha} \Delta^{\alpha}(a_{1\uparrow}^{\alpha} a_{2\downarrow}^{\alpha} - a_{2\uparrow}^{\alpha} a_{3\downarrow}^{\alpha} + a_{3\uparrow}^{\alpha} a_{4\downarrow}^{\alpha} - a_{4\uparrow}^{\alpha} a_{1\downarrow}^{\alpha} \nonumber \\
&& + a_{2\uparrow}^{\alpha} a_{1\downarrow}^{\alpha} -
a_{3\uparrow}^{\alpha} a_{2\downarrow}^{\alpha} +
a_{4\uparrow}^{\alpha} a_{3\downarrow}^{\alpha} -
a_{1\uparrow}^{\alpha} a_{4\downarrow}^{\alpha})+ h.c.
\label{Ham-imp-red}
\end{eqnarray}
The energy levels in the bath are grouped into multiples of the
cluster size ($N_c=4$) with the labels $m=1,\cdots,N_c$ and
$\alpha=1,2$ such that we have 8 bath energy levels $\epsilon
_{m\sigma}^{\alpha}$ coupled to the cluster via the hybridization
matrix $V_{m\mu\sigma}^{\alpha}$. Using lattice symmetries we
take $V_{m\mu\sigma}^{\alpha}\equiv V^{\alpha}\delta_{m\mu}$ and
$\epsilon^{\alpha}_{m\sigma}\equiv \epsilon^{\alpha}$.
$\Delta^{\alpha}$ represents the amplitude of superconducting
correlations in the bath. No static mean-field order parameter
acts directly on the cluster sites \cite{venky05,poilblanc}.
$\epsilon^{\alpha}$, $V^{\alpha}$ and $\Delta^{\alpha}$ are
determined by imposing the self-consistency condition in
eq.~\ref{selfconsistency} using a conjugate gradient minimization
algorithm with a distance function that emphasizes the lowest
frequencies of the Weiss field\cite{marce} (see eq. \ref{fdist}).
The reduced form eq. \ref{Ham-imp-red} is in fact a sub-case of
the more general Hamiltonian eq. \ref{Ham-imp-apx}, and a
canonical transformation connects eq. \ref{Ham-imp-red} to eq.
\ref{Ham-imp-apx}. In order to see this, it is most convenient to
express $\mathcal{H}$ in a Nambu's form, introducing
cluster-plaquette spinors:
\begin{eqnarray}
\Psi_{c}^{\dagger } \equiv
(c_{1\uparrow}^{\dagger},\dots,c_{4\uparrow}^{\dagger},c_{1\downarrow},\dots,c_{4\downarrow})
\nonumber \\
\Phi_{a}^{\dagger } \equiv
(a_{1\uparrow}^{\dagger},\dots,a_{4\uparrow}^{\dagger},a_{1\downarrow},\dots,a_{4\downarrow})
 \label{spinors}
\end{eqnarray}
and recast the Hamiltonian:
\begin{eqnarray}
\mathcal{H}_{\rm imp}&=&  \Psi^{\dagger}\, \mathbf{E} \, \Psi+ U
\sum_{\mu}
n_{\mu\uparrow}n_{\mu\downarrow}+ \\
&&+ \sum_{\alpha} \, \Phi^{\dagger}_{\alpha}\,
\mathbf{E^{\alpha}_{B}}\, \Phi_{\alpha}+ \Phi^{\dagger}_{\alpha}\,
\mathbf{V}_{\alpha}\, \Psi+ \Psi^{\dagger}\,
\mathbf{V}^{\dagger}_{\alpha}\, \Phi_{\alpha} \nonumber
\label{Ham-imp-nambu}
\end{eqnarray}
where the Hamiltonian coupling constants are expressed my 8X8
matrices (leaving for convenience's sake the multi-bath index
$\alpha$ implicit):
$$\mathbf{E}=\, \left( \begin{array}{cc}
E_{\mu,\nu,\uparrow}(4{\rm X}4) & 0                 \\
0                    & E_{\mu,\nu,\downarrow} (4{\rm X}4)
\end{array} \right) $$

$$\mathbf{E_{B}}=\, \left( \begin{array}{cc}
  \mathbf{E_{B}}_{\uparrow}(4{\rm X}4)  &  \mathbf{E^{S}_{B}}(4{\rm X}4)   \\
  \mathbf{E^{S\dagger}_{B}}(4{\rm X}4)  & -\mathbf{E_{B}}_{\downarrow}(4{\rm X}4)
\end{array} \right) $$
 and
$$\mathbf{V}=\, \left( \begin{array}{cc}
  \mathbf{V}_{\uparrow}(4{\rm X}4) &   \mathbf{V}^{S}_{\uparrow}(4{\rm X}4) \\
  -\mathbf{V}^{S\dagger}_{\uparrow}(4{\rm X}4) &  -\mathbf{V}_{\downarrow}(4{\rm X}4)
\end{array} \right) $$
In the more general case $\mathcal{H}_{imp}^{\ref{Ham-imp-apx}}$,
the bath matrix is diagonal, so that
$\mathbf{E_{B\sigma}^{kk'}}^{\ref{Ham-imp-apx}}= \delta_{kk'}
\varepsilon_{k\sigma}$, and
$\mathbf{E^{S}_{B}}^{\ref{Ham-imp-apx}}=0$. $\mathbf{V}_{\sigma}$
and $\mathbf{V}_{\sigma}^{S}$ have generally non-zero elements.
In the reduced parameterization case instead (eq.
\ref{Ham-imp-red}), the bath matrix is not diagonal but chosen to
mimic a cluster-plaquette, by introducing in each multi-bath
$\alpha$ the same bath-energy $\varepsilon_{\alpha\sigma}$ on
every bath-site, a next-neighbor hopping $t^{\alpha}_{b}$ and a
nearest next-neighbor hopping $t'^{\alpha}_{b}$:
$$
\mathbf{E^{\alpha}_{B}}^{\ref{Ham-imp-red}}_{\sigma}=\, \left(
\begin{array}{cccc}
\varepsilon_{\alpha\sigma} & t^{\alpha}_{b}                     &  t'^{\alpha}_{b}                  &  t^{\alpha}_{b} \\
t^{\alpha}_{b}                     & \varepsilon_{\alpha\sigma} &   t^{\alpha}_{b}                  &  t'^{\alpha}_{b} \\
t'^{\alpha}_{b}                    &  t^{\alpha}_{b}                    &\varepsilon_{\alpha\sigma} &  t^{\alpha}_{b} \\
t^{\alpha}_{b} & t'^{\alpha}_{b}                   &
t^{\alpha}_{b} & \varepsilon_{\alpha\sigma}
\end{array} \right)
$$
Moreover superconductive d-wave pairing terms $\Delta_{\alpha}$
appear in the bath:
$$
\mathbf{E^{S\alpha}_{B}}^{\ref{Ham-imp-red}}=\, \left(
\begin{array}{cccc}
0                         & \Delta_{\alpha}                     &  0                  &  -\Delta_{\alpha} \\
\Delta_{\alpha}                    & 0                          &  -\Delta_{\alpha}            &  0        \\
0                         & -\Delta_{\alpha}                    & 0                   &  \Delta_{\alpha} \\
-\Delta_{\alpha}                   & 0                          &
\Delta_{\alpha} & 0
\end{array} \right)
$$
while the hybridization between the cluster and the bath is only
normal $\mathbf{V}_{\sigma}^{S}=0$ and simplified by connecting
each site of the cluster $\mu$ with only one site
$k_{\mu}^{\alpha}$ of the multi-bath $\alpha$ with the same
coupling constant $V_{\sigma\alpha}$:
$$
\mathbf{V}_{\sigma\alpha}^{\ref{Ham-imp-red}}=\, \left(
\begin{array}{cccc}
V_{\sigma\alpha}          & 0                     &  0                  & 0       \\
0                         & V_{\sigma\alpha}      &  0                  & 0        \\
0                         & 0                     &   V_{\sigma\alpha}  & 0         \\
0                         & 0                     &  0 &
V_{\sigma\alpha}
\end{array} \right)
$$
To connect the reduced parameterization eq. \ref{Ham-imp-red} to
the general eq. \ref{Ham-imp-apx} it is sufficient to diagonalize
the hermitian bath-matrix $\mathbf{E^{\alpha}_{B}}$ via a unitary
transformation\cite{civelli-2007}:
\begin{eqnarray}
\mathbf{E^{\alpha}_{B}}^{\ref{Ham-imp-apx}}=\,
\mathbf{S_{\alpha}^{\dagger}}\,\mathbf{E^{\alpha}_{B}}^{\ref{Ham-imp-red}}\,
\mathbf{S_{\alpha}} \label{S-transform}
\end{eqnarray}
and
%where by congruence it is intended
$\mathbf{S_{\alpha}^{T}}\,\mathbf{S_{\alpha}}= \mathbf{1}$. This
is in fact the requirement needed to have a canonical
transformation which preserves the fermionic
commutation-relations. In fact if we apply the transformation to
the vectors:
\begin{equation}
\left\{
\begin{array}{c}
\mathbf{\Phi_{\alpha}}^{\ref{Ham-imp-red}}=\,
\mathbf{S_{\alpha}} \,\mathbf{\Phi_{\alpha}}^{\ref{Ham-imp-apx}}\\
\mathbf{\Phi^{\dagger}_{\alpha}}^{\ref{Ham-imp-red}}=
\,\mathbf{\Phi^{\dagger}_{\alpha}}^{\ref{Ham-imp-apx}} \,
\mathbf{S^{\dagger}_{\alpha}}
\end{array}
\right. \label{S-transform2}
\end{equation}
and we have
\begin{equation}
\left\{ \mathbf{\Phi_{\alpha}}^{\ref{Ham-imp-red}},
\mathbf{\Phi^{\dagger}_{\alpha}}^{\ref{Ham-imp-red}} \right\}=\,
\mathbf{1} \label{commutatio}
\end{equation}
it is
\begin{eqnarray}
\left\{ \Phi_{q\alpha}^{\ref{Ham-imp-red}},
\Phi^{\dagger\ref{Ham-imp-red}}_{q\alpha} \right\}=\,  &  \\
&\sum_{kk'}\, S^{\dagger}_{qk} S_{k'q}\, \left\{
\Phi_{k\alpha}^{\ref{Ham-imp-red}},
\Phi^{\dagger\ref{Ham-imp-red}}_{k'\alpha} \right\}=\,    \\
&\sum_{k}\, S^{\dagger}_{qk} S_{kq}=\, 1 \label{commutatio2}
\end{eqnarray}
%i.e., $\mathbf{S}$ has to be congruent.
Finally
$\mathbf{V}_{\alpha}$ is subjected to the same transformation:
\begin{equation}
\mathbf{V}_{\alpha}^{\ref{Ham-imp-apx}}=\,
\mathbf{S}^{\dagger}_{\alpha}\,
\mathbf{V}_{\alpha}^{\ref{Ham-imp-red}}
\end{equation}

%%%%%%%%%%%%%%%%%%%%%%%%%%%%%%%%%%%%%%%%%%%%%%%%%%%%%%%%%%%%%%%
%----------------------------------------------------------------
%%%%%%%%%%%%%%%%%%%%%%%%%%%%%%%%%%%%%%%%%%%%%%%%%%%%%%%%%%%%%%%%%%%%%%%%%%
\begin{acknowledgments}
We thank G. Kotliar, who inspired many of the ideas in this work.
We thank O. Parcollet and K. Haule for sharing their QMC results.
We also acknowledge the fruitful discussion with A. Georges, M.
Capone and T. D. Stanescu. We enjoyed the exchange of ideas and
the hospitality of A. Sacuto and the SQUAP group at the
University of Paris 7. We acknowledge A.-M. S. Tremblay, S. S.
Kancharla, B. Kyung, I. Paul, A. Cano, E. Kats and P. Nozi\`eres
for their useful comments.
\end{acknowledgments}

\newcommand{{{\PRB}}}{{{Phys. Rev. B}}}\newcommand{{{\PRL}}}{{{Phys. Rev. Lett.}}}\newcommand{{{\NPB}}}{{{Nucl. Phys.}}}\newcommand{{{\RMP}}}{{{Rev. Mod. Phys.}}}\newcommand{{{\ADV}}}{{{Adv. Phys.}}}


\begin{thebibliography}{79}
\expandafter\ifx\csname natexlab\endcsname\relax\def\natexlab#1{#1}\fi
\expandafter\ifx\csname bibnamefont\endcsname\relax
\def\bibnamefont#1{#1}\fi
\expandafter\ifx\csname bibfnamefont\endcsname\relax
\def\bibfnamefont#1{#1}\fi
\expandafter\ifx\csname citenamefont\endcsname\relax
\def\citenamefont#1{#1}\fi
\expandafter\ifx\csname url\endcsname\relax
\def\url#1{\texttt{#1}}\fi
\expandafter\ifx\csname urlprefix\endcsname\relax\def\urlprefix{URL }\fi
\providecommand{\bibinfo}[2]{#2}
\providecommand{\eprint}[2][]{\url{#2}} 
\bibitem{bednorz86}\bibinfo{author}{\bibfnamefont{J. G. Bednorz}} and  \bibinfo{author}{\bibfnamefont{K. A. Muller}}, \bibinfo{journal}{Z. Phys. B } \textbf{\bibinfo{volume}{64} }\bibinfo{page}{189} (\bibinfo{year}{1986}).
\bibitem{kamihara08}\bibinfo{author}{\bibfnamefont{Y. Kamihara, T. Watanabe, M. Hirano}} and  \bibinfo{author}{\bibfnamefont{H. Hosono}}, \bibinfo{journal}{J. Am. Chem. Soc. } \textbf{\bibinfo{volume}{130  } }\bibinfo{page}{3296} (\bibinfo{year}{2008}).
\bibitem{day08}\bibinfo{author}{\bibfnamefont{C. Day}}, \bibinfo{journal}{Physics Today  } \textbf{\bibinfo{volume}{61  } }\bibinfo{page}{11-12} (\bibinfo{year}{2008}).
\bibitem{anderson}\bibinfo{author}{\bibfnamefont{P.~W. Anderson}}, \bibinfo{journal}{Science} \textbf{\bibinfo{volume}{235} }\bibinfo{page}{1196} (\bibinfo{year}{1987}).
\bibitem{revmodmft}\bibinfo{author}{\bibfnamefont{A. Georges}}, \bibinfo{author}{\bibfnamefont{G. Kotliar}}, \bibinfo{author}{\bibfnamefont{W. Krauth}} and  \bibinfo{author}{\bibfnamefont{M. J. Rozenberg}}, \bibinfo{journal}{ \RMP} \textbf{\bibinfo{volume}{68 } }\bibinfo{page}{13} (\bibinfo{year}{1996}).
\bibitem{zein05}\bibinfo{author}{\bibfnamefont{N. E. Zein}}, \bibinfo{author}{\bibfnamefont{S. Y. Savrasov}} and  \bibinfo{author}{\bibfnamefont{G. Kotliar}}, \bibinfo{journal}{ \PRL } \textbf{\bibinfo{volume}{ 96 } }\bibinfo{page}{226403} (\bibinfo{year}{2006}).
\bibitem{lich02}"Ruthenate and Rutheno-Cuprate Materials: Unconventional Superconductivity, Magnetism and Quantum Phase Transitions" \bibinfo{author}{\bibfnamefont{A. Lichteinstein}} and  \bibinfo{author}{\bibfnamefont{A. Liebsch}}, {\sl Springer-Verlag, Berlin, Germany} (\bibinfo{year}{2002}).
\bibitem{held01}\bibinfo{author}{\bibfnamefont{K. Held}}, \bibinfo{author}{\bibfnamefont{A. K. McMahan}} and  \bibinfo{author}{\bibfnamefont{R. T. Scalettar}}, \bibinfo{journal}{ \PRL } \textbf{\bibinfo{volume}{ 87 } }\bibinfo{page}{276404} (\bibinfo{year}{2001}).
\bibitem{haule05}\bibinfo{author}{\bibfnamefont{K. Haule}}, \bibinfo{author}{\bibfnamefont{V. Oudovenko}}, \bibinfo{author}{\bibfnamefont{S. Y. Savrasov}} and  \bibinfo{author}{\bibfnamefont{G. Kotliar}}, \bibinfo{journal}{ \PRL } \textbf{\bibinfo{volume}{ 94 } }\bibinfo{page}{036401} (\bibinfo{year}{2005}).
\bibitem{nature01}\bibinfo{author}{\bibfnamefont{S. Savrasov}}, \bibinfo{author}{\bibfnamefont{G. Kotliar}} and  \bibinfo{author}{\bibfnamefont{E. Abrahams}}, \bibinfo{journal}{ Nature } \textbf{\bibinfo{volume}{ 410 } }\bibinfo{page}{793} (\bibinfo{year}{2001}).
\bibitem{rmp06}\bibinfo{author}{\bibfnamefont{G. Kotliar}}, \bibinfo{author}{\bibfnamefont{S. Savrasov}}, \bibinfo{author}{\bibfnamefont{K. Haule}}, \bibinfo{author}{\bibfnamefont{V. Oudovenko}}, \bibinfo{author}{\bibfnamefont{O. Parcollet}} and  \bibinfo{author}{\bibfnamefont{C. Marianetti}}, \bibinfo{journal}{ Rev. Mod. Phys. } \textbf{\bibinfo{volume}{ 78 } }\bibinfo{page}{000865} (\bibinfo{year}{2006}).
\bibitem{damascelli}\bibinfo{author}{\bibfnamefont{A. Damascelli}}, \bibinfo{author}{\bibfnamefont{Z.~X. Shen}} and  \bibinfo{author}{\bibfnamefont{Z. Hussain}}, \bibinfo{journal}{\RMP} \textbf{\bibinfo{volume}{75 } }\bibinfo{page}{473} (\bibinfo{year}{2003}).
\bibitem{campuzano}"Physics of Superconductors II" \bibinfo{author}{\bibfnamefont{J. C. Campuzano}}, \bibinfo{author}{\bibfnamefont{M. R Norman}} and  \bibinfo{author}{\bibfnamefont{M. Randeria}}, {\sl K. H. Bennemann and J. B. Ketterson } (\bibinfo{year}{2004}) \bibinfo{note}{167-273}.
\bibitem{jarrell-rmp04}\bibinfo{author}{\bibfnamefont{Th. Maier}}, \bibinfo{author}{\bibfnamefont{M. Jarrell}}, \bibinfo{author}{\bibfnamefont{Th. Pruschke}} and  \bibinfo{author}{\bibfnamefont{M. Hettler}}, \bibinfo{journal}{ Rev. Mod. Phys. } \textbf{\bibinfo{volume}{ 77 } }\bibinfo{page}{1027-1080} (\bibinfo{year}{2005}).
\bibitem{biroli04}\bibinfo{author}{\bibfnamefont{G. Biroli}}, \bibinfo{author}{\bibfnamefont{O. Parcollet}} and  \bibinfo{author}{\bibfnamefont{G. Kotliar}}, \bibinfo{journal}{ \PRB } \textbf{\bibinfo{volume}{ 69 } }\bibinfo{page}{205108} (\bibinfo{year}{2004}).
\bibitem{lich00}\bibinfo{author}{\bibfnamefont{A. I. Lichtenstein}} and  \bibinfo{author}{\bibfnamefont{M. I. Katsnelson}}, \bibinfo{journal}{ \PRB  } \textbf{\bibinfo{volume}{ 62  } }\bibinfo{page}{R9283} (\bibinfo{year}{2000}).
\bibitem{maier00}\bibinfo{author}{\bibfnamefont{Th. Maier}}, \bibinfo{author}{\bibfnamefont{M. Jarrell}}, \bibinfo{author}{\bibfnamefont{Th. Pruschke}} and  \bibinfo{author}{\bibfnamefont{J. Keller}}, \bibinfo{journal}{ \PRL  } \textbf{\bibinfo{volume}{ 85  } }\bibinfo{page}{1524} (\bibinfo{year}{2000}).
\bibitem{tremblay06}\bibinfo{author}{\bibfnamefont{A.-M. S. Tremblay}}, \bibinfo{author}{\bibfnamefont{B. Kyung}} and  \bibinfo{author}{\bibfnamefont{D. S\'en\'echal}}, \bibinfo{journal}{Low Temperature Physics  } \textbf{\bibinfo{volume}{32  } }\bibinfo{page}{424} (\bibinfo{year}{2006}).
\bibitem{cdmft}\bibinfo{author}{\bibfnamefont{G. Kotliar}}, \bibinfo{author}{\bibfnamefont{S.~Y. Savrasov}}, \bibinfo{author}{\bibfnamefont{G. Palsson}} and  \bibinfo{author}{\bibfnamefont{G. Biroli}}, \bibinfo{journal}{\PRL } \textbf{\bibinfo{volume}{87 } }\bibinfo{page}{186401} (\bibinfo{year}{2001}).
\bibitem{fye-hirsch86}\bibinfo{author}{\bibfnamefont{R. M. Fye}} and  \bibinfo{author}{\bibfnamefont{J. E. Hirsch}}, \bibinfo{journal}{ \PRL } \textbf{\bibinfo{volume}{ 56 } }\bibinfo{page}{2521} (\bibinfo{year}{1986}).
\bibitem{fye-hirsch89}\bibinfo{author}{\bibfnamefont{R. M. Fye}} and  \bibinfo{author}{\bibfnamefont{J. E. Hirsch}}, \bibinfo{journal}{ \PRB } \textbf{\bibinfo{volume}{ 40 } }\bibinfo{page}{47804796} (\bibinfo{year}{1989}).
\bibitem{jarrell05}\bibinfo{author}{\bibfnamefont{T. A. Maier}}, \bibinfo{author}{\bibfnamefont{M. Jarrell}}, \bibinfo{author}{\bibfnamefont{T. C. Schulthess}}, \bibinfo{author}{\bibfnamefont{P. R. C. Kent}} and  \bibinfo{author}{\bibfnamefont{J. B. White}}, \bibinfo{journal}{ \PRL  } \textbf{\bibinfo{volume}{ 95 } }\bibinfo{page}{237001} (\bibinfo{year}{2005}).
\bibitem{jarrell-scalapino06}\bibinfo{author}{\bibfnamefont{T. A. Maier}}, \bibinfo{author}{\bibfnamefont{M. Jarrell}} and  \bibinfo{author}{\bibfnamefont{D. J. Scalapino}}, \bibinfo{journal}{\PRL } \textbf{\bibinfo{volume}{96  } }\bibinfo{page}{047005} (\bibinfo{year}{2006}).
\bibitem{senechal04}\bibinfo{author}{\bibfnamefont{D. S\'en\'echal}} and  \bibinfo{author}{\bibfnamefont{A.-M. S. Tremblay}}, \bibinfo{journal}{\PRL  } \textbf{\bibinfo{volume}{92  } }\bibinfo{page}{126401} (\bibinfo{year}{2004}).
\bibitem{bpk}\bibinfo{author}{\bibfnamefont{O. Parcollet}}, \bibinfo{author}{\bibfnamefont{G. Biroli}} and  \bibinfo{author}{\bibfnamefont{G. Kotliar}}, \bibinfo{journal}{\PRL } \textbf{\bibinfo{volume}{92 } }\bibinfo{page}{226402} (\bibinfo{year}{2004}).
\bibitem{marce05}\bibinfo{author}{\bibfnamefont{M. Civelli}}, \bibinfo{author}{\bibfnamefont{M. Capone}}, \bibinfo{author}{\bibfnamefont{S. S. Kancharla}}, \bibinfo{author}{\bibfnamefont{O. Parcollet}} and  \bibinfo{author}{\bibfnamefont{G. Kotliar}}, \bibinfo{journal}{\PRL } \textbf{\bibinfo{volume}{ 95 } }\bibinfo{page}{106402} (\bibinfo{year}{2005}).
\bibitem{bumsoo06}\bibinfo{author}{\bibfnamefont{B. Kyung}}, \bibinfo{author}{\bibfnamefont{S. S. Kancharla}}, \bibinfo{author}{\bibfnamefont{D. S\'{e}n\'{e}chal}}, \bibinfo{author}{\bibfnamefont{A.-M. S. Tremblay}}, \bibinfo{author}{\bibfnamefont{M. Civelli}} and  \bibinfo{author}{\bibfnamefont{G. Kotliar}}, \bibinfo{journal}{ \PRB  } \textbf{\bibinfo{volume}{ 73  } }\bibinfo{page}{165114} (\bibinfo{year}{2006}).
\bibitem{venky05}\bibinfo{author}{\bibfnamefont{S.~S. Kancharla}}, \bibinfo{author}{\bibfnamefont{B. Kyung}}, \bibinfo{author}{\bibfnamefont{D. S\'en\'echal}}, \bibinfo{author}{\bibfnamefont{M. Civelli}}, \bibinfo{author}{\bibfnamefont{M. Capone}}, \bibinfo{author}{\bibfnamefont{G. Kotliar}} and  \bibinfo{author}{\bibfnamefont{A.-M.~S. Tremblay}}, \bibinfo{journal}{\PRB} \textbf{\bibinfo{volume}{77} }\bibinfo{page}{184516} (\bibinfo{year}{2008}).
\bibitem{massimo06}\bibinfo{author}{\bibfnamefont{M. Capone}} and  \bibinfo{author}{\bibfnamefont{G. Kotliar}}, \bibinfo{journal}{ \PRB } \textbf{\bibinfo{volume}{ 74 } }\bibinfo{page}{054513} (\bibinfo{year}{2006}).
\bibitem{aichhorn06}\bibinfo{author}{\bibfnamefont{M. Aichhorn}}, \bibinfo{author}{\bibfnamefont{E. Arrigoni}}, \bibinfo{author}{\bibfnamefont{M. Potthoff}} and  \bibinfo{author}{\bibfnamefont{W. Hanke}}, \bibinfo{journal}{ \PRB } \textbf{\bibinfo{volume}{ 74 } }\bibinfo{page}{235117} (\bibinfo{year}{2006}).
\bibitem{aichhorn06b}\bibinfo{author}{\bibfnamefont{M. Aichhorn}}, \bibinfo{author}{\bibfnamefont{E. Arrigoni}}, \bibinfo{author}{\bibfnamefont{M. Potthoff}} and  \bibinfo{author}{\bibfnamefont{W. Hanke}}, \bibinfo{journal}{ \PRB } \textbf{\bibinfo{volume}{ 74 } }\bibinfo{page}{024508} (\bibinfo{year}{2006}).
\bibitem{senechal05}\bibinfo{author}{\bibfnamefont{D. S\'en\'echal}}, \bibinfo{author}{\bibfnamefont{P.-L. Lavertu}}, \bibinfo{author}{\bibfnamefont{M.-A. Marois}} and  \bibinfo{author}{\bibfnamefont{A.-M. S. Tremblay}}, \bibinfo{journal}{\PRL  } \textbf{\bibinfo{volume}{94  } }\bibinfo{page}{156404} (\bibinfo{year}{2005}).
\bibitem{huscroft-jarrell01}\bibinfo{author}{\bibfnamefont{C. Huscroft}}, \bibinfo{author}{\bibfnamefont{M. Jarrell}}, \bibinfo{author}{\bibfnamefont{Th. Maier}}, \bibinfo{author}{\bibfnamefont{S. Moukouri}} and  \bibinfo{author}{\bibfnamefont{A. N. Tahvildarzadeh}}, \bibinfo{journal}{ \PRL } \textbf{\bibinfo{volume}{ 86 } }\bibinfo{page}{139} (\bibinfo{year}{2001}).
\bibitem{stanescu03}\bibinfo{author}{\bibfnamefont{T. D. Stanescu}} and  \bibinfo{author}{\bibfnamefont{P. Phillips}}, \bibinfo{journal}{ \PRL  } \textbf{\bibinfo{volume}{ 91 } }\bibinfo{page}{017002} (\bibinfo{year}{2003}).
\bibitem{tsvelik}\bibinfo{author}{\bibfnamefont{F.~H.~L. Essler}} and  \bibinfo{author}{\bibfnamefont{A.~M. Tsvelik}}, \bibinfo{journal}{ \PRB} \textbf{\bibinfo{volume}{65 } }\bibinfo{page}{115117} (\bibinfo{year}{2002}).
\bibitem{tudor}\bibinfo{author}{\bibfnamefont{T. D. Stanescu}} and  \bibinfo{author}{\bibfnamefont{G. Kotliar}}, \bibinfo{journal}{ \PRB} \textbf{\bibinfo{volume}{ 74 } }\bibinfo{page}{125110} (\bibinfo{year}{2006}).
\bibitem{berthod}\bibinfo{author}{\bibfnamefont{C. Berthod}}, \bibinfo{author}{\bibfnamefont{T. Giamarchi}}, \bibinfo{author}{\bibfnamefont{S. Biermann}} and  \bibinfo{author}{\bibfnamefont{A. Georges}}, \bibinfo{journal}{ \PRL} \textbf{\bibinfo{volume}{97 } }\bibinfo{page}{136401} (\bibinfo{year}{2006}).
\bibitem{stanescu-2007-75}\bibinfo{author}{\bibfnamefont{T. D. Stanescu}}, \bibinfo{author}{\bibfnamefont{P. W. Phillips}} and  \bibinfo{author}{\bibfnamefont{T.-P. Choy}}, \bibinfo{journal}{\PRB} \textbf{\bibinfo{volume}{75} }\bibinfo{page}{104503} (\bibinfo{year}{2007}).
\bibitem{haule06}\bibinfo{author}{\bibfnamefont{K. Haule}} and  \bibinfo{author}{\bibfnamefont{G. Kotliar}}, \bibinfo{journal}{\PRB} \textbf{\bibinfo{volume}{76} }\bibinfo{page}{092503} (\bibinfo{year}{2007}).
\bibitem{krauth-caffarel}\bibinfo{author}{\bibfnamefont{M. Caffarel}} and  \bibinfo{author}{\bibfnamefont{W. Krauth}}, \bibinfo{journal}{\PRL} \textbf{\bibinfo{volume}{72 } }\bibinfo{page}{1545} (\bibinfo{year}{1994}).
\bibitem{haule-ctqmc}\bibinfo{author}{\bibfnamefont{K. Haule}} and  \bibinfo{author}{\bibfnamefont{G. Kotliar}}, \bibinfo{journal}{\PRB} \textbf{\bibinfo{volume}{76} }\bibinfo{page}{104509} (\bibinfo{year}{2007}).
\bibitem{marce}\bibinfo{author}{\bibfnamefont{M. Capone}}, \bibinfo{author}{\bibfnamefont{M. Civelli}}, \bibinfo{author}{\bibfnamefont{S.~S.~Kancharla}}, \bibinfo{author}{\bibfnamefont{C. Castellani}} and  \bibinfo{author}{\bibfnamefont{G. Kotliar}}, \bibinfo{journal}{\PRB } \textbf{\bibinfo{volume}{69 } }\bibinfo{page}{195105} (\bibinfo{year}{2004}).
\bibitem{civelli-2007}\bibinfo{author}{\bibfnamefont{M. Civelli}}, \bibinfo{journal}{PhD Thesis, arXiv.org:0710.2802} (\bibinfo{year}{2007}).
\bibitem{marce08}\bibinfo{author}{\bibfnamefont{M. Civelli}}, \bibinfo{author}{\bibfnamefont{M. Capone}}, \bibinfo{author}{\bibfnamefont{A. Georges}}, \bibinfo{author}{\bibfnamefont{K. Haule}}, \bibinfo{author}{\bibfnamefont{O. Parcollet}}, \bibinfo{author}{\bibfnamefont{T.~D. Stanescu}} and  \bibinfo{author}{\bibfnamefont{G. Kotliar}}, \bibinfo{journal}{\PRL} \textbf{\bibinfo{volume}{100} }\bibinfo{page}{046402} (\bibinfo{year}{2008}).
\bibitem{patrickrmp}\bibinfo{author}{\bibfnamefont{P. A. Lee}}, \bibinfo{author}{\bibfnamefont{N. Nagaosa}} and  \bibinfo{author}{\bibfnamefont{X.-G. Wen}}, \bibinfo{journal}{ \RMP } \textbf{\bibinfo{volume}{ 78 } }\bibinfo{page}{17} (\bibinfo{year}{2006}).
\bibitem{biroli02}\bibinfo{author}{\bibfnamefont{G. Biroli}} and  \bibinfo{author}{\bibfnamefont{G. Kotliar}}, \bibinfo{journal}{ \PRB } \textbf{\bibinfo{volume}{ 65 } }\bibinfo{page}{155112} (\bibinfo{year}{2002}).
\bibitem{tudor06}\bibinfo{author}{\bibfnamefont{T. D. Stanescu}}, \bibinfo{author}{\bibfnamefont{M. Civelli}}, \bibinfo{author}{\bibfnamefont{K. Haule}} and  \bibinfo{author}{\bibfnamefont{G. Kotliar}}, \bibinfo{journal}{ An. of Phys.} \textbf{\bibinfo{volume}{ 321 } }\bibinfo{page}{1682} (\bibinfo{year}{2006}).
\bibitem{PSK}\bibinfo{author}{\bibfnamefont{A. Perali}}, \bibinfo{author}{\bibfnamefont{M. Sindel}} and  \bibinfo{author}{\bibfnamefont{G. Kotliar}}, \bibinfo{journal}{Eur. Phys. J. B} \textbf{\bibinfo{volume}{24} }\bibinfo{page}{487} (\bibinfo{year}{2002}).
\bibitem{hubbard63}\bibinfo{author}{\bibfnamefont{J. Hubbard}}, \bibinfo{journal}{Proc. Roy. Soc. A  } \textbf{\bibinfo{volume}{276  } }\bibinfo{page}{238} (\bibinfo{year}{1963}).
\bibitem{venky}\bibinfo{author}{\bibfnamefont{C.~J. Bolech}}, \bibinfo{author}{\bibfnamefont{S.~S. Kancharla}} and  \bibinfo{author}{\bibfnamefont{G. Kotliar}}, \bibinfo{journal}{ \PRB} \textbf{\bibinfo{volume}{67 } }\bibinfo{page}{075110} (\bibinfo{year}{2003}).
\bibitem{haule06b}\bibinfo{author}{\bibfnamefont{K. Haule}}, \bibinfo{journal}{\PRB} \textbf{\bibinfo{volume}{75} }\bibinfo{page}{155113} (\bibinfo{year}{2007}).
\bibitem{liu}\bibinfo{author}{\bibfnamefont{G. Kotliar}} and  \bibinfo{author}{\bibfnamefont{J. Liu}}, \bibinfo{journal}{ \PRB   } \textbf{\bibinfo{volume}{ 38  } }\bibinfo{page}{R5142} (\bibinfo{year}{1988}).
\bibitem{davis05}\bibinfo{author}{\bibfnamefont{K. McElroy, D.-H. Lee, J. E. Hoffman, K. M. Lang, J. Lee, E. W. Hudson, H. Eisaki, S. Uchida}} and  \bibinfo{author}{\bibfnamefont{J. C. Davis}}, \bibinfo{journal}{ \PRL } \textbf{\bibinfo{volume}{ 94 } }\bibinfo{page}{197005} (\bibinfo{year}{2005}).
\bibitem{tacon06}\bibinfo{author}{\bibfnamefont{M. Le Tacon}}, \bibinfo{author}{\bibfnamefont{A. Sacuto}}, \bibinfo{author}{\bibfnamefont{A. Georges}}, \bibinfo{author}{\bibfnamefont{G. Kotliar}}, \bibinfo{author}{\bibfnamefont{Y. Gallais}}, \bibinfo{author}{\bibfnamefont{D. Colson}} and  \bibinfo{author}{\bibfnamefont{A. Forget}}, \bibinfo{journal}{ Natur. Phys. } \textbf{\bibinfo{volume}{ 2 } }\bibinfo{page}{537} (\bibinfo{year}{2006}).
\bibitem{tanaka06}\bibinfo{author}{\bibfnamefont{K. Tanaka, W. S. Lee, D. H. Lu, A. Fujimori, T. Fujii, Risdiana, I. Terasaki, D. J. Scalapino, T. P. Devereaux, Z. Hussain}} and  \bibinfo{author}{\bibfnamefont{Z.-X. Shen}}, \bibinfo{journal}{ Science } \textbf{\bibinfo{volume}{ 314 } }\bibinfo{page}{1910} (\bibinfo{year}{2006}).
\bibitem{cho06}\bibinfo{author}{\bibfnamefont{A. Cho}}, \bibinfo{journal}{ Science } \textbf{\bibinfo{volume}{ 314 } }\bibinfo{page}{1072} (\bibinfo{year}{2006}).
\bibitem{millis06}\bibinfo{author}{\bibfnamefont{A. J. Millis}}, \bibinfo{journal}{ Science } \textbf{\bibinfo{volume}{ 314 } }\bibinfo{page}{1888} (\bibinfo{year}{2006}).
\bibitem{deutscher99}\bibinfo{author}{\bibfnamefont{G. Deutscher}}, \bibinfo{journal}{ Nature } \textbf{\bibinfo{volume}{ 397 } }\bibinfo{page}{410} (\bibinfo{year}{1999}).
\bibitem{huefner-2008-71}\bibinfo{author}{\bibfnamefont{S. Huefner}}, \bibinfo{author}{\bibfnamefont{M.~A. Hossain}}, \bibinfo{author}{\bibfnamefont{A. Damascelli}} and  \bibinfo{author}{\bibfnamefont{G.~A. Sawatzky}}, \bibinfo{journal}{Rep. Prog. Phys.} \textbf{\bibinfo{volume}{71} }\bibinfo{page}{062501} (\bibinfo{year}{2008}).
\bibitem{kondo07}\bibinfo{author}{\bibfnamefont{T. Kondo, T. Takeuchi, A. Kaminski, S. Tsuda}} and  \bibinfo{author}{\bibfnamefont{S. Shin}}, \bibinfo{journal}{\PRL } \textbf{\bibinfo{volume}{98 } }\bibinfo{page}{267004} (\bibinfo{year}{2007}).
\bibitem{gomes2007}\bibinfo{author}{\bibfnamefont{K. K. Gomes, A. N. Pasupathy, A. Pushp, S. Ono, Y. Ando}} and  \bibinfo{author}{\bibfnamefont{A. Yazdani}}, \bibinfo{journal}{Nature } \textbf{\bibinfo{volume}{447 } }\bibinfo{page}{569} (\bibinfo{year}{2007}).
\bibitem{Millis1998}\bibinfo{author}{\bibfnamefont{L. B. Ioffe}} and  \bibinfo{author}{\bibfnamefont{A. J. Millis}}, \bibinfo{journal}{\PRB  } \textbf{\bibinfo{volume}{58  } }\bibinfo{page}{11631} (\bibinfo{year}{1998}).
\bibitem{Zheleznyak1998}\bibinfo{author}{\bibfnamefont{A. T. Zheleznyak}}, \bibinfo{author}{\bibfnamefont{V. M. Yakovenko}}, \bibinfo{author}{\bibfnamefont{H. D. Drew}} and  \bibinfo{author}{\bibfnamefont{I. I. Mazin}}, \bibinfo{journal}{\PRB } \textbf{\bibinfo{volume}{57   } }\bibinfo{page}{3089} (\bibinfo{year}{1998}).
\bibitem{Hlubina1995}\bibinfo{author}{\bibfnamefont{R. Hlubina}} and  \bibinfo{author}{\bibfnamefont{T. M. Rice}}, \bibinfo{journal}{\PRB  } \textbf{\bibinfo{volume}{51  } }\bibinfo{page}{9253} (\bibinfo{year}{1995}).
\bibitem{olivier-note}\bibinfo{author}{\bibfnamefont{O. Parcollet}}, \bibinfo{journal}{private communication } \textbf{\bibinfo{volume}{  } }\bibinfo{page}{}.
\bibitem{ding}\bibinfo{author}{\bibfnamefont{H. Matsui}}, \bibinfo{author}{\bibfnamefont{T. Sato}}, \bibinfo{author}{\bibfnamefont{T. Takahahi}}, \bibinfo{author}{\bibfnamefont{S.-C. Wang}}, \bibinfo{author}{\bibfnamefont{H.-B. Yang}}, \bibinfo{author}{\bibfnamefont{H. Ding}}, \bibinfo{author}{\bibfnamefont{T. Fujii}}, \bibinfo{author}{\bibfnamefont{T. Watanabe}} and  \bibinfo{author}{\bibfnamefont{A. Matsuda}}, \bibinfo{journal}{\PRL} \textbf{\bibinfo{volume}{ 90 } }\bibinfo{page}{217002} (\bibinfo{year}{2003}).
\bibitem{Luttinger60}\bibinfo{author}{\bibfnamefont{J. M. Luttinger}}, \bibinfo{journal}{Phys. Rev.  } \textbf{\bibinfo{volume}{119  } }\bibinfo{page}{1153} (\bibinfo{year}{1960}).
\bibitem{deleo-2008}\bibinfo{author}{\bibfnamefont{L. De Leo}}, \bibinfo{author}{\bibfnamefont{M. Civelli}} and  \bibinfo{author}{\bibfnamefont{G. Kotliar}}, \bibinfo{journal}{arXiv.org:0804.3314} (\bibinfo{year}{2008}).
\bibitem{ferrero-2005-72}\bibinfo{author}{\bibfnamefont{M. Ferrero}}, \bibinfo{author}{\bibfnamefont{F. Becca}}, \bibinfo{author}{\bibfnamefont{M. Fabrizio}} and  \bibinfo{author}{\bibfnamefont{M. Capone}}, \bibinfo{journal}{\PRB} \textbf{\bibinfo{volume}{72} }\bibinfo{page}{205126} (\bibinfo{year}{2005}).
\bibitem{medici-2005-72}\bibinfo{author}{\bibfnamefont{L.~de' Medici}}, \bibinfo{author}{\bibfnamefont{A. Georges}} and  \bibinfo{author}{\bibfnamefont{S. Biermann}}, \bibinfo{journal}{\PRB} \textbf{\bibinfo{volume}{72} }\bibinfo{page}{205124} (\bibinfo{year}{2005}).
\bibitem{Shen-Nature03}\bibinfo{author}{\bibfnamefont{X. J. Zhou, T. Yoshida, A. Lanzara, P. V. Bogdanov, S. A. Kellar, K. M. Shen, W. L. Yang, F. Ronning, T. Sasagawa, T. Kakeshita, T. Noda, H. Eisaki, S. Uchida, C. T. Lin, F. Zhou, J. W. Xiong, W. X. Ti, Z. X. Zhao, A. Fujimori, Z. Hussain}} and  \bibinfo{author}{\bibfnamefont{Z.-X. Shen}}, \bibinfo{journal}{Nature} \textbf{\bibinfo{volume}{423} }\bibinfo{page}{398} (\bibinfo{year}{2003}).
\bibitem{Bonn96}\bibinfo{author}{\bibfnamefont{D. A. Bonn}}, \bibinfo{journal}{  Czech. J. Phys. } \textbf{\bibinfo{volume}{ 46 } }\bibinfo{page}{3195} (\bibinfo{year}{1996}).
\bibitem{Panagopoulos98}\bibinfo{author}{\bibfnamefont{C. Panagopoulos}} and  \bibinfo{author}{\bibfnamefont{T. Xiang}}, \bibinfo{journal}{ Phys. Rev. Lett. } \textbf{\bibinfo{volume}{ 81 } }\bibinfo{page}{2336} (\bibinfo{year}{1998}).
\bibitem{kyung02}\bibinfo{author}{\bibfnamefont{B. Kyung}} and  \bibinfo{author}{\bibfnamefont{A.-M. S. Tremblay}}, \bibinfo{journal}{cond-mat/0204500} \textbf{\bibinfo{volume}{  } }\bibinfo{page}{} (\bibinfo{year}{2002}).
\bibitem{yang-2006-73}\bibinfo{author}{\bibfnamefont{Kai-Yu Yang}}, \bibinfo{author}{\bibfnamefont{T.~M. Rice}} and  \bibinfo{author}{\bibfnamefont{Fu-Chun Zhang}}, \bibinfo{journal}{\PRB} \textbf{\bibinfo{volume}{73} }\bibinfo{page}{174501} (\bibinfo{year}{2006}).
\bibitem{valenzuela07}\bibinfo{author}{\bibfnamefont{B. Valenzuela}} and  \bibinfo{author}{\bibfnamefont{E. Bascones}}, \bibinfo{journal}{ Phys. Rev. Lett. } \textbf{\bibinfo{volume}{ 98 } }\bibinfo{page}{227002} (\bibinfo{year}{2007}).
\bibitem{aichhorn-2007-99}\bibinfo{author}{\bibfnamefont{M. Aichhorn}}, \bibinfo{author}{\bibfnamefont{E. Arrigoni}}, \bibinfo{author}{\bibfnamefont{Z.~B. Huang}} and  \bibinfo{author}{\bibfnamefont{W. Hanke}}, \bibinfo{journal}{\PRL} \textbf{\bibinfo{volume}{99} }\bibinfo{page}{257002} (\bibinfo{year}{2007}).
\bibitem{guyard-2008}\bibinfo{author}{\bibfnamefont{W. Guyard}}, \bibinfo{author}{\bibfnamefont{A. Sacuto}}, \bibinfo{author}{\bibfnamefont{M. Cazayous}}, \bibinfo{author}{\bibfnamefont{Y. Gallais}}, \bibinfo{author}{\bibfnamefont{M.~Le Tacon}}, \bibinfo{author}{\bibfnamefont{D. Colson}} and  \bibinfo{author}{\bibfnamefont{A. Forget}}, \bibinfo{journal}{arXiv.org:0802.3166} (\bibinfo{year}{2008}).
\bibitem{Doiron-Leyraud07}\bibinfo{author}{\bibfnamefont{N. Doiron-Leyraud, C. Proust, D. LeBoeuf, J. Levallois, J.-B. Bonnemaison, R. Liang, D.A. Bonn, W.N. Hardy}} and  \bibinfo{author}{\bibfnamefont{L. Taillefer}}, \bibinfo{journal}{Nature  } \textbf{\bibinfo{volume}{447  } }\bibinfo{page}{565} (\bibinfo{year}{2007}).
\bibitem{daou-2008}\bibinfo{author}{\bibfnamefont{R. Daou}}, \bibinfo{author}{\bibfnamefont{D. LeBoeuf}}, \bibinfo{author}{\bibfnamefont{N. Doiron-Leyraud}}, \bibinfo{author}{\bibfnamefont{S.~Y. Li}}, \bibinfo{author}{\bibfnamefont{F. Laliberte}}, \bibinfo{author}{\bibfnamefont{O. Cyr-Choiniere}}, \bibinfo{author}{\bibfnamefont{Y.~J. Jo}}, \bibinfo{author}{\bibfnamefont{L. Balicas}}, \bibinfo{author}{\bibfnamefont{J.~-Q. Yan}}, \bibinfo{author}{\bibfnamefont{J.~-S. Zhou}}, \bibinfo{author}{\bibfnamefont{J.~B. Goodenough}} and  \bibinfo{author}{\bibfnamefont{L. Taillefer}}, \bibinfo{journal}{arXiv.org:0806.2881} (\bibinfo{year}{2008}).
\bibitem{balakirev-2007}\bibinfo{author}{\bibfnamefont{F.~F. Balakirev}}, \bibinfo{author}{\bibfnamefont{J.~B. Betts}}, \bibinfo{author}{\bibfnamefont{A. Migliori}}, \bibinfo{author}{\bibfnamefont{I. Tsukada}}, \bibinfo{author}{\bibfnamefont{Yoichi Ando}} and  \bibinfo{author}{\bibfnamefont{G.~S. Boebinger}}, \bibinfo{journal}{arXiv.org:0710.4612} (\bibinfo{year}{2007}).
\bibitem{balakirev-2003}\bibinfo{author}{\bibfnamefont{F. F. Balakirev}}, \bibinfo{author}{\bibfnamefont{J. B. Betts}}, \bibinfo{author}{\bibfnamefont{A. Migliori}}, \bibinfo{author}{\bibfnamefont{S. Ono, Yoichi Ando}} and  \bibinfo{author}{\bibfnamefont{G. S. Boebinger}}, \bibinfo{journal}{Nature  } \textbf{\bibinfo{volume}{424  } }\bibinfo{page}{912} (\bibinfo{year}{2003}).
\bibitem{Plakida07}\bibinfo{author}{\bibfnamefont{N. M. Plakida}} and  \bibinfo{author}{\bibfnamefont{V. S. Oudovenko}}, \bibinfo{journal}{JETP  } \textbf{\bibinfo{volume}{104  } }\bibinfo{page}{230} (\bibinfo{year}{2007}).
\bibitem{avella07}\bibinfo{author}{\bibfnamefont{A. Avella}} and  \bibinfo{author}{\bibfnamefont{F. Mancini}}, \bibinfo{journal}{\PRB} \textbf{\bibinfo{volume}{75} }\bibinfo{page}{134518} (\bibinfo{year}{2007}).
\bibitem{Dzyaloshinskii03}\bibinfo{author}{\bibfnamefont{I. E. Dzyaloshinskii}}, \bibinfo{journal}{\PRB  } \textbf{\bibinfo{volume}{68  } }\bibinfo{page}{085113} (\bibinfo{year}{2003}).
\bibitem{essler-2003-90}\bibinfo{author}{\bibfnamefont{F.~H.~L. Essler}} and  \bibinfo{author}{\bibfnamefont{A.~M. Tsvelik}}, \bibinfo{journal}{\PRL} \textbf{\bibinfo{volume}{90} }\bibinfo{page}{126401} (\bibinfo{year}{2003}).
\bibitem{konik-2005}\bibinfo{author}{\bibfnamefont{R.~M. Konik}}, \bibinfo{author}{\bibfnamefont{T.~M. Rice}} and  \bibinfo{author}{\bibfnamefont{A.~M. Tsvelik}}, \bibinfo{journal}{arXiv.org:cond-mat/0511268} (\bibinfo{year}{2005}).
\bibitem{rosch-2007-59}\bibinfo{author}{\bibfnamefont{A. Rosch}}, \bibinfo{journal}{Eur. Phys. Jour. B} \textbf{\bibinfo{volume}{59} }\bibinfo{page}{495} (\bibinfo{year}{2007}).
\bibitem{Lanzara01}\bibinfo{author}{\bibfnamefont{A. Lanzara}}, \bibinfo{author}{\bibfnamefont{P. V. Bogdanov}}, \bibinfo{author}{\bibfnamefont{X. J. Zhou}}, \bibinfo{author}{\bibfnamefont{S. A. Kellar}}, \bibinfo{author}{\bibfnamefont{D. L. Feng}}, \bibinfo{author}{\bibfnamefont{E. D. Lu}}, \bibinfo{author}{\bibfnamefont{T. Yoshida}}, \bibinfo{author}{\bibfnamefont{H. Eisaki}}, \bibinfo{author}{\bibfnamefont{A. Fujimori}}, \bibinfo{author}{\bibfnamefont{K. Kishio, J.-I. Shimoyama}}, \bibinfo{author}{\bibfnamefont{T. Noda}}, \bibinfo{author}{\bibfnamefont{S. Uchida}}, \bibinfo{author}{\bibfnamefont{Z. Hussain}} and  \bibinfo{author}{\bibfnamefont{Z.-X. Shen}}, \bibinfo{journal}{Nature  } \textbf{\bibinfo{volume}{412  } }\bibinfo{page}{510} (\bibinfo{year}{2001}).
\bibitem{Z-XShen02}\bibinfo{author}{\bibfnamefont{Z.-X. Shen}}, \bibinfo{author}{\bibfnamefont{A. Lanzara}}, \bibinfo{author}{\bibfnamefont{S. Ishihara}} and  \bibinfo{author}{\bibfnamefont{N. Nagaosa}}, \bibinfo{journal}{Philos. Mag. B  } \textbf{\bibinfo{volume}{82 } }\bibinfo{page}{1349} (\bibinfo{year}{2002}).
\bibitem{he-2001-86}\bibinfo{author}{\bibfnamefont{H. He}}, \bibinfo{author}{\bibfnamefont{Y. Sidis}}, \bibinfo{author}{\bibfnamefont{P. Bourges}}, \bibinfo{author}{\bibfnamefont{G.~D. Gu}}, \bibinfo{author}{\bibfnamefont{A. Ivanov}}, \bibinfo{author}{\bibfnamefont{N. Koshizuka}}, \bibinfo{author}{\bibfnamefont{B. Liang}}, \bibinfo{author}{\bibfnamefont{C.~T. Lin}}, \bibinfo{author}{\bibfnamefont{L.~P. Regnault}}, \bibinfo{author}{\bibfnamefont{E. Schoenherr}} and  \bibinfo{author}{\bibfnamefont{B. Keimer}}, \bibinfo{journal}{\PRL} \textbf{\bibinfo{volume}{86} }\bibinfo{page}{1610} (\bibinfo{year}{2001}).
\bibitem{Hwang04}\bibinfo{author}{\bibfnamefont{J. Hwang}}, \bibinfo{author}{\bibfnamefont{T. Timusk}} and  \bibinfo{author}{\bibfnamefont{G. D. Gu}}, \bibinfo{journal}{Nature  } \textbf{\bibinfo{volume}{427  } }\bibinfo{page}{714} (\bibinfo{year}{2004}).
\bibitem{cuk-2004-94}\bibinfo{author}{\bibfnamefont{T. Cuk}}, \bibinfo{author}{\bibfnamefont{F. Baumberger}}, \bibinfo{author}{\bibfnamefont{D. H. Lu}}, \bibinfo{author}{\bibfnamefont{N. Ingle}}, \bibinfo{author}{\bibfnamefont{X. J. Zhou}}, \bibinfo{author}{\bibfnamefont{H. Eisaki}}, \bibinfo{author}{\bibfnamefont{N. Kaneko}}, \bibinfo{author}{\bibfnamefont{Z. Hussain}}, \bibinfo{author}{\bibfnamefont{T. P. Devereaux}}, \bibinfo{author}{\bibfnamefont{N. Nagaosa}} and  \bibinfo{author}{\bibfnamefont{Z.-X. Shen}}, \bibinfo{journal}{\PRL} \textbf{\bibinfo{volume}{94} }\bibinfo{page}{117003} (\bibinfo{year}{2004}).
\bibitem{kaminski}\bibinfo{author}{\bibfnamefont{A. Kaminski}}, \bibinfo{author}{\bibfnamefont{M. Randeira}}, \bibinfo{author}{\bibfnamefont{J. C. Campuzano}}, \bibinfo{author}{\bibfnamefont{M. R. Norman}}, \bibinfo{author}{\bibfnamefont{H. Fretwell}}, \bibinfo{author}{\bibfnamefont{J. Mesot}}, \bibinfo{author}{\bibfnamefont{T. Sato}}, \bibinfo{author}{\bibfnamefont{T. Takahasci}} and  \bibinfo{author}{\bibfnamefont{K. Kadowaki}}, \bibinfo{journal}{ \PRL} \textbf{\bibinfo{volume}{ 86 } }\bibinfo{page}{1070} (\bibinfo{year}{2001}).
\bibitem{byczuk-2007-3}\bibinfo{author}{\bibfnamefont{K. Byczuk}}, \bibinfo{author}{\bibfnamefont{M. Kollar}}, \bibinfo{author}{\bibfnamefont{K. Held}}, \bibinfo{author}{\bibfnamefont{Y.~-F. Yang}}, \bibinfo{author}{\bibfnamefont{I.~A. Nekrasov}}, \bibinfo{author}{\bibfnamefont{Th. Pruschke}} and  \bibinfo{author}{\bibfnamefont{D. Vollhardt}}, \bibinfo{journal}{Nature Physics} \textbf{\bibinfo{volume}{3} }\bibinfo{page}{168} (\bibinfo{year}{2007}).
\bibitem{chakraborty-2007}\bibinfo{author}{\bibfnamefont{S. Chakraborty}}, \bibinfo{author}{\bibfnamefont{D. Galanakis}} and  \bibinfo{author}{\bibfnamefont{P. Phillips}}, \bibinfo{journal}{arXiv.org:0712.2838} (\bibinfo{year}{2007}).
\bibitem{graf-2007-98}\bibinfo{author}{\bibfnamefont{J. Graf}}, \bibinfo{author}{\bibfnamefont{G.-H. Gweon}}, \bibinfo{author}{\bibfnamefont{K. McElroy}}, \bibinfo{author}{\bibfnamefont{S.~Y. Zhou}}, \bibinfo{author}{\bibfnamefont{C. Jozwiak}}, \bibinfo{author}{\bibfnamefont{E. Rotenberg}}, \bibinfo{author}{\bibfnamefont{A. Bill}}, \bibinfo{author}{\bibfnamefont{T. Sasagawa}}, \bibinfo{author}{\bibfnamefont{H. Eisaki}}, \bibinfo{author}{\bibfnamefont{S. Uchida}}, \bibinfo{author}{\bibfnamefont{H. Takagi}}, \bibinfo{author}{\bibfnamefont{D.-H. Lee}} and  \bibinfo{author}{\bibfnamefont{A. Lanzara}}, \bibinfo{journal}{\PRL} \textbf{\bibinfo{volume}{98} }\bibinfo{page}{067004} (\bibinfo{year}{2007}).
\bibitem{valla-2007-98}\bibinfo{author}{\bibfnamefont{T. Valla}}, \bibinfo{author}{\bibfnamefont{T.~E. Kidd}}, \bibinfo{author}{\bibfnamefont{Z.-H. Pan}}, \bibinfo{author}{\bibfnamefont{A.~V. Fedorov}}, \bibinfo{author}{\bibfnamefont{W.-G. Yin}}, \bibinfo{author}{\bibfnamefont{G.~D. Gu}} and  \bibinfo{author}{\bibfnamefont{P.~D. Johnson}}, \bibinfo{journal}{\PRL} \textbf{\bibinfo{volume}{98} }\bibinfo{page}{167003} (\bibinfo{year}{2007}).
\bibitem{cieplak06}\bibinfo{author}{\bibfnamefont{M. Z. Cieplak}}, \bibinfo{author}{\bibfnamefont{A. Abal'oshev}}, \bibinfo{author}{\bibfnamefont{I. Zaytseva}}, \bibinfo{author}{\bibfnamefont{M. Berkowski}}, \bibinfo{author}{\bibfnamefont{S. Guha}} and  \bibinfo{author}{\bibfnamefont{Q. Wu}}, \bibinfo{journal}{Acta Physica Polonica A} \textbf{\bibinfo{volume}{109  } }\bibinfo{page}{573} (\bibinfo{year}{2006}).
\bibitem{bza}\bibinfo{author}{\bibfnamefont{G. Baskaran}}, \bibinfo{author}{\bibfnamefont{Z. Zou}} and  \bibinfo{author}{\bibfnamefont{P.~W. Anderson}}, \bibinfo{journal}{ Solid State Com.  } \textbf{\bibinfo{volume}{63 } }\bibinfo{page}{973} (\bibinfo{year}{1987}).
\bibitem{ruckestein87}\bibinfo{author}{\bibfnamefont{A. E. Ruckenstein}}, \bibinfo{author}{\bibfnamefont{P. J. Hirschfeld}} and  \bibinfo{author}{\bibfnamefont{J. Appel}}, \bibinfo{journal}{ \PRB } \textbf{\bibinfo{volume}{ 36 } }\bibinfo{page}{857} (\bibinfo{year}{1987}).
\bibitem{stanescu04}\bibinfo{author}{\bibfnamefont{T. D. Stanescu}} and  \bibinfo{author}{\bibfnamefont{G. Kotliar}}, \bibinfo{journal}{ \PRB  } \textbf{\bibinfo{volume}{ 70 } }\bibinfo{page}{205112} (\bibinfo{year}{2004}).
\bibitem{campuzano07}\bibinfo{author}{\bibfnamefont{M. Shi}}, \bibinfo{author}{\bibfnamefont{J. Chang}}, \bibinfo{author}{\bibfnamefont{S. Pailh\'es, M. R. Norman}}, \bibinfo{author}{\bibfnamefont{J. C. Campuzano}}, \bibinfo{author}{\bibfnamefont{M. Mansson}}, \bibinfo{author}{\bibfnamefont{T. Claesson}}, \bibinfo{author}{\bibfnamefont{O. Tjernberg}}, \bibinfo{author}{\bibfnamefont{A. Bendounan}}, \bibinfo{author}{\bibfnamefont{L. Patthey}}, \bibinfo{author}{\bibfnamefont{N. Momono}}, \bibinfo{author}{\bibfnamefont{M. Oda}}, \bibinfo{author}{\bibfnamefont{M. Ido}}, \bibinfo{author}{\bibfnamefont{C. Mudry}} and  \bibinfo{author}{\bibfnamefont{J. Mesot}}, \bibinfo{journal}{\PRL  } \textbf{\bibinfo{volume}{101  } }\bibinfo{page}{047002} (\bibinfo{year}{2007}).
\bibitem{campuzano07b}\bibinfo{author}{\bibfnamefont{A. Kanigel}}, \bibinfo{author}{\bibfnamefont{U. Chatterjee}}, \bibinfo{author}{\bibfnamefont{M. Randeria}}, \bibinfo{author}{\bibfnamefont{M. R. Norman}}, \bibinfo{author}{\bibfnamefont{S. Souma}}, \bibinfo{author}{\bibfnamefont{M. Shi}}, \bibinfo{author}{\bibfnamefont{Z. Z. Li}}, \bibinfo{author}{\bibfnamefont{H. Raffy}} and  \bibinfo{author}{\bibfnamefont{J. C. Campuzano}}, \bibinfo{journal}{\PRL  } \textbf{\bibinfo{volume}{99  } }\bibinfo{page}{157001} (\bibinfo{year}{2007}).
\bibitem{campuzano08}\bibinfo{author}{\bibfnamefont{A. Kanigel}}, \bibinfo{author}{\bibfnamefont{U. Chatterjee}}, \bibinfo{author}{\bibfnamefont{M. Randeria}}, \bibinfo{author}{\bibfnamefont{M. R. Norman}}, \bibinfo{author}{\bibfnamefont{G. Koren}}, \bibinfo{author}{\bibfnamefont{K. Kadowaki}} and  \bibinfo{author}{\bibfnamefont{J. C. Campuzano}}, \bibinfo{journal}{arXiv:0803.3052v1  } (\bibinfo{year}{2008}).
\bibitem{sachdev92}\bibinfo{author}{\bibfnamefont{S. Sachdev}} and  \bibinfo{author}{\bibfnamefont{J. Ye}}, \bibinfo{journal}{\PRL} \textbf{\bibinfo{volume}{69} }\bibinfo{page}{2411} (\bibinfo{year}{1992}).
\bibitem{perali96}\bibinfo{author}{\bibfnamefont{A. Perali}}, \bibinfo{author}{\bibfnamefont{C. Castellani}}, \bibinfo{author}{\bibfnamefont{C. Di Castro}} and  \bibinfo{author}{\bibfnamefont{M. Grilli}}, \bibinfo{journal}{Phys. Rev. B} \textbf{\bibinfo{volume}{54} }\bibinfo{page}{16216--16225} (\bibinfo{year}{1996}).
\bibitem{kivelson98}\bibinfo{author}{\bibfnamefont{S. A. Kivelson}}, \bibinfo{author}{\bibfnamefont{E. Fradkin}} and  \bibinfo{author}{\bibfnamefont{V. J. Emery}}, \bibinfo{journal}{Nature  } \textbf{\bibinfo{volume}{393  } }\bibinfo{page}{550} (\bibinfo{year}{1998}).
\bibitem{varma99}\bibinfo{author}{\bibfnamefont{C. M. Varma}}, \bibinfo{journal}{Phys. Rev. Lett.} \textbf{\bibinfo{volume}{83} }\bibinfo{page}{3538} (\bibinfo{year}{1999}).
\bibitem{ando95}\bibinfo{author}{\bibfnamefont{Yoichi Ando}}, \bibinfo{author}{\bibfnamefont{G. S. Boebinger}}, \bibinfo{author}{\bibfnamefont{A. Passner}}, \bibinfo{author}{\bibfnamefont{T. Kimura}} and  \bibinfo{author}{\bibfnamefont{K. Kishio}}, \bibinfo{journal}{Phys. Rev. Lett.} \textbf{\bibinfo{volume}{75} }\bibinfo{page}{4662--4665} (\bibinfo{year}{1995}).
\bibitem{boebinger96}\bibinfo{author}{\bibfnamefont{G. S. Boebinger}}, \bibinfo{author}{\bibfnamefont{Yoichi Ando}}, \bibinfo{author}{\bibfnamefont{A. Passner}}, \bibinfo{author}{\bibfnamefont{T. Kimura}}, \bibinfo{author}{\bibfnamefont{M. Okuya}}, \bibinfo{author}{\bibfnamefont{J. Shimoyama}}, \bibinfo{author}{\bibfnamefont{K. Kishio}}, \bibinfo{author}{\bibfnamefont{K. Tamasaku}}, \bibinfo{author}{\bibfnamefont{N. Ichikawa}} and  \bibinfo{author}{\bibfnamefont{S. Uchida}}, \bibinfo{journal}{Phys. Rev. Lett.} \textbf{\bibinfo{volume}{77} }\bibinfo{page}{5417--5420} (\bibinfo{year}{1996}).
\bibitem{ferrero-2008}\bibinfo{author}{\bibfnamefont{M. Ferrero}}, \bibinfo{author}{\bibfnamefont{P.~S. Cornaglia}}, \bibinfo{author}{\bibfnamefont{L. De Leo}}, \bibinfo{author}{\bibfnamefont{O. Parcollet}}, \bibinfo{author}{\bibfnamefont{G. Kotliar}} and  \bibinfo{author}{\bibfnamefont{A. Georges}}, \bibinfo{journal}{arXiv.org:0806.4383} (\bibinfo{year}{2008}).
\bibitem{haydock1975}\bibinfo{author}{\bibfnamefont{R. Haydock}}, \bibinfo{author}{\bibfnamefont{V. Heine}} and  \bibinfo{author}{\bibfnamefont{M. J. Kelly}}, \bibinfo{journal}{J. Phys. C} \textbf{\bibinfo{volume}{8} }\bibinfo{page}{2591} (\bibinfo{year}{1975}).
\bibitem{poilblanc}\bibinfo{author}{\bibfnamefont{D. Poilblanc}} and  \bibinfo{author}{\bibfnamefont{D. J. Scalapino}}, \bibinfo{journal}{\PRB} \textbf{\bibinfo{volume}{66} }\bibinfo{page}{052513} (\bibinfo{year}{2002}) \bibinfo{note}{}.
 \end{thebibliography}
\end{document}